\documentclass[twocolumn,epjc3]{svjour3}          

\smartqed  
\usepackage{amsmath}
\usepackage{widetext}
\usepackage{euscript}

\usepackage{mathtools}
\usepackage{graphicx}
\usepackage{mathptmx}      
\usepackage{amssymb}
\usepackage{flushend}
\usepackage[numbers,sort&compress]{natbib}
\usepackage[colorlinks,citecolor=blue,urlcolor=blue,linkcolor=blue]{hyperref}
\usepackage{caption}
\usepackage{subcaption}

\journalname{Eur. Phys. J. C}

\begin{document}

\title{Studies on gluon evolution and geometrical scaling in kinematic constrained unitarized BFKL equation: application to high-precision HERA DIS data}

\author{P. Phukan\thanksref{e1,addr1}
        \and
        M. Lalung\thanksref{e2,addr1}
        \and
        J. K. Sarma\thanksref{e3,addr1} 
}

\thankstext{e1}{e-mail: pragyanp@tezu.ernet.in}
\thankstext{e2}{e-mail: mlalung@tezu.ernet.in}
\thankstext{e3}{e-mail: jks@tezu.ernet.in}

\institute{Department of Physics, Tezpur University, Sonitpur, Assam, India\label{addr1}
}

\date{}

\maketitle

\begin{abstract}
We suggest a modified form of a unitarized BFKL equation imposing the so-called kinematic constraint on the gluon evolution in multi-Regge kinematics. The underlying nonlinear effects on the gluon evolution are investigated by solving the unitarized BFKL equation analytically. We obtain an equation of the critical boundary between dilute and dense partonic system, following a new differential geometric approach and sketch a phenomenological insight on geometrical scaling. Later we illustrate the phenomenological implication of our solution for unintegrated gluon distribution $f(x,k_T^2)$ towards exploring high precision HERA DIS data by theoretical prediction of proton structure functions ($F_2$ and $F_L$) as well as double differential reduced cross section $(\sigma_r)$. The validity of our theory in the low $Q^2$ transition region is established by studying virtual photon-proton cross section in light of HERA data.
\end{abstract}
\setcounter{tocdepth}{5}
\tableofcontents

\section{Introduction}\label{intro}
In perturbative QCD the two well-known parton evolution equations DGLAP \cite{63,64,65} and BFKL \cite{66,67} serve as the basic tools for prediction of parton distribution functions (PDFs) on their respective kinematic domains. The DGLAP approach is valid for the large interaction scale $Q^2$ since it sums higher order $\alpha_s$ contributions enhanced by the leading logarithmic powers of $\ln Q^2$. On the other hand, BFKL evolution deals with the small-$x$ and semi-hard $Q^2$ kinematic region by summing the leading logarithmic contributions of $(\alpha_s \ln 1/x)^n$. However, both the linear evolutions turn out to be inadequate to comply with the unitary bound or the conventional Froissart bound \cite{68}. To restore the unitarity, a higher order pQCD correction is required to shadow the rapid parton growth which eventually leads to the saturation phenomena at very small-$x$. In this respect, several nonlinear equations have been proposed in recent years to entertain this shadowing correction in DGLAP and BFKL equation.
\par In pQCD interactions, the origin of the shadowing correction is primarily considered as the gluon recombination ($g+g\rightarrow  g$) which is simply the inverse process of gluon splitting ($g\rightarrow g+g$). The modification of gluon recombination to the original DGLAP equation was first performed in GLR-MQ equation \cite{16,75,86,87} proposed by Gribov-Levin-Ryskin and Mueller-Qui. It is considered that the interferant cut diagrams of the recombination amplitude are the origin  for the negative shadowing effects in the gluon recombination processes \cite{59,69}. The GLR-MQ equation was originally derived using AGK-cutting rules \cite{71} to compute the contributions from these interference processes. However, GLR-MQ does not possess any antishadowing correction term since in this equation momentum conservation is violated in the gluon recombination processes. The shadowing and antishadowing effects are expected to happen in a complementary fashion which in fact ensure the momentum conservation during the process of gluon recombination \cite{70}. In consequence of gluon recombination, a part of gluon momentum lost due to shadowing is believed to be compensated in terms of new gluons with comparatively larger $x$. This causes an enhanced density of larger momentum gluons responsible for the antishadowing effect. In this respect, to take care of momentum conservation and antishadowing effects in gluon evolution, a modified DGLAP equation was proposed \cite{69} to replace the GLR-MQ equation. In the derivation of the modified DGLAP equation, the author has used TOPT-cutting rules based on time ordered perturbation theory (TOPT) to calculate the contribution of the gluon recombination, instead of using AGK-cutting rules unlike in the derivation of GLR-MQ equation. This equation naturally comes with two separate nonlinear terms, quadratic in gluon distribution function which are identified as the positive antishadowing and negative shadowing components in the gluon evolution. This suggests that negative shadowing effects are suppressed by antishadowing effects in the gluon evolution process.
\par   However, above mentioned modified DGLAP approach cannot predict the evolution of unintegrated gluon distribution because DGLAP is based on collinear factorization scheme while the later deals with $k_T$-factorization. In this respect, a new unitarized BFKL equation was proposed by Ruan, Shen, Yang and Zhu \cite{1} incorporating both shadowing and antishadowing correction. This modified BFKL (MD-BFKL) equation allows one to study antishadowing effects in unintegrated gluon distribution platform.
\par However, there are several other BFKL based nonlinear evolution equations that consider corrections from gluon fusion. One of the most widely studied models is Balitsky-Kovchegov (BK) equation \cite{10,11} which is originally derived in terms of scattering amplitude. It can be explained by the dipole model in which the nonlinear term comes from the dipole splitting while the screening effect origins from the double scattering of the dipole on the target \cite{85}. The solution of the BK equation suggests an intense shadowing leading to the so-called saturation of the scattering amplitude showing a complete flat spectrum. However, like GLR-MQ equation the BK equation is also irrelevant to the gluon antishadowing. On the other hand, besides inclusion of antishadowing correction term, the nonlinear terms in MD-BFKL equation hold a simple form which is quadratic in unintegrated gluon distribution as well as running coupling constant with some constant coefficients. These unique features of MD-BFKL equation motivate our current studies on small-$x$ physics.
\par A numerical solution of MD-BFKL equation suggests a sizable impact of antishadowing effect on gluon evolution particularly in the pre-asymptotic regime \cite{1}. In the literature \cite{73} the phenomenology of the MD-BFKL equation is extended to study the implicit nuclear shadowing and antishadowing effects. In our current work, we try to construct a modified evolution equation by implanting the so-called kinematic constraint or consistency constraint on the MD-BFKL equation. The kinematic constraint (see Fig.~\ref{2x} ) is implemented in different forms:
\begin{align}
	\label{1.1}
	q_T^2&<\frac{k_T^2}{z} \text{     } \text{ LDC \cite{6,74}},\\
	\label{1.2}
	q_T^2&<\frac{(1-z)k_T^2}{z}\text{  \cite{3}},\\
	\label{1.3}
	k_T^{'2}&<\frac{k_T^2}{z}\text{      }\text{  BFKL \cite{3}}.
\end{align}
This constraint arises as a consequence of BFKL multi-Regge kinematics which suggests the exchanged gluon virtuality is dominated by transverse components while the longitudinal components of the gluon momentum are required to be small i.e. $k^{'2}\approx k_T^{'2}$. The kinematic constraint gives an implicit cutoff on $k_T^{'2}$ as depicted by \eqref{1.3}. The inequality \eqref{1.3} can be considered as a special case of \eqref{1.1} recalling the fact that for a given value of $k_T^2$, a high $q_T^2$ implies an equally high $k_T^{'2}$ \cite{3}. Although there are other constraints coming from energy-momentum conservation \cite{7}, this bound is considerably tighter than the later \cite{3}. Besides the introduction of this cutoff on the upper limit of integration found to preserve the scale invariance of the BFKL equation.
\par Our primary goal of this work is to investigate the nonlinear effects on gluon evolution in terms of MD-BFKL equation supplemented by the kinematic constraint. Our studies are particularly focused on the near saturation region where shadowing effects are dominant over antishadowing effects. The content of the paper is organized as follows. In Sect. \ref{analytic} we present our kinematic constraint improved MD-BFKL equation starting from the construction of the equation and a brief discussion on some of its main features (Sect. \ref{analytic1}). We suggest an analytical solution to this equation and sketch a particular solution in terms of $x$ and $k_T^2$ evolution followed by three-dimensional realization of gluon evolution in $x\text{-}k_T^2$ phase space (Sect. \ref{analytic2}). We have also extracted collinear gluon distribution $xg(x,Q^2)$ from unintegrated gluon distribution $f(x,k_T^2)$ and compared our prediction with that of global parameterization groups NNPDF 3.1sx and CT 14. In Sect. \ref{critical} we suggest a differential geometric approach towards finding the equation of the critical boundary between high and low gluon density regime. In this section, a phenomenological insight of the geometrical scaling is discussed as well.
\par In Sect. \ref{hera} we explore the phenomenological implication of our solution for unintegrated gluon distribution towards prediction of DIS structure function $F_2$ and longitudinal structure function $F_L$ at HERA. We show the comparisons of reduced cross sections (Sect. \ref{hera1}) as well as virtual photon-proton cross sections in the transition region (Sect. \ref{hera2}) with HERA H1 and ZEUS data. Finally, in Sect. \ref{con} we summarize and outline our conclusion as well as possible future prospects.

\section{Nonlinear effects in kinematic constraint improved MD-BFKL equation}\label{analytic}

\subsection{Construction of kinematic constraint improved MD-BFKL}\label{analytic1}
\begin{figure*}[tbp]
	\centering 
	\includegraphics[width=.21\textwidth,clip]{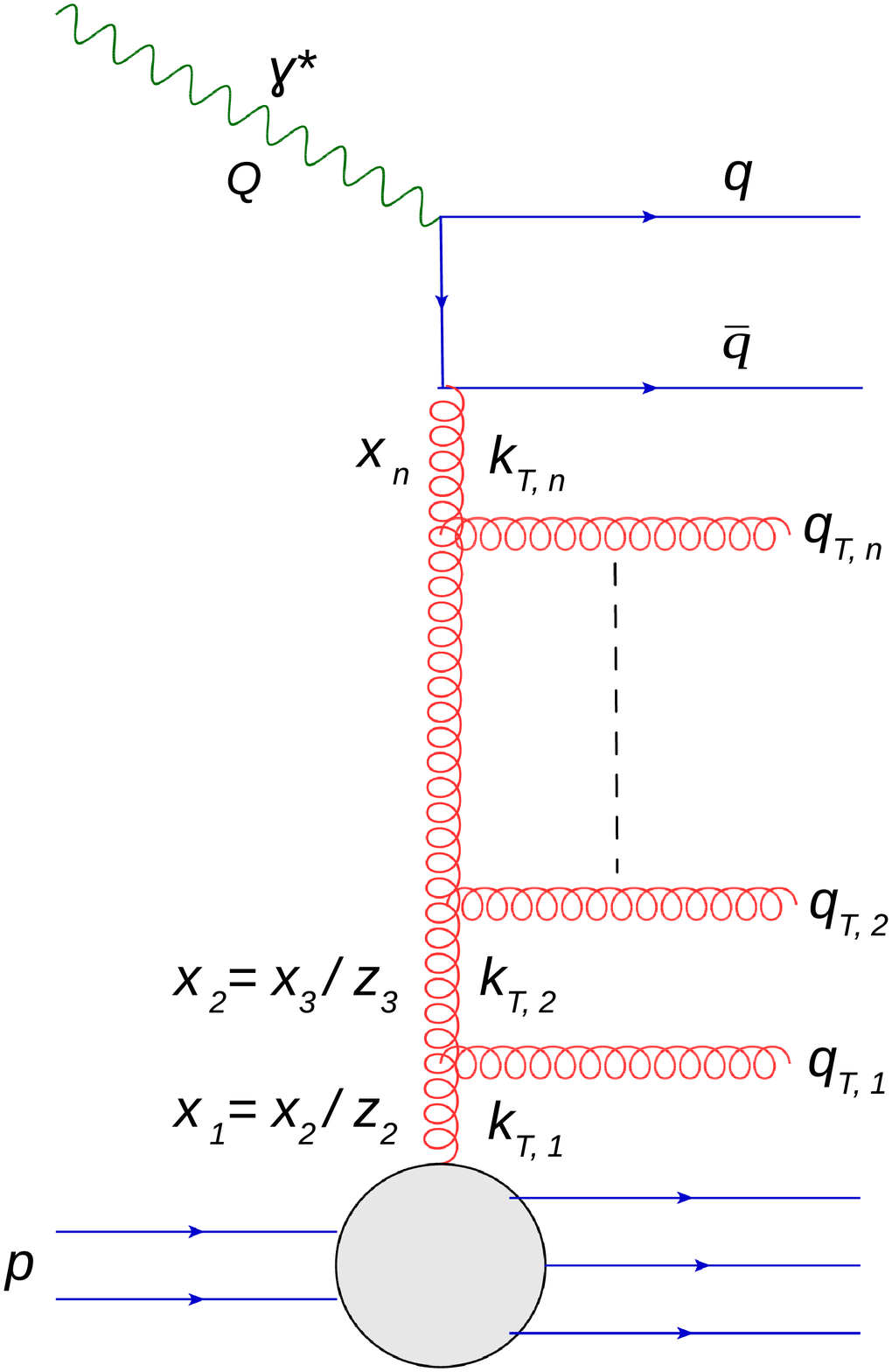}\hspace{5mm}
	\includegraphics[width=.23\textwidth,clip]{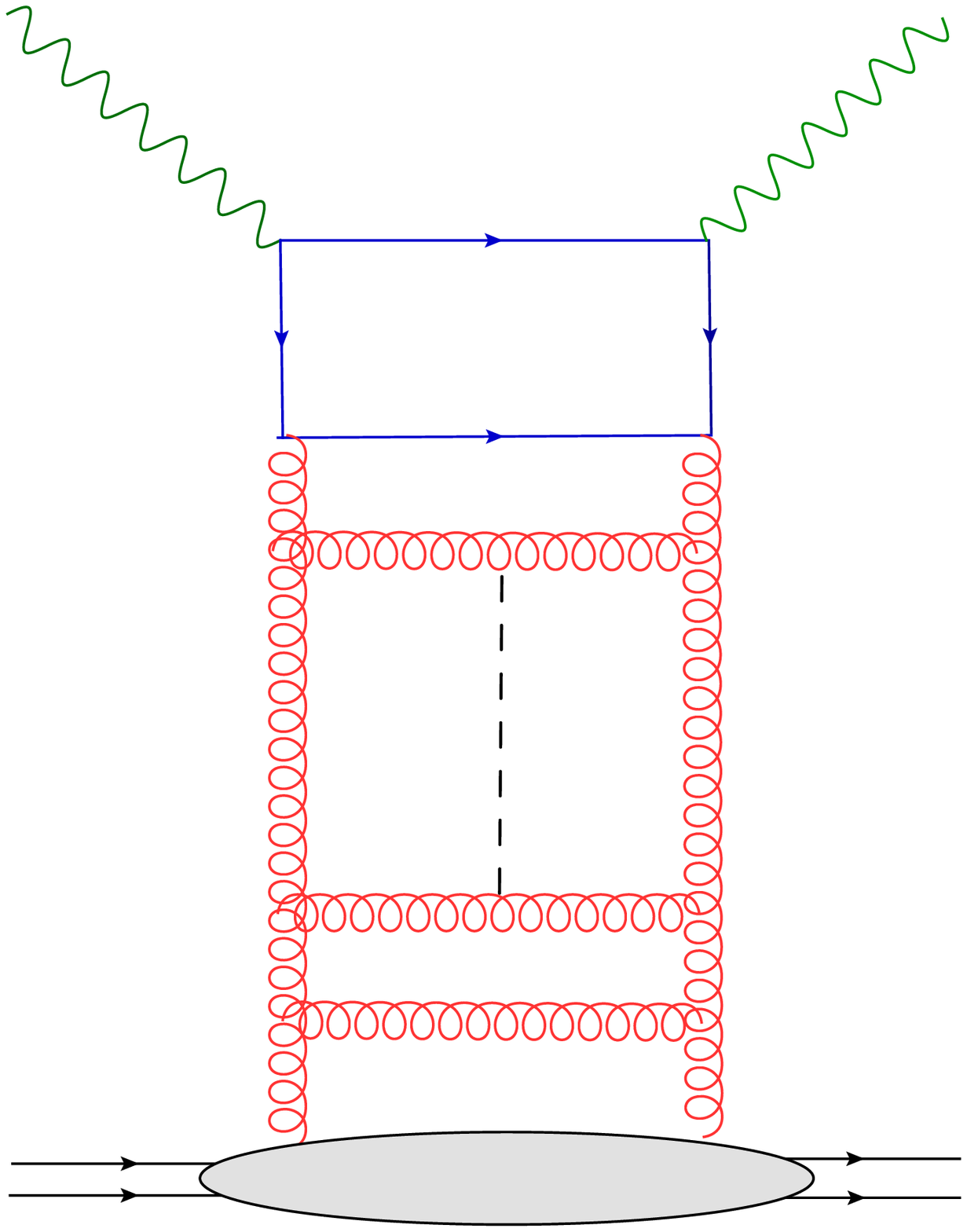}\hspace{5mm}
	\includegraphics[width=.15\textwidth,clip]{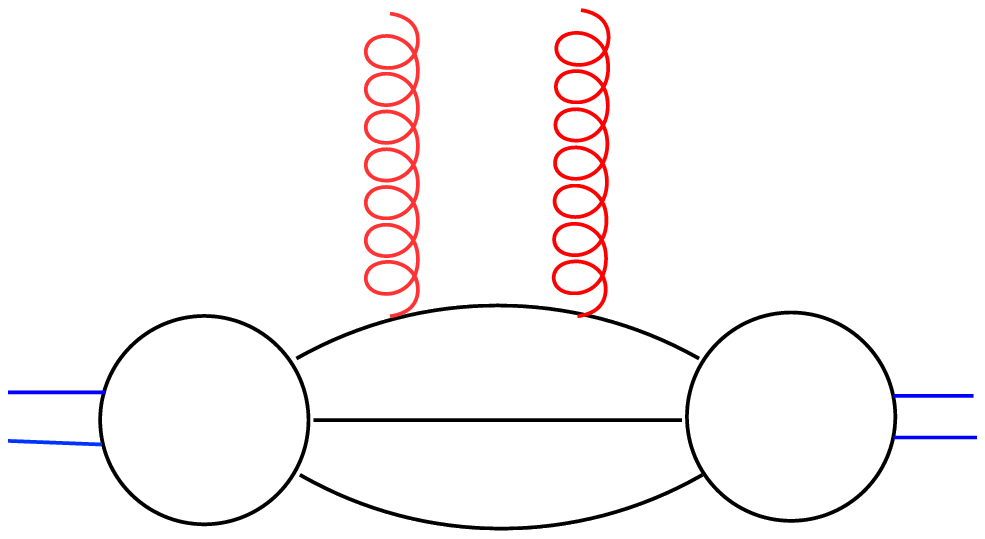}\hspace{5mm}
	\includegraphics[width=.15\textwidth,clip]{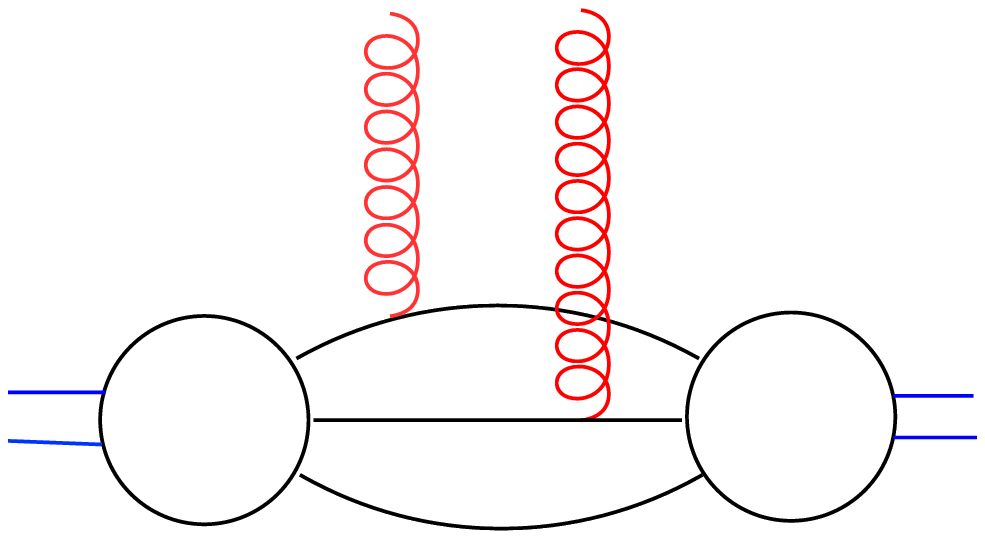}
	
	\caption{\label{2x} From left: (a) chain of sequential gluon emission which forms the basis of BFKL equation. On squaring the amplitude of (a) the ladder diagram (b) is generated which when summed gives the BFKL kernel. (c) and (d) are two simple examples of the inhomogeneous driving term of \eqref{4} which correspond to the shaded region of gluon-virtual photon coupling in (b). }
\end{figure*}

The modified BFKL (MD-BFKL) equation reads \cite{1},
\begin{equation}
	\label{1}
	\begin{split}
		&-x\frac{\partial f \left(x, k_T^2\right)}{\partial x}=\\&\frac{\alpha_s N_c k_T^2}{\pi}\int_{k_{T_{\min }}^{'2}}^{\infty}\frac{dk_T^{'2}}{k_T^{'2}}\left[\frac{f (x, k_T^{'2})-f \left(x, k_T^2\right)}{|k_T^{'2}-k_T^2|} + \frac{f \left(x, k_T^2\right)}{\sqrt{k_T^4+4k_T^{'4}}}\right]\\
		&-\frac{36\alpha_s^2}{\pi k_T^2 R^2}\frac{N_c^2}{N_c^2-1}f^2(x,k_T^2)+\frac{18\alpha_s^2}{\pi k_T^2 R^2}\frac{N_c^2}{N_c^2-1}f^2(\frac{x}{2},k_T^2),
	\end{split}
\end{equation}
where $f \left(x, k_T^2\right)$ denotes the gluon distribution unintegrated over the transverse momentum of gluon $k_T$ and $k_{T_{\min }}^{'2}$ is the infrared cutoff of the evolution. The first part of \eqref{1} linear in $f(x,k_T^{'2})$ (or $f(x,k_T^2)$) is the BFKL kernel at leading logarithmic of $1/x$ (LLx) accuracy. A diagrammatic representation for probing gluon content of the proton at high photon virtuality $Q^2$ is sketched in Fig.~\ref{2x} (a). The BFKL kernel corresponds to the sum of gluon ladder diagram Fig.~\ref{2x}(b) generated by squaring the amplitude of Fig.~\ref{2x}(a). The 1st term in the BFKL kernel, involving $f(x,k_T^{'2})$   corresponds to diagrams with real gluon emission while the second term takes care of diagrams with virtual corrections. Note that the apparent singularity that is observed at $k_T^{'2} =k_T^2$ cancels between real and virtual contributions. 

\par
\begin{figure}[tbp]
	\centering 
	\includegraphics[width=.27\textwidth,clip]{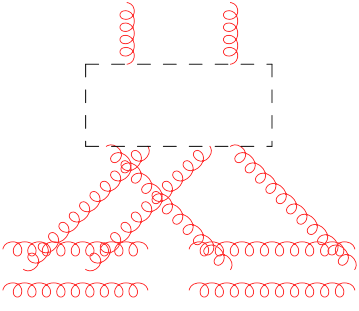}	
	\caption{\label{3x} QCD cut diagram representing nonlinear terms in \eqref{1} with $4\rightarrow 2$ gluon recombination kernel (dashed box). The dashed box includes all possible pQCD diagrams which couple 4 gluons to 2 gluons.}
\end{figure}

 The quadratic terms in \eqref{1} viz.  
$\frac{\partial f_{\text{shad}} }{\partial \ln{1/x}}=-\frac{36\alpha_s^2}{\pi k_T^2 R^2}\frac{N_c^2}{N_c^2-1}f^2$
and
$\frac{\partial f_{\text{antishad}} }{\partial \ln{1/x}}$ $=\frac{18\alpha_s^2}{\pi k_T^2 R^2}\frac{N_c^2}{N_c^2-1}f^2$ depict shadowing and antishadowing correction to the original BFKL equation respectively. The factor $\pi R^2$ represents the transverse area populated by gluons. The radius parameter $R$ arises in the QCD cut diagram coupling 4 gluons to 2 gluons (see Fig.~\ref{3x}), the value of which depends on how exactly the gluon ladders couple to hadron. If the gluon ladder couples to two different constituents of hadron, $R$ is characterized by hadronic radius i.e. $R\approx R_H=5 \text{ GeV}^{-1}$ whereas if we consider the possibility of ladder coupling to the same parton then the appropriate choice of $R$ is the radius of valence quark i.e. $R=2\text{ GeV}^{-1}$. In the latter case $(R=2\text{ GeV}^{-1})$ we have "hot spots" \cite{75,72}.

\par The interpretation of the real emission term in the BFKL kernel is that we have a splitting $k^{'}\rightarrow k + q$ inside hadron resulting an infinite chain of reggeized gluons labeled as $k_1,k_2,k_3,\text{...}k_n$ (Fig.~\ref{2x}(a)). In high energy limit, the longitudinal components of the gluon momentum are strongly ordered while there is no ordering on the transverse components of the gluon momentum i.e.
	\begin{align}
		\label{2.2a}
		&x_1\ll x_2\ll x_3\text{.....}\ll x_{n-1}\ll x_n,\\
		\label{2.2b}
		&k_{1 T}\sim k_{2 T}\sim k_{3 T}\sim \text{...}.\sim k_{n-1 T}\sim k_{\text{nT}}.
	\end{align}
In addition, the small-$x$ regime where the BFKL is valid, the gluon virtuality along the chain must be dominated by the transverse components of the gluon momentum, 
\begin{align}
	\label{2.3}
	k^2=2 k^+ k^--k_T^2\approx k_T^2.
\end{align}\\
The above kinematics corresponding to \eqref{2.2a},\eqref{2.2b} and \eqref{2.3} is referred as multi-Regge kinematics. The inequality \eqref{2.2a} implies $z=\frac{x}{\frac{x}{z}}=\frac{x_{n-1}}{x_n}\ll1$ and since transverse momenta are of the same order: $k_T\simeq k_T{}^{'}$, depict a cutoff 
\begin{equation}
	\label{2.4}
	k_T{}^{'}{}^2< \frac{k_T{}^2}{z},
\end{equation}
which is so-called  kinematic constraint or consistency constraint for real gluon emission \cite{3,4,5,6}. The BFKL kernel in \eqref{1} is at leading logarithmic $1/x$ (LLx) accuracy. However, higher order corrections to the BFKL equation are already been evaluated up to NLLx accuracy, which turns out to be quite large \cite{2}. Implementation of the constraint \eqref{2.4} on the evolution of BFKL based equations makes the theory more realistic in the sense that this ensures the participation of only the LLx part of higher order correction in the evolution.  This, in fact, portrays the importance NLLx correction in BFKL kernel.

\par Recalling that BFKL equation can be written as an integral equation of  $f(x,k_T^2)$ \cite{15,16}, the kinematic  constraint \eqref{2.4} can be imposed onto the real emission part of the BFKL kernel as follows
\begin{equation}
	\label{2.5}
	\begin{split}
		\int_{x}^{1}\frac{dz}{z}\int _{k_{T_{\min }}^{'2}}\frac{dk_T^{'2}}{k_T^{'2}}\bigg[&\frac{\Theta \left(\frac{k_T^2}{z}-k_T^{'2}\right)f\left(\frac{x}{z},k_T^{'2}\right)-f\left(\frac{x}{z},k_T^2\right)}{\left| k_T^{'}{}^2-k_T^2\right| }\\
		&+\frac{f\left(\frac{x}{z},k_T^2\right)}{\sqrt{k_T^4+4k_T^{'4}}}\bigg],
	\end{split}
\end{equation}
where the heaviside function $\Theta \left(\frac{k^2}{z}-k^{'}{}^2\right)$ in \eqref{2.5} serves the purpose of the kinematic cutoff. The upper limit of the integration over $k_T^{'2}$ is implicit in $\Theta$. The BFKL kernel gains one more degree of freedom after implementation of kinematic constraint for real emission. Now expressing  the MD-BFKL equation \eqref{1} in terms of the KC improved BFKL kernel \eqref{2.5} yields
\begin{equation}
	\label{4}
	\begin{split}
		&f\left(x,k_T^2\right)=f^{(0)}\left(x,k_T^2\right)+\\&\frac{\alpha _s k_T^2 N_c }{\pi }{\int _x^1\frac{dz}{z}\int _{k_{T_{\min }}^{'2}}}\frac{dk_T^{'2}}{k_T^{'2}}\bigg[\frac{\Theta \left(\frac{k_T^2}{z}-k_T^{'2}\right)f\left(\frac{x}{z},k_T^{'2}\right)-f\left(\frac{x}{z},k_T^2\right)}{\left| k_T^{'2}-k_T^2\right| }\\
		&+\frac{f\left(\frac{x}{z},k_T^2\right)}{\sqrt{k_T^4+4k_T^{'4}}}\bigg]-\frac{36 \alpha _s^2}{\pi  k_T^2 R^2}\frac{N_c^2}{N_c^2-1}\int _x^1\frac{dz}{z}f^2\left(x,k_T^2\right)\\
		&+\frac{18 \alpha _s^2}{\pi  k_T^2 R^2}\frac{N_c^2}{N_c^2-1}\int _x^1\frac{dz}{z}f^2\left(\frac{x}{2},k_T^2\right).
	\end{split}
\end{equation}
The above equation \eqref{4} is the integral form of our KC improved MD-BFKL equation. The inhomogeneous driving term $f^{(0)}(x,k_T^2)$ depicts gluon-proton coupling corresponding to the shaded region in Fig.~\ref{2x}(b). The Fig.~\ref{2x}(c, d) represents two simple possible contributions to $f^{0}(x,k_T^2)$ which indicates radiation of gluons from valence quark.

\par To sketch an integro-differential form of KC improved MD-BFKL equation out of \eqref{4} we differentiate \eqref{4} w.r.t. $\ln(1/x)$, then using properties of $\Theta$ and Dirac-$\delta$ function, to be specific  $\Theta '(t)=\delta  (t)$ and $f( t)\delta  (t-a)=f(a)\delta  (t-a)$ and doing some algebra one can show

\begin{equation}
	\label{5}
	\begin{split}
		\frac{\partial}{\partial \ln\frac{1}{x}}\int_{x}^{1}\frac{\text{d}z}{z}\Theta \left(\frac{k_T^2}{z}-k_T^{'2}\right)&f(x,k_T^{'2})\longrightarrow          \Theta(k_T^2-k_T^{'2})f(x,k_T^{'2})\\
		&+\Theta(k_T^{'2}-k_T^{2})f\left(\frac{k_T^{'2}}{k_T^{2}}x,k_T^{'2}\right).
	\end{split}
\end{equation}
Above prescription allows one to express \eqref{4} in the following integro-differential form:
\begin{align}
	\label{2.8}
	\begin{split}
		&-x\frac{\partial f\left(x,k_T^2\right)}{\partial x}=\frac{\alpha _s k_T^2 N_c }{\pi }{\int _{k_{T_{\min }}^{'2}}}\frac{dk_T^{'2}}{k_T^{'2}}\\&\times\bigg[\frac{\Theta \left(k_T^2-k_T^{'2}\right)f\left(x,k_T^{'2}\right)+\Theta (k_T^{'2}-k_T^{2})f\left(\frac{k_T^{'2}}{k_T^{2}}x,k_T^2\right)}{\left| k_T^{'2}-k_T^2\right| }\\&-\frac{f(x,k_T^2)}{|k_T^{'2}-k_T^2|}+\frac{f\left(x,k_T^2\right)}{\sqrt{k_T^4+4k_T^{'4}}}\bigg]-\frac{36 \alpha _s^2}{\pi  k_T^2 R^2}\frac{N_c^2}{N_c^2-1}f^2\left(x,k_T^2\right)\\&+\frac{18 \alpha _s^2}{\pi  k_T^2 R^2}\frac{N_c^2}{N_c^2-1}f^2\left(\frac{x}{2},k_T^2\right).
	\end{split}
\end{align}
In derivation of \eqref{2.8} we have neglected the term $x\frac{\partial f^{(0)}}{\partial x}$. This is justified in the sense that $x\frac{\partial f^{(0)}}{\partial x}$ is much less singular than $x\frac{\partial f}{\partial x}$ at small-$x$. Moreover if we consider $f^{(0)}$ is particularly contributed by the diagrams shown in Fig.~\ref{2x} (c, d) then $f^{(0)}$ would be independent of $x$ (or, $\frac{\partial f^{(0)}}{\partial x}=0$ ) at small-$x$ limit \cite{6,15}.

\par To simplify the distribution function $f(\frac{k_T^{'2}}{k_T^{'2}}x,k_T^{2})$ corresponding to real emission term in \eqref{2.8} we have incorporated Regge like behavior of gluon distribution. Before the advent of pQCD, Regge theory is been considered as a very successful theory regarding the phenomenological analysis of total hadronic and  photoproduction cross sections. These non-perturbative processes can economically described by two Regge contributions namely pomerons with intercept slightly above unity (${\alpha_{P}(0)\approx 1.08}$) and Reggons with intercept $\alpha_{R}(0)\approx 0.5$. The Regge behavior corresponding to sea quark and anti-quark distributions is given by $q_{sea}\sim x^{-\alpha_{P}}$ whereas that of valence quark distribution is given by $q_{v}\sim x^{-\alpha_{R}}$. On the other hand, in pQCD also, particularly the small-$x$ region is considered to have higher possibility towards exploring Regge limit of pQCD. Note that the BFKL dynamics itself is based on the concept of pomeranchuk theorem or pomeron: the Regge-pole carrying the quantum-numbers of the vacuum. However, the BFKL pomeron (also called hard pomeron) should be contrasted with non perturbative description of pomeron (soft pomeron) in the sense that they pose relatively different magnitude of intercept. For BFKL-pomeron the intercept is
\begin{equation*}
	\alpha_{P}^{BFKL}(0) = 1 + \lambda_{\text{BFKL}},
\end{equation*}
where  $\lambda_{\text{BFKL}}= \frac{3 \alpha_s}{\pi} 28  \zeta(3)$, $\zeta$ being Reiman zeta function \cite{9}. The typical value of $\lambda_{\text{BFKL}}$ for $\alpha_s$=0.2 is $\sim 0.5$ implies $\alpha_{P}^{BFKL}(0)\approx1.5$ which is potentially large in magnitude compared to soft pomeron $(\alpha_{P}(0)=1.08)$.
\par In pQCD the small-$x$ behavior of the parton distribution is supposed to be controlled by intercept of appropriate Regge trajectory. Regge model provides parametrizations of DIS distribution functions, $f_i(x,Q^2)=A_i(Q^2)x^{-\lambda_i}$ (i=$\sum$ (singlet structure function)and g (gluon distribution)), where$\lambda_i$ is the pomeron intercept minus one $(\alpha_{P}(0)-1)$.  At small-$x$, the leading order calculations in \(\text{ln}(1/x)\) with fixed \(\alpha _s\) predicts a steep power law behavior of \(\textit{f}\left(x,k^2\right)\sim x^{-\lambda_\text{BFKL}}\) \cite{7,8}. This motivates us to consider a simple form of Regge factorization as follows,
\begin{equation}
	\label{6}
	\begin{split}
		f\left(\frac{k_T^{'2}}{k_T^2}x,k_T^2\right)&\simeq x^{-\lambda_{\text{BFKL}}}\left(\frac{k_T^{2}}{k_T^{'2}}\right)^{\lambda_{\text{BFKL}}} H(k_T^2)\\
		&= \left(\frac{k_T^2}{k_T^{'2}}\right)^{\lambda }f\left(x,k_T^2\right),
	\end{split}
\end{equation}
where we have dropped the subscript on $\lambda$ and in the rest of the text we will follow this notation. Similar way we can express 
\begin{equation}
	\label{2.10}
	\begin{split}
		f\left(\frac{x}{2},k^2\right)&\simeq 2^{\lambda }f\left(x,k^2\right).
	\end{split}
\end{equation}
This type of Regge like form considered in \eqref{6} and \eqref{2.10} is supported by various literatures \cite{76,77,78,79}. But note that Regge factorization can't be taken as good ansatz for the entire kinematic region of $x$ and $k_T^2$ \cite{80}. This type of Regge behavior is considered to be valid only in the vicinity of the saturation scale, $Q_s^2$ where scattering amplitude depends only on a single dimensionless variable, $Q^2/Q_s^2$. The Regge theory is applicable if the quantity invariant mass, $W$ $(=\sqrt{Q^2(1-x)/x})$ is much greater than all other variables. Therefore, we expect it to be valid if $x$ is enough small, for any value of $Q^2$.

\par Now recalling $\Theta(t)=1-\Theta(-t)$ and substituting \eqref{6} and \eqref{2.10} in \eqref{2.8} we get
\begin{equation}
	\label{7}
	\begin{split}
		&-x\frac{\partial f\left(x,k_T^2\right)}{\partial x}=\\&\frac{\alpha _s N_c k_T^2 }{\pi }\int _{k_{T_{\min }}^{'2}}\frac{dk'^2}{k_T^{'}{}^2}\bigg[\Theta \left(k_T^2-k_T^{'2}\right)\frac{1-\left(\frac{k_T^2}{k_T^{'2}}\right)^{\lambda }}{\left| k_T^{'2}-k_T^2\right| }f\left(x,k_T^{'2}\right)\\
		&+\left(\frac{k_T^2}{k_T^{'2}}\right)^{\lambda }\frac{f\left(x,k_T^2\right)}{\left| k_T^{'2}-k^2\right|}-\frac{f\left(x,k_T^2\right)}{\left| k_T^{'2}-k^2\right|} +\frac{f\left(x,k_T^2\right)}{\sqrt{k_T^4+4k_T^{'4}}}\bigg]\\
		&-\frac{36 \alpha _s^2}{\pi  k_T^2 R^2}\frac{N_c^2}{N_c^2-1}(1-2^{2 \lambda -1})f^2\left(x,k_T^2\right).
	\end{split}
\end{equation}
The above equation \eqref{7} is our kinematic constraint improved MD-BFKL equation in LLx accuracy.

\par Before going into in depth phenomenological analysis and consistency of the equation towards experimental result, we want to highlight two important features of the evolution equation \eqref{7}. As we follow, in the region far below saturation limit, for the canonical choice of $\lambda\sim0.5$, the quadratic term in \eqref{7} tends to vanish since $(1-2^{2\lambda-1}) \rightarrow0$. This indicates that below saturation region both the contribution from shadowing correction $\frac{\partial f_{\text{shad}}}{\partial \ln{\frac{1}{x}}}$ and antishadowing correction $\frac{\partial f_{\text{antishad}}}{\partial \ln{\frac{1}{x}}}$ coexists but seem to balance each other for which nonlinear effect becomes negligible. Therefore, in the below saturation regime, we can revert  \eqref{7} back to the original BFKL equation choosing $\lambda = 0.5$.

On the other hand, in the vicinity of saturation limit, our interpretation of gluon distribution for the nonlinear term has to be reviewed. In this region, the gluon distribution becomes flat which makes our Regge factorization for the nonlinear term in  \eqref{2.10} invalid. Rather essentially, we should take the approximation
\begin{equation*}
	f\left(\frac{x}{2},k_T^2\right) \simeq f(x,k_T^2),
\end{equation*} 
which is supposed to justify our understanding of the saturation region. Taking this approximation in \eqref{2.8} one can arrive at
\begin{equation}
	\label{8}
	\begin{split}
		&-x\frac{\partial f\left(x,k_T^2\right)}{\partial x}\\
		&=\frac{N_c \alpha _sk_T^2 }{\pi }\int _{k_{T_{\min }}^{'2}}\frac{dk'^2}{k_T^{'}{}^2}\bigg[\Theta \left(k_T^2-k_T^{'2}\right)\frac{1-\left(\frac{k_T^2}{k_T^{'2}}\right)^{\lambda }}{\left| k_T^{'2}-k_T^2\right| }f\left(x,k_T^{'2}\right)\\
		&+\left(\frac{k_T^2}{k_T^{'2}}\right)^{\lambda }\frac{f\left(x,k_T^{'2}\right)}{\left| k_T^{'2}-k^2\right|}-\frac{f\left(x,k_T^2\right)}{\left| k_T^{'2}-k^2\right|} +\frac{f\left(x,k_T^2\right)}{\sqrt{k_T^4+4k_T^{'4}}}\bigg]\\
		&-\frac{18 \alpha _s^2}{\pi  k_T^2 R^2}\frac{N_c^2}{N_c^2-1}f^2\left(x,k_T^2\right).
	\end{split}
\end{equation}
Equation \eqref{8} serves the evolution of gluon near the saturation limit. Here the $\lambda$ dependence of the nonlinear term has been completely wiped up, unlike \eqref{7}. In this regime, the shadowing contribution becomes twice as that of antishadowing effect forecasting a net shadowing effect in the evolution.

\subsection{Analytical solution of KC improved MD-BFKL}\label{analytic2}

In this section, we present an analytical solution of our KC improved MD-BFKL equation, particularly in the saturation region where shadowing effect is the dominant one. Since our calculations are limited to fixed strong coupling $\alpha_s$, so do fix $\lambda$, therefore,  the solution for both \eqref{7} and \eqref{8} will exhibit the same form  only differ by the coefficients of their respective quadratic terms.

\par Recalling that there is no ordering for transverse momenta $k_T^{'}\backsimeq k_T$ in BFKL multi-Regge kinematics, this allows us to write the gluon distribution in Taylor series,
\begin{equation}
	\label{9}
	f\left(x,k_T^{'2}\right)=f\left(x,k_T^2\right)+\frac{\partial f\left(x,k_T^2\right)}{\partial k_T^2}\left(k_T^{'2}-k_T^2\right) +  \EuScript{O}\left(k_T^{'2}-k_T^2\right),
\end{equation}
where  $\EuScript{O}\left(k_T^{'2}-k_T^2\right)$ denotes the higher order terms. Above series is a convergent series in $\left(k_T^{'2}-k_T^2\right)$ as no ordering of the transverse momenta in BFKL kinematics $k_T^{'}- k_T\backsimeq 0$ implies the higher order terms  $\EuScript{O}\left(k_T^{'2}-k_T^2\right)\rightarrow 0$, thereby ensures the higher order terms to become insignificant and can be neglected. Thus this assumption would hold good as long as no ordering condition of the transverse momenta in BFKL kinematics is concerned.  This type of series expansion of distribution function is well supported in the literature \cite{12}.
\par Now neglecting these higher order terms  we can express \eqref{8} as
\begin{equation}
	\label{10}
	\begin{split}
		-x\frac{\partial f\left(x,k_T^2\right)}{\partial x}=&\xi(k_T^2) \frac{\partial f\left(x,k_T^2\right)}{\partial k_T^2}
		+\zeta(k_T^2) f\left(x,k_T^2\right)\\&-\Delta(k_T^2) f^2\left(x,k_T^2\right),
	\end{split}
\end{equation}
where
\begin{equation}
	\label{11}
	\begin{split}
		\xi(k_T^2)=\frac{\alpha _s N_c }{\pi }k_T^2\bigg[&\int _{k_{T_{\min }}^{'2}}^{k_T^2}\frac{dk_T^{'2}}{k_T^{'2}}\frac{k_T^{'2}-k_T^2}{\left| k_T^{'2}-k_T^2\right| }\\&+\int _{k_T^{2}}^\infty\frac{dk_T^{'2}}{k_T^{'2}}\left(\frac{k_T^2}{k_T^{'2}}\right)^{\lambda }\frac{k_T^{'2}-k_T^2}{\left| k_T^{'2}-k_T^2\right| }\bigg],\\
	\end{split}
\end{equation}

\begin{equation}
	\label{12}
	\begin{split}
		\zeta(k_T^2)&=\frac{\alpha _s N_c }{\pi }k_T^2\bigg[-\int _{k_T^2}^{\infty }\frac{dk_T^{'2}}{k_T^{'2}}\frac{1}{\left| k_T^{'2}-k_T^2\right| }\\&+\int _{k_T^2}^\infty\frac{dk_T^{'2}}{k_T^{'2}}\left(\frac{k_T^2}{k_T^{'2}}\right)^{\lambda }\frac{1}{\left| k_T^{'2}-k_T^2\right| }+\int _{k_{T_{\min }}^{'2}}^{\infty }\frac{dk_T^{'2}}{k_T^{'2}}\frac{1}{\sqrt{k_T^4+4k_T^{'4}}}\bigg],
	\end{split}
\end{equation}
and $\Delta(k_T^2)=\frac{18 \alpha _s^2}{\pi  k_T^2 R^2}\frac{N_c^2}{N_c^2-1}$.
\par Note that the integrals in \eqref{12}, $I_1=\int _{k_T^2}^{\infty }\frac{dk_T^{'2}}{k_T^{'2}}\frac{1}{\left| k_T^{'2}-k_T^2\right| }$ and $I_2=\int _{k_T^2}\frac{dk_T^{'2}}{k_T^{'2}}\left(\frac{k_T^2}{k_T^{'2}}\right)^{\lambda }\frac{1}{\left| k_T^{'2}-k_T^2\right| }$ are improper since both blow-up at the lower limit $k_T^2$.
To get rid of the singularity in $I_1$ we have performed some angular integral prescription. To be specific we have used the standard trigonometric integral

\begin{equation}
	\label{13}
	\int_{0}^{2\pi}\frac{\text{d}\theta}{p+q\cos\theta}=\frac{2\pi}{\sqrt{p^2+q^2}},
\end{equation} 
valid for $p+q>0$. Replacing the variable $k_T^{'} \rightarrow k_T^{'} +k_T$ one can write
\begin{equation}
	\label{14}
	\int\frac{\text{d}^2k_T^{'}}{(k_T^{'}-k_T)^2}\frac{1}{(k_T^{'}-k_T)^2+k_T^{'2}}=\int\frac{\text{d}^2k_T^{'}}{k_T^{'2}}\frac{1}{k_T^{'2}+(k_T^{'}+k_T)^2}.
\end{equation}
Now using \eqref{13} in the r.h.s. of \eqref{14} we get
\begin{equation}
	\label{15}
	\begin{split}
		\int\frac{\text{d}k_T^{'}  k_T^{'} \text{d}\theta}{k_T^{'2}(2k_T^{'2}+k_T^2+2k_T^{'}k_T\cos\theta)}&=2\pi\int\frac{\text{d}k_T^{'} k_T^{'}}{k_T^{'2} \sqrt{4k_T^{'4}+k_T^4}}\\&=\pi\int\frac{\text{d}k_T^{'2}}{k_T^{'2} \sqrt{4k_T^{'4}+k_T^4}}.
	\end{split}
\end{equation}
Similarly using \eqref{13} in the l.h.s. of \eqref{14} we get

\begin{equation}
	\label{16}
	\begin{split}
		\int\frac{\text{d}^2k_T^{'}}{(k_T^{'}-k_T)^2}\frac{1}{(k_T^{'}-k_T)^2+k_T^{'2}} &=\pi \int \frac{dk_T^{'2}}{k_T^{'2}}\frac{1}{\left| k_T^{'2}-k_T^2\right| }\\& -\pi\int\frac{\text{d}k_T^{'2}}{k_T^{'2} \sqrt{4k_T^{'4}+k_T^4}}.
	\end{split}
\end{equation}
Equating \eqref{15} and \eqref{16} we obtain

\begin{equation}
	\label{17}
	\int\frac{dk_T^{'2}}{k_T^{'2}}\frac{1}{\left| k_T^{'2}-k_T^2\right| }= 2 \int\frac{\text{d}k_T^{'2} }{k_T^{'2} \sqrt{4k_T^{'4}+k_T^4}},
\end{equation}
which turns out to be well-behaved for both integration limit for $I_1$. On the other hand, in the limit $k_{T_{\text{min}}}^{'2} \leq k_T^{'2}\leq k_T^2$ i.e. for  not too large $k_T^{'2}$, the contribution from the longitudinal component to the gluon virtuality $k^{'2}$ becomes negligible which in turn preserves the no ordering condition of transverse momentum of BFKL kinematics i.e. $k_T^{'2}\approx k_T^2$ very strictly. Therefore, in this limit $k_{T_{\text{min}}}^{'2} \leq k_T^{'2}\leq k_T^2$, it is justified to implant a factor $\left(k_T^{2}/k_T^{'2}\right)^\lambda$ inside the integrals which will in fact make our calculation simpler without altering the underlying physics. Moreover, this will fix the infrared divergence problem of the second improper integral $I_2$ by allowing us to evaluate the integral down to $k_{T_\text{min}}^{'2}$.
\par
\begin{figure*}[tbp]
	\centering 
	\includegraphics[width=.38\textwidth,clip]{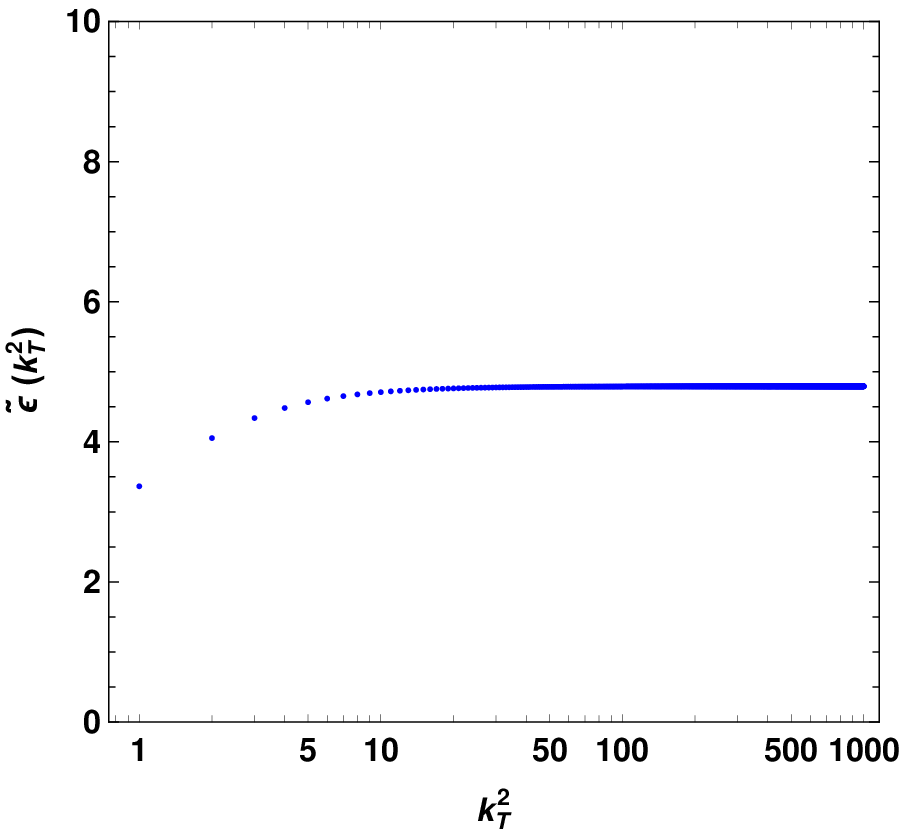}\hspace{1cm}
	\includegraphics[width=.38\textwidth,clip]{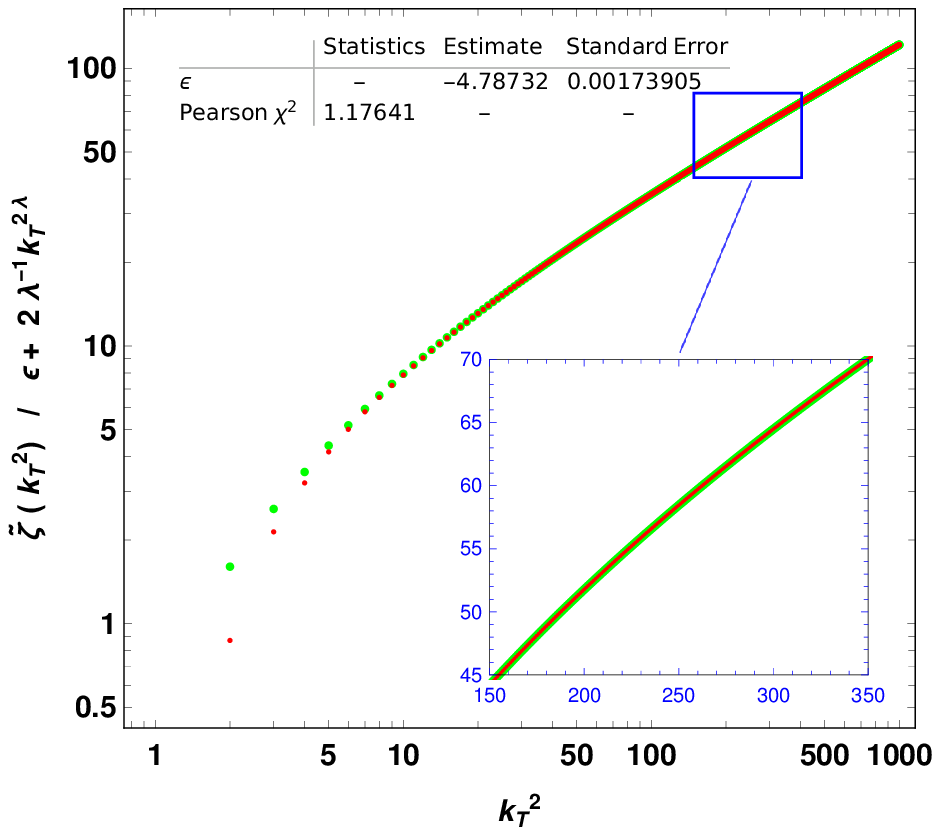}
	\caption{\label{reg} (a) $k_T^2$ dependence of the small perturbation $\tilde{\epsilon}=|\epsilon|$ corresponding to the term with the hypergeometric function in \eqref{9} (left). (b) A phenomenological calculation of $\epsilon$ using standard non-linear regression (right). Comparison between $\tilde{\zeta}=\frac{\pi}{\alpha_sN_c}\zeta$ (green dotted line) and $\epsilon+2\lambda^{-1}k_T^{2\lambda}$ (red dotted line) is shown.}
\end{figure*}

 Now considering these approximations in \eqref{11} and \eqref{12} and putting $I_1$ in \eqref{12} we obtain
\begin{align}
	\label{18}
	\begin{split}
		\xi(k_T^2)=&\frac{\alpha _s N_c k_T^2}{\pi }\bigg[\int _{k_{T_{\min }}^{'2}}^{k_T^2}\frac{dk_T^{'2}}{k_T^{'2}}\left(\frac{k_T^2}{k_T^{'2}}\right)^\lambda\frac{k_T^{'2}-k_T^2}{\left| k_T^{'2}-k_T^2\right| }\\&+\int _{k_T^{2}}^{\infty}\frac{dk_T^{'2}}{k_T^{'2}}\left(\frac{k_T^2}{k_T^{'2}}\right)^{\lambda }\frac{k_T^{'2}-k_T^2}{\left| k_T^{'2}-k_T^2\right| }\bigg]\\=&\frac{\alpha_s N_c k_T^2}{\pi}\frac{1}{\lambda}\left(2-k_T^{2\lambda}\right)\approx-\frac{\alpha_s N_c}{\pi\lambda}(k_T^2)^{\lambda+1},
	\end{split}
\end{align}

\begin{align}
	\label{19}
	\begin{split}
		\zeta(k_T^2)=&\frac{\alpha _s N_c }{\pi }k_T^2\bigg[\int _{k_{T_\text{min}}^{'2}}^{\infty }\left(\frac{k_T^2}{k_T^{'2}}\right)^\lambda\frac{dk_T^{'2}}{k_T^{'2}}\frac{1}{\left| k_T^{'2}-k_T^2\right| }
		\\&-\int _{k_{T_{\min }}^{'2}}^{\infty }\frac{dk_T^{'2}}{k_T^{'2}}\frac{1}{\sqrt{k_T^4+4k_T^{'4}}}\bigg]\\
		=&\frac{\alpha _s N_c }{\pi }\bigg[\frac{k_T^{2\lambda}}{\lambda}- 2^{2-\frac{\lambda }{2}}  \lambda^{-1}k_T^{2\lambda}\bigg(1-\frac{\sqrt{k_T^4+4}}{k_T^2}\bigg)^{\lambda /2} \, _2F_1\\&-\lambda\ln \left(\frac{k_T^2}{2}+\sqrt{1+\frac{k_T^4}{4}}\right)\bigg]\approx \frac{\alpha _s N_c }{\pi }\left(\epsilon+\frac{(k_T^2)^\lambda }{\lambda}\right),
	\end{split}
\end{align}
where $\, _2F_1$=$\, _2\text{F}_1\left(-\frac{\lambda }{2},\frac{\lambda }{2};1-\frac{\lambda }{2};\frac{k_T^2+\sqrt{k_T^4+4}}{2 k_T^2}\right)$ is a standard hypergeometric function. In the above calculations we have taken infrared cutoff $k_{T_\text{min}}^{'2}= 1 \text{ GeV}^2$ since for unified BFKL-DGLAP framework  this provided a very consistent result towards HERA DIS  data for proton structure function $F_2$ \cite{14}.  In \eqref{19} $(k_T^2)^\lambda/\lambda$ is the only dominant term since other terms are significantly small.  The contribution from the logarithmic term in \eqref{19} is negligible in comparison to the net contribution for all $k_T^2$. However, phenomenological studies shows the term involving hypergeometric function in \eqref{19} becomes irrelevant towards change in $k_T^2$ i.e. it possess a constant value $(\approx -4.79)$ (see Fig.~\ref{reg}(a)). This constant contribution is insignificant to the net contribution for $\zeta(k_T^2)$ if $k_T^2$ is enough high. But for small $k_T^2$ this contribution can not be neglected since at this range, the $k_T^{2\lambda}/\lambda$ contribution  itself is small. In consequence, this constant contribution can be treated as a small perturbation $\epsilon$ to the dominant term $k_T^{2\lambda}/\lambda$ . We have performed phenomenological determination of the constant perturbation parameter $\epsilon$ using standard non-linear regression method (see Fig.~\ref{reg}(b)).

Now we are set to solve our original equation \eqref{10} which is indeed a 1st order  semilinear (nonlinear) PDE. Our analytical approach of solving the same involves two steps: first we will express the nonlinear PDE in terms of a linear PDE then we will solve the linear PDE via. change of coordinates.
\par Substitution of $f(x, k_T^2)$  by its inverse function $\omega(x,k_T^2)$ i.e. $f=\omega^{-1}$ in \eqref{10} yields
\begin{equation}
	\label{20}
	-x\frac{\partial \omega }{\partial x}=\xi \frac{\partial \omega }{\partial k_T^2}-\zeta  \omega +\Delta,
\end{equation}
which is in fact a linear PDE in $\frac{\partial \omega}{\partial x}$, $\frac{\partial \omega}{\partial k_T^2}$
and $\omega$. Now we construct a new set of  co-ordinate  $\sigma\equiv\sigma(x, k_T^2)$ and $\eta\equiv\eta(x,k_T^2)$ such that it transforms \eqref{20} into an ODE. To be specific we define this transformation in  such a way that it is one to one for all $(x,k_T^2)$ in some set of points D in $x$-$k_T^2$ plane. This will allow us to solve \eqref{20} for $x$ and $k_T^2$ as functions of $\sigma$ and $\eta$. The only requirement is that we should ensure  the Jacobian of the transformation does not vanish  i.e.
$\text{J}=\left|
\begin{array}{cc}
\sigma_x & \eta_x  \\
\sigma_{k_T^2} & \eta_{k_T^2}  
\end{array} \right| \neq 0$ in D.\\
\par Next we want to recast \eqref{20} in $(\sigma,\eta)$ plane computing the derivatives via chain rule:
\begin{equation}
	\label{21}
	\begin{split}
		&\omega_x=\omega_\sigma\sigma_x+\omega_\eta\eta_x, \\
		&\omega_{k_T^2}=\omega_\sigma\sigma_{k_T^2}+\omega_\eta\eta_{k_T^2}.
	\end{split}
\end{equation}
Substitution of \eqref{21} into \eqref{20} yields
\begin{equation}
	\label{22}
	-(x \sigma_x+\xi\sigma_{k_T^2})\omega_\sigma-(x\eta_x+\xi\eta_{k_T^2})\omega_\eta+\zeta\omega-\Delta=0.
\end{equation}
Since we want above equation to be expressed as an ODE, we require either,
\begin{equation*}
	x \sigma_x+\xi\sigma_{k_T^2}=0 \text{    or  } x\eta_x+\xi\eta_{k_T^2}=0.
\end{equation*}
If we consider $x \sigma_x+\xi\sigma_{k_T^2}=0$, the solution $\sigma$ is constant along the curves that satisfy
\begin{equation}
	\label{23}
	\frac{\text{d}k_T^2}{\text{d}x}=\frac{\xi}{x} \implies \ln (x C_1)=\int\frac{\text{d} k_T^2}{\xi}
	\implies C_1=\frac{e^{-\frac{k_T^{-2\lambda}}{l \lambda}}}{x},
\end{equation}
where $l=-\frac{\alpha_s N_c}{\pi \lambda}$ and $C_1$ is the constant of integration. Now we can choose the new coordinates as $\sigma =\frac{e^{-\frac{k_T^{-2\lambda}}{l \lambda}}}{x}$ and $\eta=x$. Here $\text{J}=\left|
\begin{array}{cc}
\sigma_x & \eta_x  \\
\sigma_{k_T^2} & \eta_{k_T^2}  
\end{array} \right| \neq 0$ as required.
\par We rewrite \eqref{21} as
\begin{align}
	\label{24}
	\begin{split}
		&\omega_x=\omega_\sigma\sigma_x+\omega_\eta\eta_x=-\frac{e^{-\frac{k_T^{-2\lambda}}{l \lambda}}}{x^2}\omega_\sigma+\omega_\eta \text{,}\\
		&\omega_{k_T^2}=\omega_\sigma\sigma_{k_T^2}+\omega_\eta\eta_{k_T^2}=\frac{(k_T^2)^{-1-\lambda}e^{-\frac{k_T^{-2\lambda}}{l \lambda}}}{lx}.
	\end{split}
\end{align}
Now putting \eqref{24} in \eqref{22} we get
\begin{align}
	\label{25}
	-\eta \omega_\eta+\zeta\omega-\Delta=0.
\end{align}
Equation \eqref{25} is an ODE and it can be solved using standard ODE solving techniques. Now solving \eqref{25} and then transforming it to the original coordinates $(\sigma,\eta)\rightarrow (x,k_T^2)$ we get the general solution of the KC improved MD-BFKL equation, 
\begin{widetext}
	\begin{align}
	\begin{split}
	\label{26}
	&f(x,k_T^2)=\omega^{-1}(x,k_T^2)=\frac{k_T^{-2\frac{n}{l}} (-1)^{\frac{n}{\lambda  l}} l^{-\frac{n}{\lambda  l}} \lambda ^{-\frac{n}{\lambda  l}}}{\text{G}\left(\frac{e^{-\frac{k_T^{-2\lambda}}{l \lambda}}}{x}\right) x^m+\frac{18\alpha_s ^2 (-1)^{1/\lambda } \lambda ^{1/\lambda } N_c^2 l^{1/\lambda } e^{-\frac{m k_T^{-2\lambda }}{\lambda  l}} m^{-\frac{\lambda  l+l+n}{\lambda  l}} \Gamma \left(\frac{l+n}{l \lambda }+1,-\frac{k_T^{-2\lambda } m}{l \lambda }\right)}{\pi  R^2 \left(N_c^2-1\right)}},
	\end{split}
	\end{align}
\end{widetext}
where  $\Gamma \left(\frac{l+n}{l \lambda }+1,-\frac{k_T^{-2\lambda } m}{l \lambda }\right)$ is a standard Gamma function and $\text{G}\left(\frac{e^{-\frac{k^{-2\lambda}}{l \lambda}}}{x}\right)$ is an arbitrary continuously differentiable function. The parameters $m=\frac{\epsilon\alpha_sN_c }{\pi}$ and $n=\frac{\alpha_sN_c}{\pi\lambda}$ are coming from \eqref{19}. In the following section, we try to get particular solutions for the PDE applying some initial boundary condition and present an analysis of the phenomenological aspects of \eqref{26}.

\subsubsection{$x$ and $k_T^2$ evolution}

In this section, we study the small-$x$ dependence of gluon distribution by picking an appropriate input distribution at some high $x$, as well as the $k_T^2$ dependence of gluon distribution setting input distribution at some low $k_T^2$.
\par Let us rewrite \eqref{26} rearranging a bit 
\begin{widetext}
\begin{align}
	\label{27}
	\begin{split}
		\text{G}\left(\frac{e^{-\frac{k_T^{-2\lambda}}{l \lambda}}}{x}\right)=\frac{1}{N_c^2-1}\Bigg[&x^{-m}N_C^2 \bigg(\frac{18\alpha_s ^2 (-1)^{\frac{1}{\lambda }+1} \lambda ^{1/\lambda }  l^{1/\lambda } e^{-\frac{m k_T^{-2\lambda }}{\lambda  l}} m^{-\frac{\lambda  l+l+n}{\lambda  l}} \Gamma \left(\frac{l+n}{l \lambda }+1,-\frac{k_T^{-2\lambda } m}{l \lambda }\right)}{\pi  R^2}\\
		&+\frac{1}{f(x,k_T^2)} k_T^{-2\frac{n}{l}} (-1)^{\frac{n}{\lambda  l}} l^{-\frac{n}{\lambda  l}} \lambda ^{-\frac{n}{\lambda  l}}\bigg)+\frac{1}{f(x,k_T^2)}x^{-m} k_T^{-2\frac{n}{l}} (-1)^{\frac{n}{\lambda  l}+1} l^{-\frac{n}{\lambda  l}} \lambda ^{-\frac{n}{\lambda  l}}\Bigg],
	\end{split}
\end{align}
\end{widetext}First we will try to evaluate the functional form of the arbitrary differentiable function G applying some initial boundary condition on it. For $k_T^2$ evolution we set the initial distribution at $(x,k_0^2)$ where $x$ is fixed throughout the evolution. At $(x,k_0^2)$  let the argument of G be denoted by
$t=\frac{e^{-\frac{k_0^{-2\lambda}}{l \lambda}}}{x}$.

Now writing \eqref{27} for $(x,k_0^2)$ and putting $x=\frac{e^{-\frac{k_0^{-2\lambda}}{l \lambda}}}{t}$ we get
\begin{widetext}
\begin{equation}
	\label{28}
	\begin{split}
		\text{G}\left(t\right)=\frac{1}{N_c^2-1}\Bigg[&\frac{e^{\frac{mk_0^{-2\lambda}}{l \lambda}}}{t^{-m}}N_c^2 \bigg(\frac{18\alpha_s ^2 (-1)^{\frac{1}{\lambda }+1} \lambda ^{1/\lambda }  l^{1/\lambda } e^{-\frac{m k_0^{-2\lambda }}{\lambda  l}} m^{-\frac{\lambda  l+l+n}{\lambda  l}} \Gamma \left(\frac{l+n}{l \lambda }+1,-\frac{k_0^{-2\lambda } m}{l \lambda }\right)}{\pi  R^2}\\
		&+\frac{1}{F_0} k_0^{-2\frac{n}{l}} (-1)^{\frac{n}{\lambda  l}} l^{-\frac{n}{\lambda  l}} \lambda ^{-\frac{n}{\lambda  l}}\bigg)+\frac{1}{F_0}\frac{e^{\frac{mk_0^{-2\lambda}}{l \lambda}}}{t^{-m}} k_0^{-2\frac{n}{l}} (-1)^{\frac{n}{\lambda  l}+1} l^{-\frac{n}{\lambda  l}} \lambda ^{-\frac{n}{\lambda  l}}\Bigg],
	\end{split}
\end{equation}
\end{widetext}
where $F_0$ is the unintegrated gluon distribution at $(x,k_0^2)$.
Equation \eqref{28} is the functional form of G. Replacing $t$ with any other argument will give us the value of G at that particular argument. We put $t=\frac{e^{-\frac{k_T^{-2\lambda}}{l \lambda}}}{x}$ in \eqref{28} which gives us the l.h.s. of \eqref{27} and then we solve  \eqref{27} for $f(x,k_T^2)$. The solution for $k_T^2$ evolution turns out to be
\begin{widetext}
\begin{equation}
	\label{29}
	\begin{split}
		f(x,k_T^2)=\dfrac{(-1)^{\frac{n}{\lambda  l}} \big(N_c^2-1\big) k_T^{-2\frac{n}{l}} k_0^{2n/l}   \big( e^{\frac{k_T^{-2\lambda }-k_0^{-2\lambda }}{\lambda  l}}\big)^m\text{A}[x,k_T^2]}{(-1)^{\frac{1}{\lambda }+1}N_c^2 \text{B}[ k_0^2 ] +(-1)^{\frac{n}{\lambda  l}}N_c^2 F_0 \text{A}[x,k_T^2] +(-1)^{1/\lambda } N_c^2 \text{B}[k_T^2]+ (-1)^{\frac{n}{\lambda  l}+1}F_0 \text{A}[x,k_T^2]},
	\end{split}
\end{equation}
\end{widetext}
where
\begin{align*}
	\begin{split}
		&\text{A}[x,k_T^2]=\pi  R^2 x^m m^{\frac{\lambda  l+l+n}{\lambda  l}} e^{\frac{m \big(k_T^{-2\lambda }+k_0^{-2\lambda }\big)}{\lambda  l}}; \\
		&\text{B}[i]=18 \alpha_s ^2 x^m l^{\frac{l+n}{\lambda  l}} \lambda ^{\frac{l+n}{\lambda  l}} k_0^{2n/l} \Gamma\left. \left(\frac{l+n}{l \lambda }+1,-\frac{m (i)^{-\lambda }}{l \lambda }\right)\right\vert_{i=k_T^2 , k_0^2}.
	\end{split}
\end{align*}
\begin{figure*}[t]
	\centering 
	\includegraphics[trim=25 5 21.5 0,width=.46\textwidth,clip]{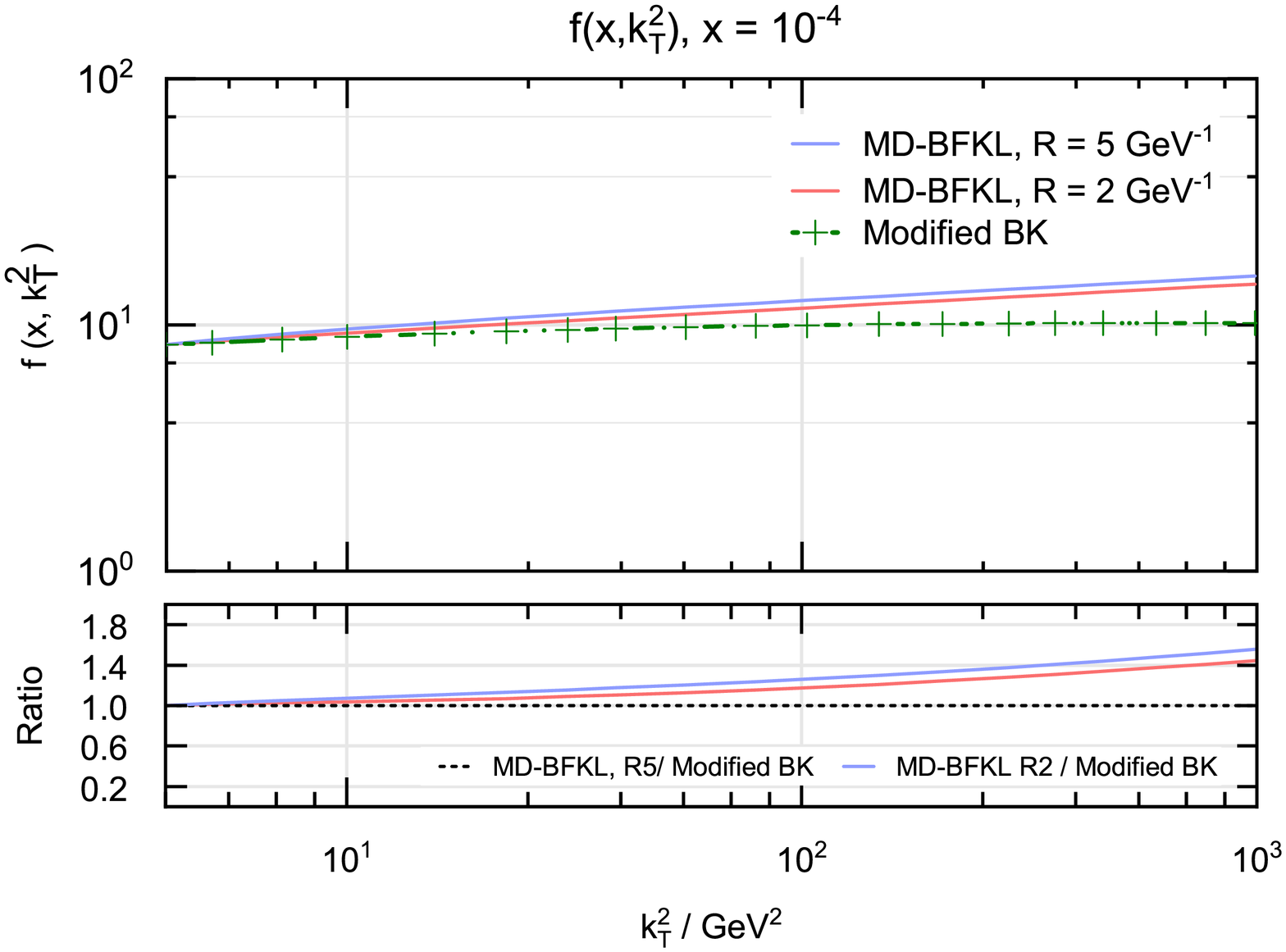}\hspace{5mm}
	\includegraphics[trim=25 5 21.5 0,width=.46\textwidth,clip]{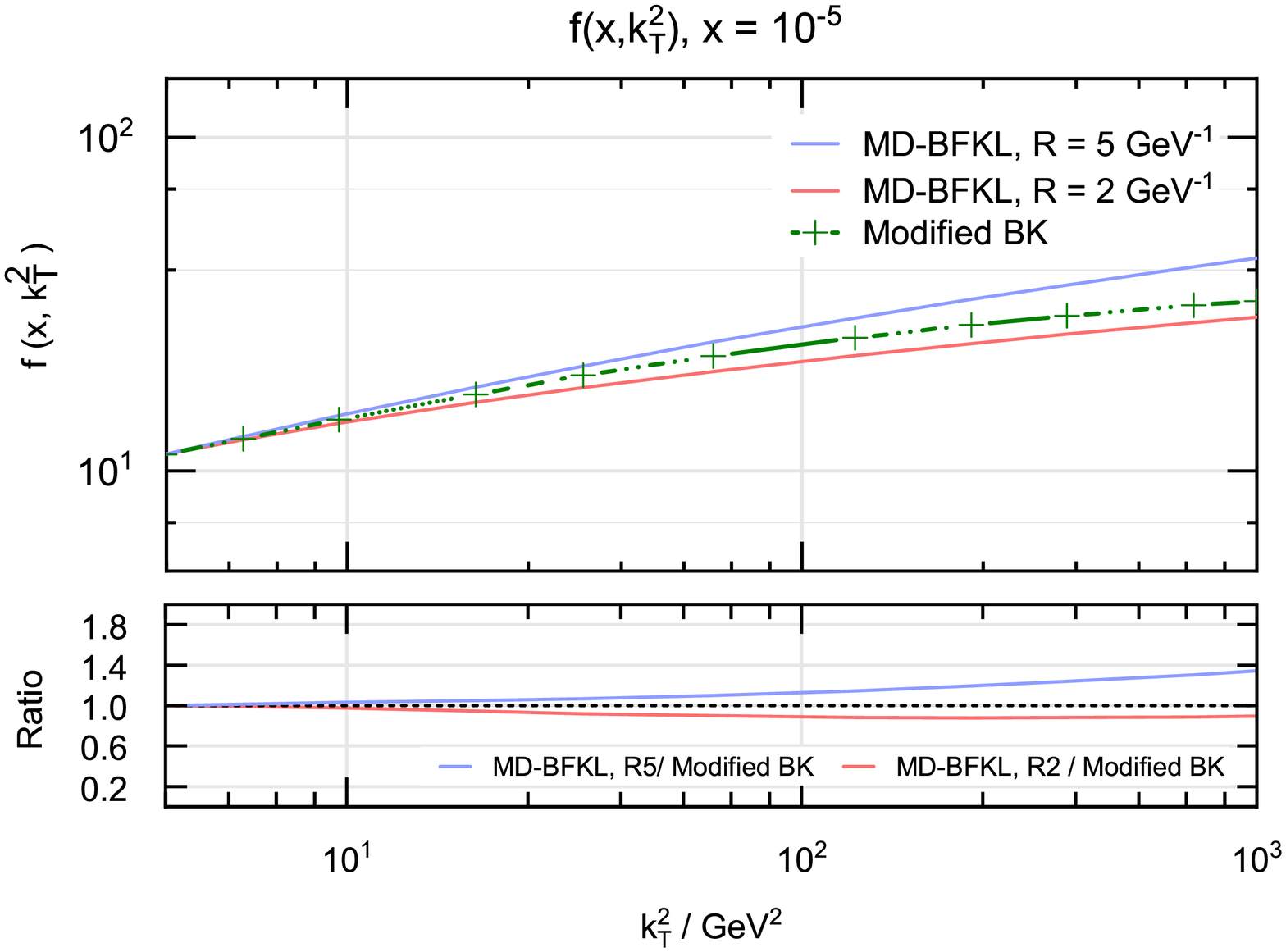}
	
	\caption{\label{k1}$k_T^2$ evolution of unintegrated gluon distribution $f(x,k_T^2)$. Our results of KC improved MD-BFKL are shown for conventional $R=5 \text{ GeV}^{-1}$ (gluons are distributed throughout the nucleus) and $R=2 \text{ GeV}^{-1}$(at "hot-spots" within proton disk). Prediction from modified BK equation is plotted for comparison.}
\end{figure*}
\begin{figure*}[t]
	\centering 
	\includegraphics[trim=25 5 21.5 0,width=.46\textwidth,clip]{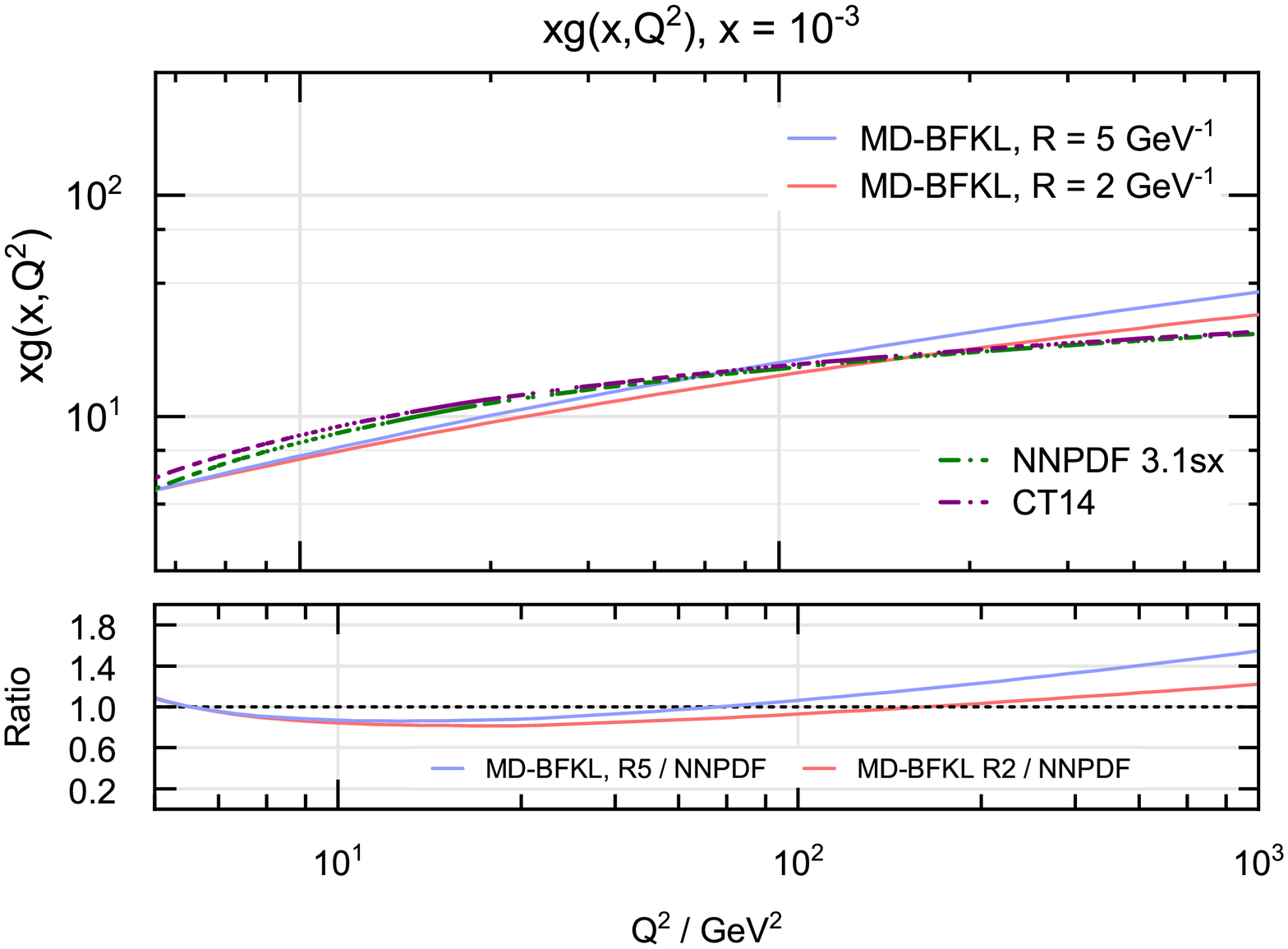}\hspace{5mm}
	\includegraphics[trim=25 5 21.5 0,width=.46\textwidth,clip]{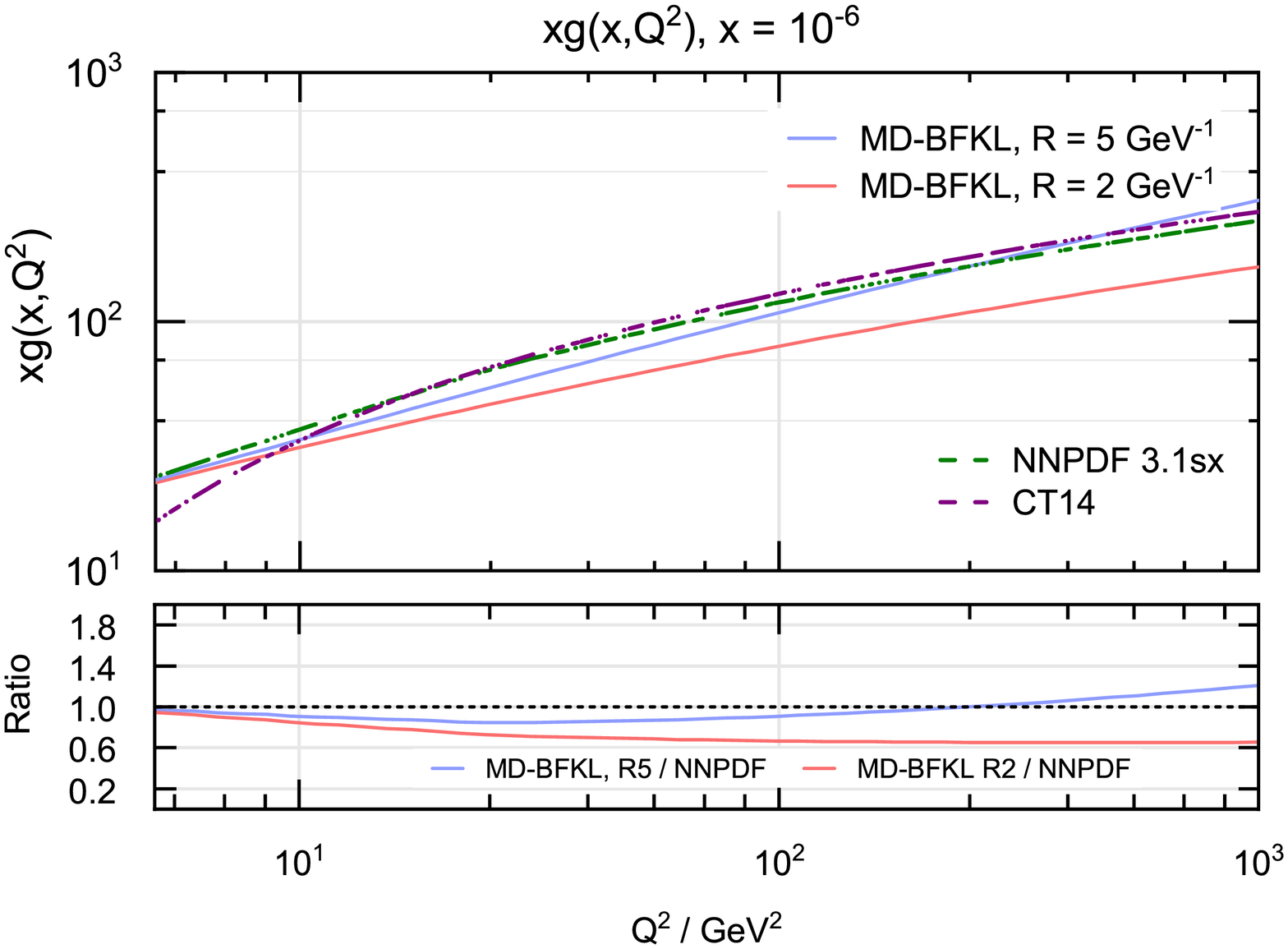}
	
	\caption{\label{k2}$Q^2$ evolution of collinear gluon distribution $xg(x,Q^2)$. Our results of KC improved MD-BFKL are shown for conventional $R=5 \text{ GeV}^{-1}$ (gluons are distributed throughout the nucleus) and $R=2 \text{ GeV}^{-1}$(at "hot-spots" within proton disk). Theoretical prediction is compared with global datasets NNPDF 3.1sx and CT 14.}
\end{figure*}
Similarly setting the initial distribution at $(x_0,k_T^2)$, we obtain the solution for x-evolution as follows
\begin{widetext}
\begin{equation}
	\label{30}
	f(x,k_T^2)=\dfrac{(-1)^{\frac{3}{\lambda }+\frac{2 n}{\lambda  l}}\left(N_c^2-1\right) (\frac{x_0}{x})^m k_T^{-2\frac{n}{l}}  \left(-\frac{k_T^{-2\lambda }}{\lambda  l}-\ln \frac{x}{x_0}\right)^{-\frac{n}{\lambda l} }\text{C}[x,k_T^2]}
	{\Bigg[\splitdfrac{(-1)^{\frac{3}{\lambda }+\frac{3 n}{\lambda  l}}N_c^2 F_0^{'} \text{C}[x,k_T^2] +\frac{ (-1)^{\frac{2}{\lambda }+\frac{n}{\lambda  l}} }{k_T^2} \Gamma_1 \text{D}[x,k_T^2]}{+
			(-1)^{\frac{2}{\lambda }+\frac{n}{\lambda  l}+1}  k_T^{2\left(\frac{n}{l}-\lambda  \left(\frac{1}{\lambda }+\frac{n}{\lambda  l}\right)\right)} \Gamma_2 \text{D}[x,k_T^2]+(-1)^{\frac{3}{\lambda }+\frac{3 n}{\lambda  l}+1}\pi  R^2 F_0^{'} x^m \text{C}[x,k_T^2]}\Bigg]},
\end{equation}
\end{widetext}
where, $F_0^{'}$ is the initial gluon distribution at $(x_0,k_T^2)$,
\begin{equation*}
	\begin{split}
		&\Gamma_1=\Gamma\left(\frac{l+n}{l \lambda }+1,-\frac{k_T^{-2\lambda } m}{l \lambda }\right),\text{     }
		\\&\Gamma_2=\Gamma\left(\frac{l+n}{l \lambda }+1,m \left(-\frac{k_T^{-2\lambda }}{l \lambda }-\ln \frac{x}{x_0}\right)\right),\\
		&\text{C}[x,k_T^2]=\pi  R^2 x^ml^{-\frac{1}{\lambda }-\frac{n}{\lambda  l}} \lambda ^{-\frac{1}{\lambda }-\frac{n}{\lambda  l}} k_T^{2\left(\frac{n}{l}-\lambda  \left(\frac{1}{\lambda }+\frac{n}{\lambda  l}\right)\right)}\\
		&\times m^{\frac{2}{\lambda }+\frac{2 n}{\lambda  l}+1} \left(-\frac{k_T^{-2\lambda }}{\lambda  l}-\ln \frac{x}{x_0}\right)^{\frac{1}{\lambda }+\frac{n}{\lambda  l}},\\
		&\text{D}[x,k_T^2]=18 \alpha_s ^2 N_c^2 x_0^m  l^{-\frac{n}{\lambda  l}} \lambda ^{-\frac{n}{\lambda  l}} e^{-\frac{m k_T^{-2\lambda }}{\lambda  l}}\\
		&\times m^{\frac{1}{\lambda }+\frac{n}{\lambda  l}} \left(-\frac{k_T^{-2\lambda }}{\lambda  l}-\ln \frac{x}{x_0}\right)^{1/\lambda }.
	\end{split}
\end{equation*}

\begin{figure*}[t]
	\centering 
	\includegraphics[trim=25 5 21.5 0,width=.46\textwidth,clip]{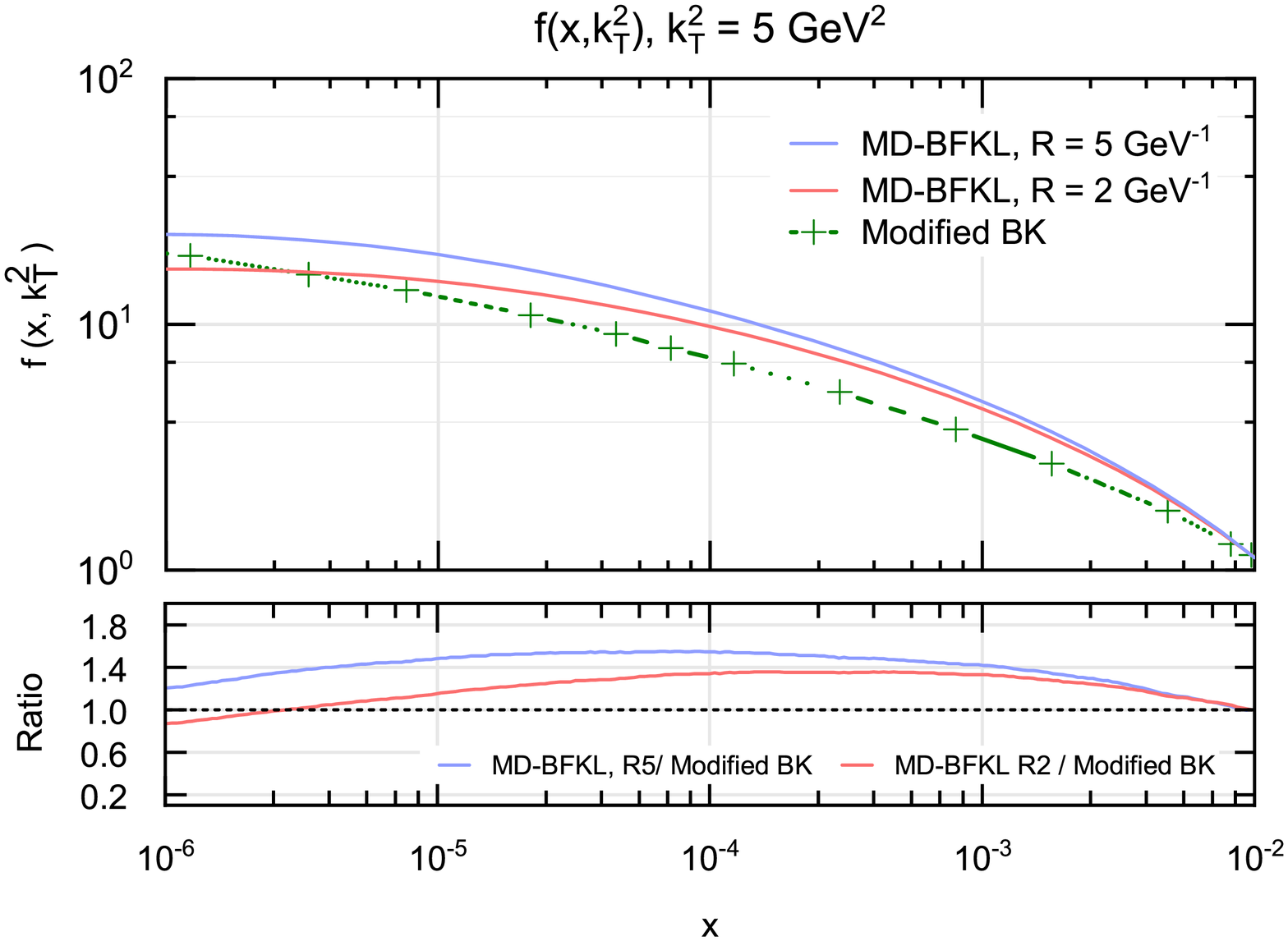}\hspace{5mm}
	\includegraphics[trim=25 5 21.5 0,width=.46\textwidth,clip]{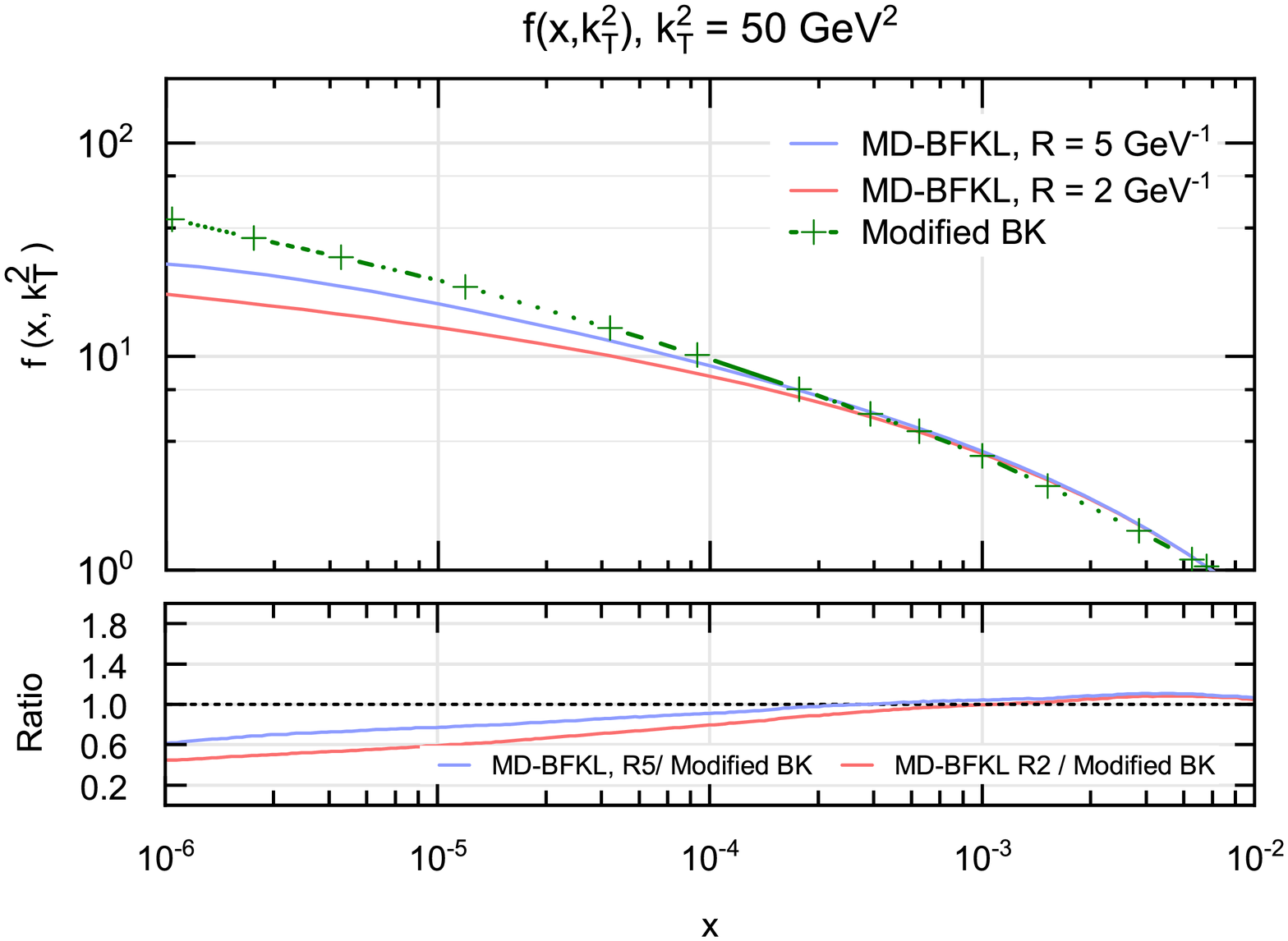}
	
	\caption{\label{x1}$x$ evolution of unintegrated gluon distribution $f(x,k_T^2)$. Our results of KC improved MD-BFKL are shown for conventional $R=5 \text{ GeV}^{-1}$ (gluons are distributed throughout the nucleus) and $R=2 \text{ GeV}^{-1}$(at "hot-spots" within proton disk). Prediction from modified BK equation is plotted for comparison.}
\end{figure*}
\begin{figure*}[t]
	\centering 
	\includegraphics[trim=25 5 21.5 0,width=.46\textwidth,clip]{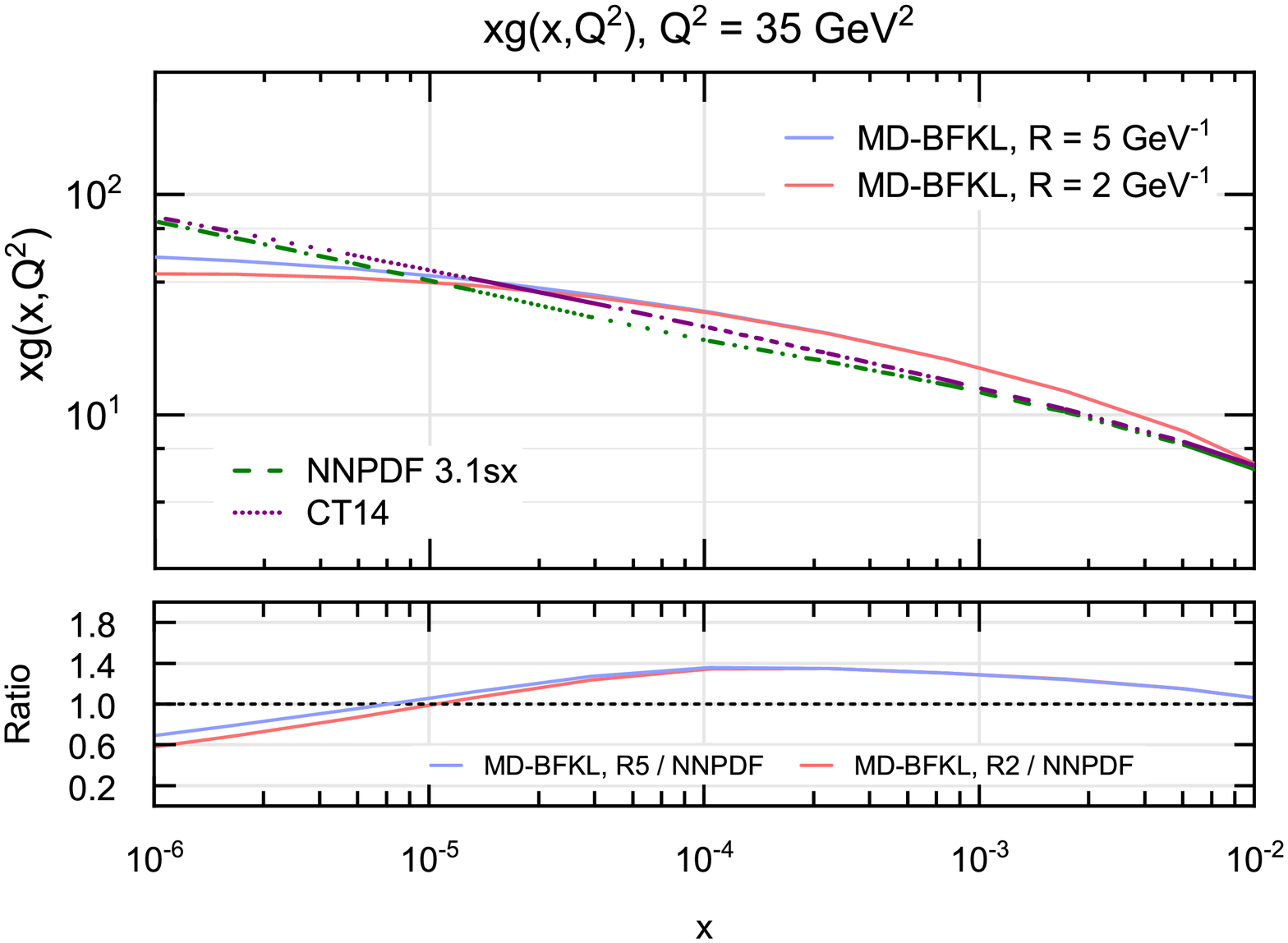}\hspace{5mm}
	\includegraphics[trim=25 5 21.5 0,width=.46\textwidth,clip]{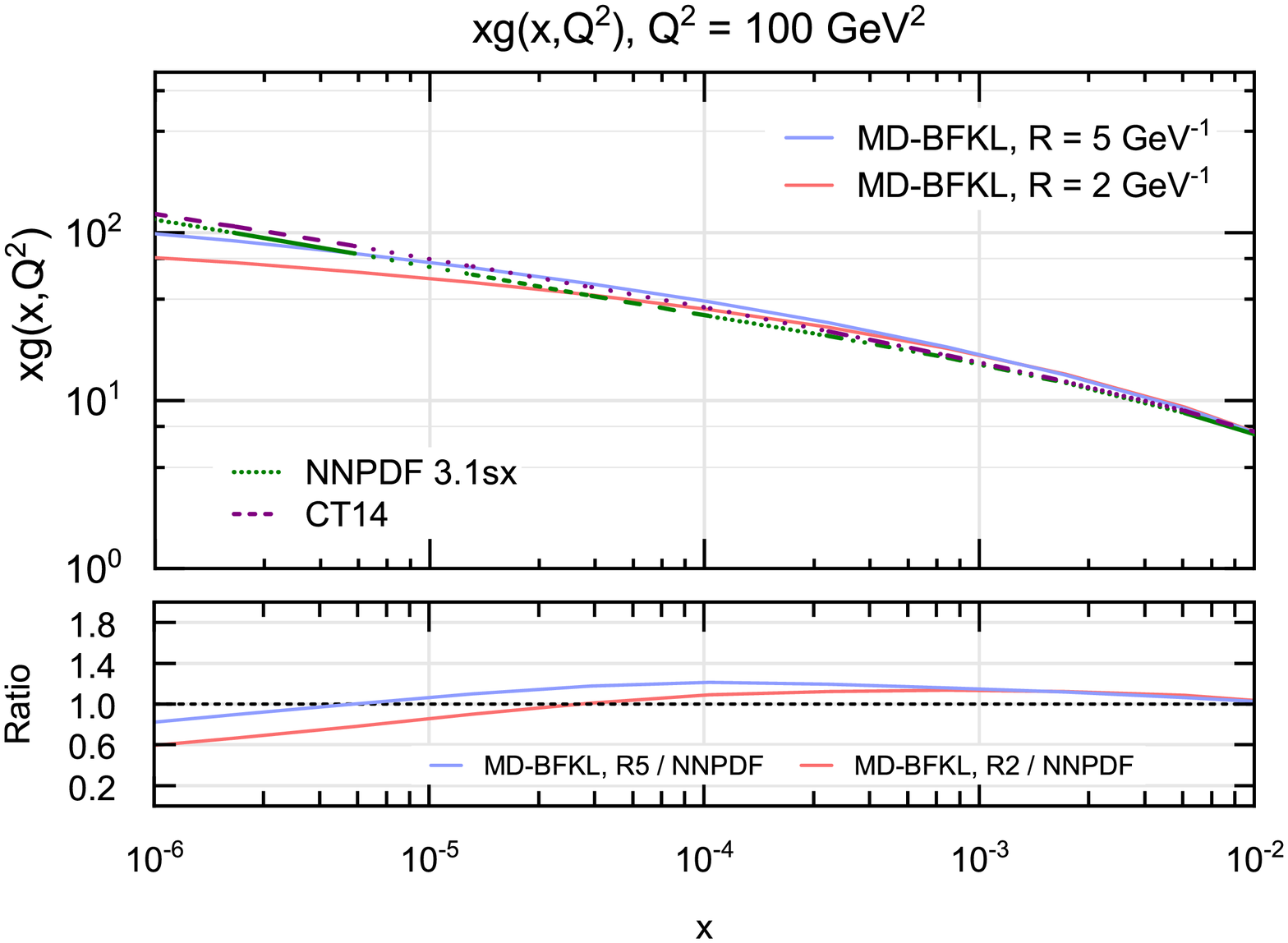}
	
	\caption{\label{x2}$x$ evolution of collinear gluon distribution $xg(x,k_T^2)$. Our results of KC improved MD-BFKL are shown for conventional $R=5 \text{ GeV}^{-1}$ (gluons are distributed throughout the nucleus) and $R=2 \text{ GeV}^{-1}$(at "hot-spots" within proton disk). Theoretical prediction is compared with global datasets NNPDF 3.1sx and CT 14.}
\end{figure*}

We have plotted the solution for both $x$ evolution \eqref{30} and $k_T^2$ evolution \eqref{29} in Fig.~\ref{k1} and Fig.~\ref{x1}. Our prediction of unintegrated gluon distribution $f(x,k_T^2)$ is contrasted with that of modified BK equation \cite{5},

\begin{equation}
\label{bk}
\begin{split}
&-x\frac{\partial f \left(x, k_T^2\right)}{\partial x}=\\&\frac{\alpha_s N_c k_T^2}{\pi}\int_{k_{T_{\min }}^{'2}}^{\infty}\frac{dk_T^{'2}}{k_T^{'2}}\left[\frac{f (x, k_T^{'2})-f \left(x, k_T^2\right)}{|k_T^{'2}-k_T^2|} + \frac{f \left(x, k_T^2\right)}{\sqrt{k_T^4+4k_T^{'4}}}\right]\\
&-\alpha_{s}\left(1-k_T^2\frac{\text{d}}{\text{d}k_T^2}\right)^2\frac{k_T^2}{R^2}\left[\int_{k_T^2}^{\infty}\frac{\text{d}k_T^{'2}}{k_T^{'2}}\ln \left(\frac{k_T^{'2}}{k_T^2}\right)f(x,k_T^2)\right]
\end{split}
\end{equation}
which is BFKL equation supplemented by the negative nonlinear term, derived in approximation of infinite and uniform target. In \cite{5} the perturbative parton saturation is studied to a vast extent, including modification of \eqref{bk} in terms of kinematic constraint, DGLAP, $\text{P}_{gg}$ splitting function and running coupling constant, then solving the same numerically. We have also extracted collinear gluon distribution from unintegrated gluon distribution using the standard relation,
\begin{equation}
\label{f2g}
xg(x,Q^2)=\int_{0}^{Q^2}\frac{\text{d}k_T^2}{k_T^2}f(x,k_T^2)
\end{equation}
sketched in Fig.~\ref{k2} and Fig.~\ref{x2}. Our predicted collinear gluon density is compared with that of LHAPDF global parameterization groups NNPDF 3.1sx \cite{36} and CT 14 \cite{53}. Both of the LHAPDF datasets  include HERA as well as recent LHC data with high precision PDF sensitive measurements.
\begin{figure*}[t]
	\centering 
	\includegraphics[trim=25 0 21.5 0,width=.47\textwidth,clip]{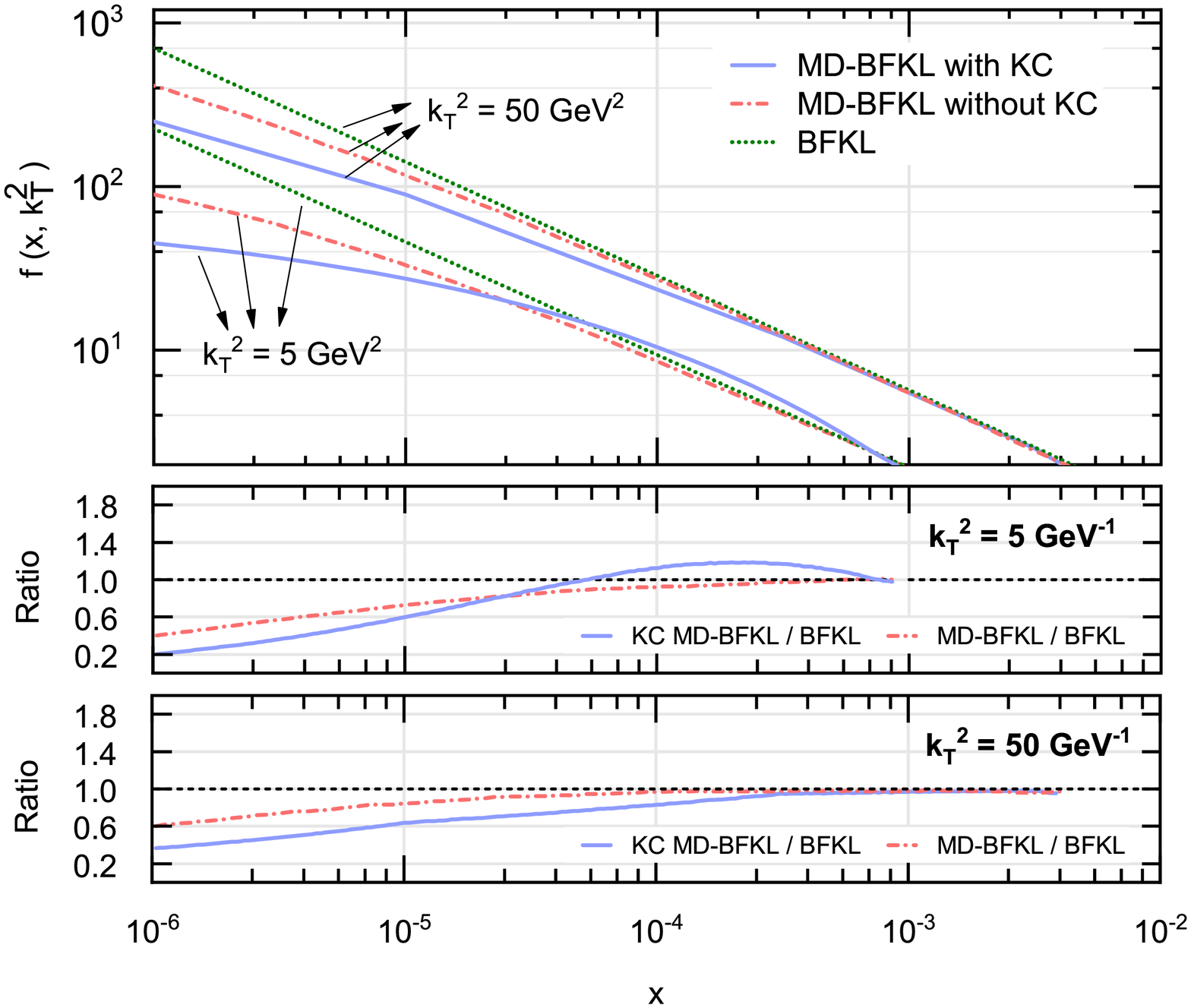}
	\hspace{1cm}
	\includegraphics[trim=25 0 21 0,width=.46\textwidth,clip]{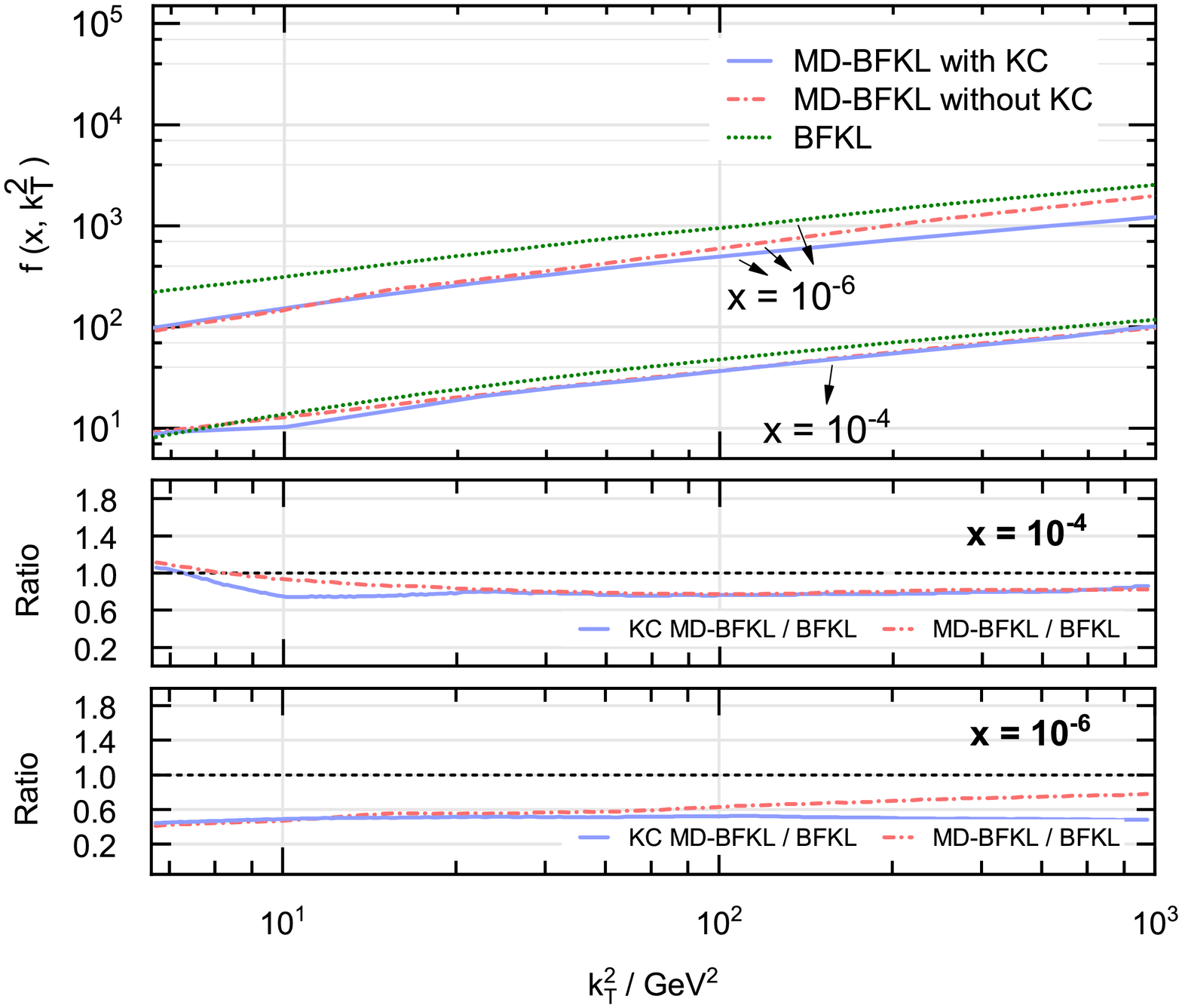}
	
	\caption{\label{3a} (a) Comparison between linear BFKL, MD-BFKL without KC and MD-BFKL with KC for $k_T^2$ evolution at $x=10^{-4}$ and $10^{-6}$ (b) Comparison between linear BFKL, MD-BFKL without KC and MD-BFKL with KC for $x$ evolution at $k_T^2=5\text{ GeV}^2$ and $50\text{ GeV}^2$. Results are shown for $R= 5 \text{ GeV}^{-1}$. }
\end{figure*} 
\par Our prediction of both unintegrated and collinear gluon distribution is obtained for  two different  form of shadowing: conventional $R= 5 \text{ GeV}^{-1}$ (order of proton radius) where gluons are spread throughout the nucleus and extreme $R= 2 \text{ GeV}^{-1}$ where gluons are expected to concentrated in hotspots within the proton disk,  recalling that  $\pi R^2$ is the transverse area within which gluons are concentrated inside proton. From Fig.~\ref{k1}-\ref{x2} it is clear that shadowing correction are more prominent when gluons are concentrated in hotspots within proton.

\par The $k_T^2$ (and $Q^2$) evolution in Fig.~\ref{k1}-\ref{k2} is studied for the kinematic range $1 \text{ GeV}^2\leq k_T^2(\text{or }Q^2)\leq 10^3 \text{ GeV}^2$ corresponding to four different values of $x$ as indicated in the figure. Our evolution for both $f(x,k_T^2)$ and $xg(x,Q^2)$ shows a similar growth as modified BK as well as datasets respectively for all $x$. It is also observed that the growth of $f(x,k_T^2)$ (or $xg(x,Q^2)$) is almost linear for the entire kinematic range of $k_T^2$ (or $Q^2$). This is expected since the net shadowing term in \eqref{8} has $1/k_T^2$ dependence, which suppresses the contribution from the shadowing term at large $k_T^2$.

\par The $x$ evolution of $f(x,k_T^2)$ and $xg(x,Q^2)$ is shown in Fig.~\ref{x1}-\ref{x2} for  two $k_T^2$ values viz. $5 \text{ GeV}^2$, $50 \text{ GeV}^2$ and two $Q^2$ values viz. $35 \text{ GeV}^2$, $100 \text{ GeV}^2$. The input is taken at higher $x$ value $x=10^{-2}$ and then evolved down to smaller $x$ value upto $x=10^{-6}$ thereby setting the kinematic range of evolution $10^{-6}\leq x\leq10^{-2}$. We observed that the singular $x^{-\lambda}$ growth of the gluon is eventually suppressed by the net shadowing effect. KC improved MD-BFKL seem to poses a more intense shadowing then modified BK. However, it is hard to establish the existence of shadowing for $x\geq 10^{-3}$. The obvious distinction between the two form of shadowing $R=  \text{5 GeV}^{-1}$ (conventional) and $R= 2\text{ GeV}^{-1}$ ("hotspot") is also observed towards small-$x$ $(x\leq 10^{-3})$.  Interestingly towards very  small-$x$ $(x\leq 10^{-5})$, at small $k_T^2$ (or $Q^2$) values (viz. $k_T^2$=$5\text{ GeV}^2$, $Q^2$=$35 \text{ GeV}^2$) the gluon distribution becomes almost irrelevant of change in $x$. This could be a strong hint for the possible saturation phenomena in the small-$x$ high density regime.

\par In Fig.~\ref{3a}(a) we have shown a comparison between the linear BFKL equation, MD-BFKL without kinematic constraint and MD-BFKL with kinematic constraint for $x$ evolution. The three solutions are compared for two different values of $k_T^2$ i.e. $ 5 \text{ GeV}^2$ and $50 \text{ GeV}^2$. Similarly in Fig.~\ref{3a}(b) we have shown similar comparison but for $k_T^2$ evolution for two different values of $x$ i.e. $x=10^{-4}$ and $x=10^{-6}$. In Fig.~\ref{3a}  the singular $x^{-\lambda}$  behavior of $f(x,k_T^2)$ is distinct for unshadowed linear BFKL equation. On the other hand, the deviation of the two different forms of the MD-BFKL equation from linear BFKL equation reflects the underlying shadowing correction. It is also observed that shadowing effect is more intense in MD-BFKL with KC than MD-BFKL without KC.

\subsubsection{Complete solution of KC improved MD-BFKL}
In this section we implant a functional form of the input distribution (more likely a dynamic one) on the general solution of our KC improved MD-BFKL equation \eqref{26} and try to obtain a parametric form of the solution. The underlying motivation towards doing so is that this allows us to evolve our solution for both $x$ and $k_T^2$ simultaneously in $x\text{-}k_T^2$ phase space which help us to portray a three-dimensional realization of the gluon evolution. 
\par Recall the well-known solution of the linear BFKL equation \cite{9}
\begin{equation}
	\label{32}
	f(x,k_T^2)=\beta \frac{x^{-\lambda}\sqrt{k_T^2}}{\sqrt{\ln\frac{1}{x}}}\exp \left(-\frac{\ln^2(k_T^2/k_s^2)}{2\Omega\ln (1/x)}\right),
\end{equation}
where $\lambda= \frac{3 \alpha_s}{\pi} 28  \zeta(3)$, $\zeta$ being Reimann zeta function and $\Omega=32.1 \alpha_s$. The nonperturbative parameter $k_s^2= 1 \text{ GeV}^2$ and the normalization constant $\beta\sim 0.01$ \cite{1}. Since far below saturation region both linear and nonlinear equation should give the same solution, therefore we can take \eqref{32} as the input distribution for our general solution of KC improved MD-BFKL equation.

First we try to find the functional form of the arbitrary differentiable function $\text{G}\left(\frac{e^{-\frac{k_T^{-2\lambda}}{l \lambda}}}{x}\right)$ present in the general solution \eqref{26} applying the initial distribution \eqref{32}.
Let us rewrite \eqref{27} which is the rearranged form of the general solution \eqref{26}

\begin{widetext}
\begin{equation}
	\label{33}
	\begin{split}
		\text{G}\left(\frac{e^{-\frac{k_T^{-2\lambda}}{l \lambda}}}{x}\right)=&\frac{1}{N_c^2-1}\Bigg[x^{-m}N_c^2 \bigg(\frac{18\alpha_s ^2 (-1)^{\frac{1}{\lambda }+1} \lambda ^{1/\lambda }  l^{1/\lambda } e^{-\frac{m k_T^{-2\lambda }}{\lambda  l}} m^{-\frac{\lambda  l+l+n}{\lambda  l}} \Gamma \left(\frac{l+n}{l \lambda }+1,-\frac{k_T^{-2\lambda } m}{l \lambda }\right)}{\pi  R^2}\\
		&+\frac{1}{f(x,k_T^2)} k_T^{-2\frac{n}{l}} (-1)^{\frac{n}{\lambda  l}} l^{-\frac{n}{\lambda  l}} \lambda ^{-\frac{n}{\lambda  l}}\bigg)+\frac{1}{f(x,k_T^2)}x^{-m} k_T^{-2\frac{n}{l}} (-1)^{\frac{n}{\lambda  l}+1} l^{-\frac{n}{\lambda  l}} \lambda ^{-\frac{n}{\lambda  l}}\Bigg].
	\end{split}
\end{equation}
\end{widetext}
Setting initial parameters at $(x_0,k_T^2)$ we get initial distribution \eqref{32} as
\begin{equation}
	\label{34}
	f(x_0,k_T^2)=\beta \frac{x_0^{-\lambda}\sqrt{k_T^2}}{\sqrt{\ln\frac{1}{x_0}}}\exp \left(-\frac{\ln^2(k_T^2/k_s^2)}{2\Omega\ln (1/x_0)}\right).
\end{equation}

We denote the argument of G at $(x_0,k_T^2)$ as  $\tau=\frac{e^{-\frac{k_T^{-2\lambda}}{l \lambda}}}{x_0}$ which implies $k_T^2=( -l\lambda  \ln (\tau  x_0))^{-1/\lambda }$. Now writing \eqref{33} for $(x_0,k_T^2)$  we obtain
\begin{widetext}
\begin{equation}
	\label{35}
	\begin{split}
		\text{G}(\tau)=&x^{-m} k_T^{-\frac{n}{l}} (-l\lambda)^{\frac{n}{\lambda  l}}  \Bigg(\frac{x^{\lambda } \sqrt{\ln \frac{1}{x}} (\lambda  (-l) \ln (\tau  x_0))^{1/\lambda } \exp \left(\frac{\ln ^2\left((\lambda  (-l) \ln (\tau  x_0))^{-1/\lambda }\right)}{2 \Omega  \ln \frac{1}{x}}\right)}{\beta }\\
		&-\frac{18\alpha_s ^2  N_c^2 \left(k_T^{-\lambda }\right)^{1/\lambda } \left(e^{-\frac{k_T^{-\lambda }}{\lambda  l}}\right)^m \left(-\frac{m k_T^{-\lambda }}{\lambda  l}\right)^{-\frac{l+n}{\lambda  l}} \Gamma \left(\frac{l+n}{l \lambda }+1,-\frac{k_T^{-\lambda } m}{l \lambda }\right)}{\pi  m R^2 \left(N_c^2-1\right)}\Bigg),
	\end{split}
\end{equation}
\end{widetext}
where, $k_T^2=( -l\lambda  \ln (\tau  x_0))^{-1/\lambda }$. Note that \eqref{35} is the functional form of G. We substitute $\tau=\frac{e^{-\frac{k_T^{-2\lambda}}{l \lambda}}}{x}$ in \eqref{35} which gives us the l.h.s. of \eqref{33} and then solve \eqref{33} for $f(x,k_T^2)$,
\begin{widetext}
\begin{equation}
	\label{36}
	\begin{split}
		f(x,k_T^2)=\frac{\beta  m q^{n/l} \left(-\frac{m k_T^{-2\lambda }}{\lambda  l}\right)^{\frac{l+n}{\lambda  l}} \left(-\frac{m q^{-\lambda }}{\lambda  l}\right)^{\frac{l+n}{\lambda  l}}}
		{\left(-\frac{m q^{-\lambda }}{\lambda  l}\right)^{\frac{l+n}{\lambda  l}} \left(\tilde{\Delta}  q^{n/l} \tilde{A}[ k_T^2 ] +\chi\right)-\tilde{\Delta} \beta   k_T^{2n/l} \tilde{A}(q) \left(-\frac{m k_T^{-2\lambda }}{\lambda  l}\right)^{\frac{l+n}{\lambda  l}}},
	\end{split}
\end{equation}

where,
\begin{equation*}
	\begin{split}
		&\chi=m x^{\lambda } \sqrt{\ln \frac{1}{x}} k_T^{2n/l} e^{\frac{\ln ^2(q)}{2 \Omega  \ln \frac{1}{x}}} \left(-\frac{m k_T^{-2\lambda }}{\lambda  l}\right)^{\frac{l+n}{\lambda  l}} q^{-1},\text{    } \text{   }\tilde{A}[i]= i^{-1}\beta  \left(e^{-\frac{q^{-\lambda }}{\lambda  l}}\right)^m  \Gamma \left[i\right], \text{    }\text{    } \text{   }
		q=\left(-l\lambda \ln \frac{x_0 e^{-\frac{k_T^{-2\lambda }}{\lambda  l}}}{x}\right)^{-1/\lambda },\\
		&\Gamma \left[i\right]=\Gamma\left(1+\frac{l+n}{l\lambda},\frac{m (i)^{-\lambda}}{l\lambda}\right), \text{  }\tilde{\Delta}=\frac{18 \alpha _s^2}{\pi  R^2}\frac{N_c^2}{N_c^2-1}.
	\end{split}
\end{equation*}
\end{widetext}
\begin{figure}[t]
	\centering
	\includegraphics[width=.40\textwidth,clip]{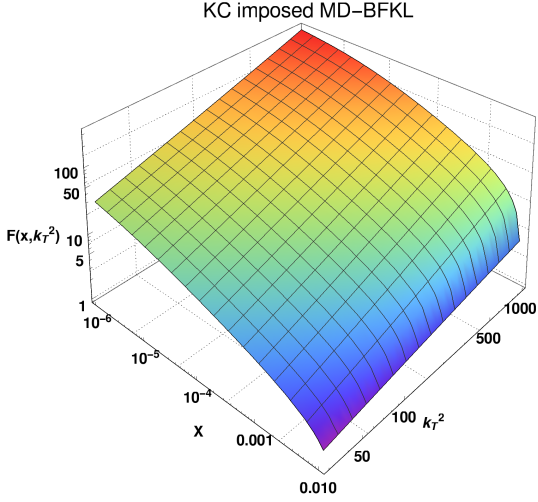}
	\caption{\label{5x} Three dimensional representation of (a) TMDlib HERA data fit: PB-NLO-HERAI+II+2018 (left)   (b) our solution for KC improved MD-BFKL in $x$-$k_T^2$ phase space (right).}
\end{figure}

For simplicity let us denote r.h.s. of \eqref{36} by $\gamma$ i.e.

\begin{equation}
	\label{37}
	f(x,k_T^2) = \gamma(x,k_T^2).
\end{equation}
At $x=x_0$ and $k_T^2=k_0^2$
\begin{equation}
	\label{38}
	f(x_0,k_0^2) = \gamma(x_0,k_0^2).
\end{equation}
Now dividing \eqref{37} by \eqref{38} we get

\begin{equation}
	\label{39}
	f(x,k_T^2) = f(x_0,k_0^2)\frac{\gamma(x,k_T^2)}{\gamma(x_0,k_0^2)}.
\end{equation}
From \eqref{32} we have the input distribution
\begin{equation}
	\label{40}
	f(x_0,k_0^2)=\beta \frac{x_0^{-\lambda}\sqrt{k_0^2}}{\sqrt{\ln\frac{1}{x_0}}}\exp \left(-\frac{\ln^2(k_0^2/k_s^2)}{2\Omega\ln (1/x_0)}\right).
\end{equation}
\begin{figure*}[t]
	\centering 
	\includegraphics[width=.29\textwidth,clip]{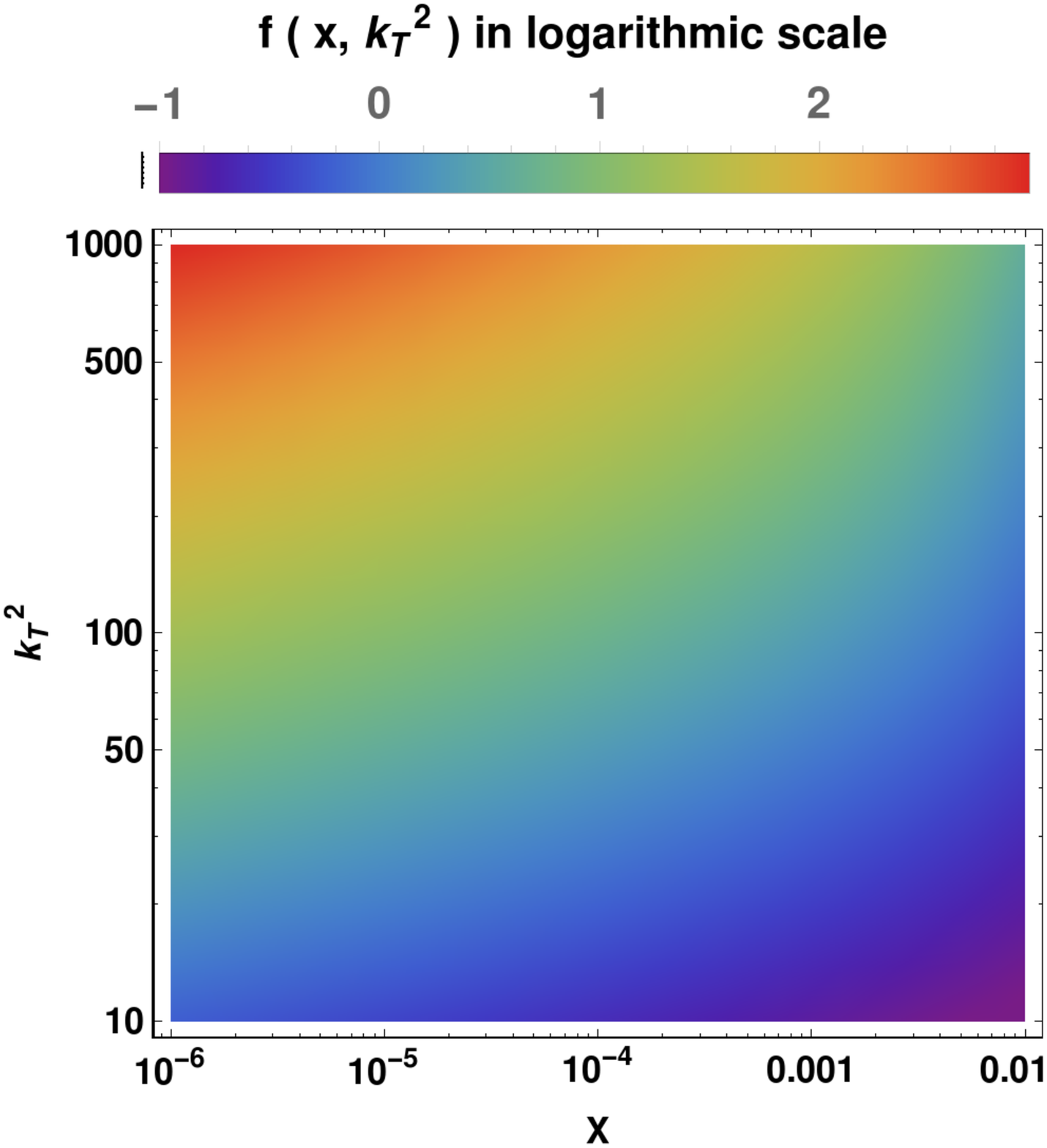}\hspace{1cm}
	\includegraphics[width=.29\textwidth,clip]{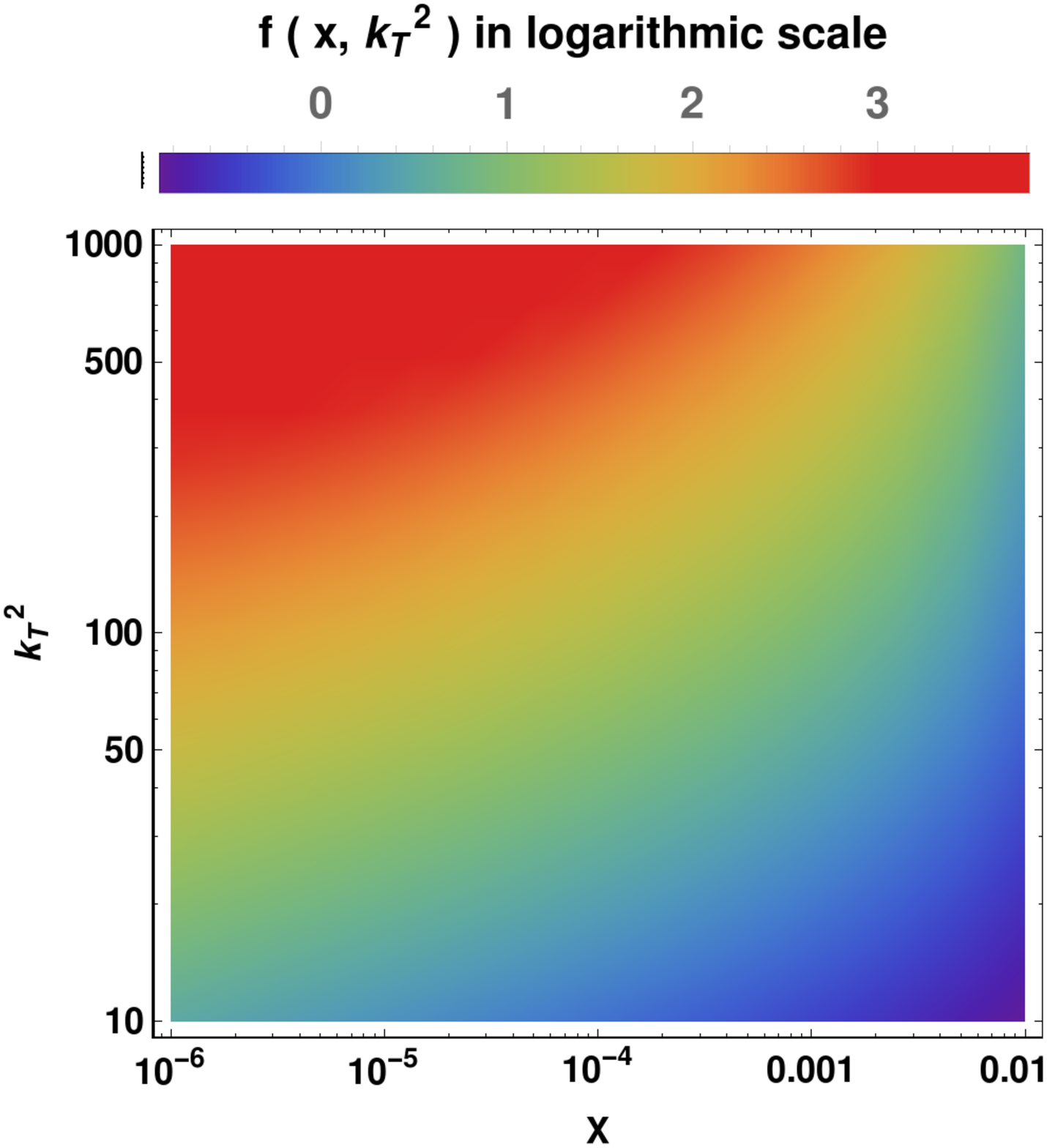}
	\caption{\label{6y} Density plots showing $\lambda$ sensitivity of unintegrated gluon distribution $f(x,k_T^2)$, sketched for two canonical choice of $\lambda$ viz.  $\lambda=0.4$ (left) and  $\lambda=0.6$ (right). }
\end{figure*}
Now substituting $f(x,k_0^2)$ from \eqref{40} into \eqref{39} we obtain
\begin{equation}
	\label{41}
	\begin{split}
		&f(x,k_T^2)=\\&\frac{(\ln \frac{1}{x_0})^{-\frac{1}{2}}q^{n/l} k_0^{-2\frac{n}{l}+4\frac{l+n}{ l}-2}\beta  x_0^{-\lambda }  e^{-\frac{\ln ^2(k_0^2)}{2 \Omega  \ln \frac{1}{x_0}}} \left(k_T^2q\right)^{-\frac{l+n}{l}} \tilde{\phi}}
		{\left(-\frac{m q^{-\lambda }}{\lambda  l}\right)^{\frac{l+n}{\lambda  l}} \left(\tilde{\Delta}  q^{n/l} \tilde{A}[k_T^2] +\chi\right)-\tilde{\Delta} k_T^{2n/l} \tilde{A}[q] \left(-\frac{m k_T^{-2\lambda }}{\lambda  l}\right)^{\frac{l+n}{\lambda  l}}},
	\end{split}
\end{equation}
where
\begin{equation*}
	\begin{split}
		\delta=&m x_0^{\lambda } \sqrt{\ln \frac{1}{x_0}} k_0^{2n/l-2} e^{\frac{\ln ^2(k_0^2)}{2 \Omega  \ln \frac{1}{x_0}}}   \left(-\frac{m k_0^{-2\lambda }}{\lambda  l}\right)^{\frac{l+n}{\lambda  l}}, \\
		\tilde{\phi}=&\left(-\frac{m k_0^{-2\lambda }}{\lambda  l}\right)^{\frac{l+n}{\lambda  l}} \left(\tilde{\Delta}k_0^{2n/l} \tilde{A}[k_0^2] +\delta[x,k_T^2]\right)\\
		&-\tilde{\Delta} k_0^{2n/l}\tilde{A}[k_0^2]\left(-\frac{m k_0^{-2\lambda }}{\lambda  l}\right)^{\frac{l+n}{\lambda  l}}.	
	\end{split}
\end{equation*}

\begin{figure*}[tbp]
	\centering 
	\includegraphics[width=.29\textwidth,clip]{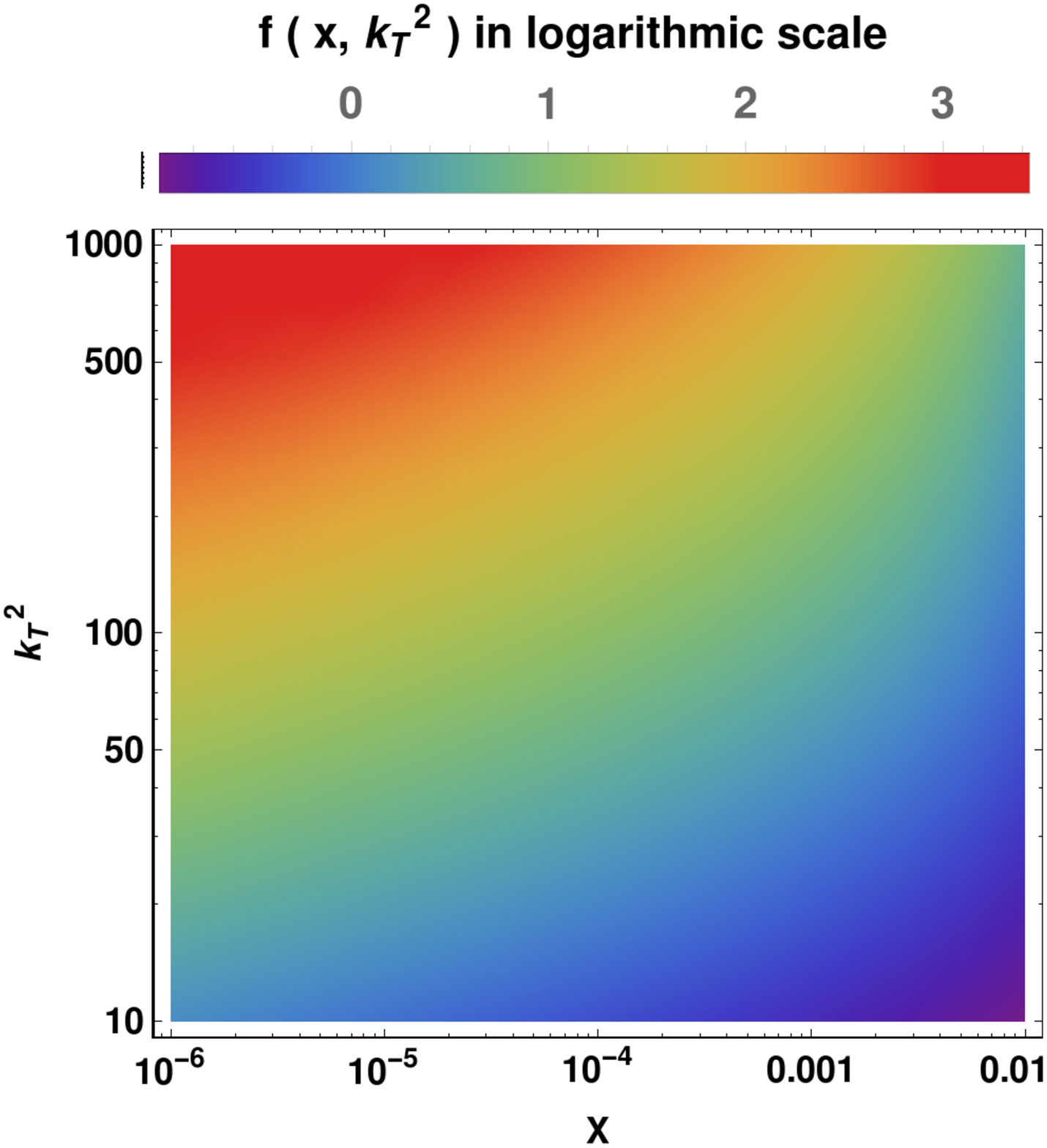}\hspace{1cm}
	\includegraphics[width=.29\textwidth,clip]{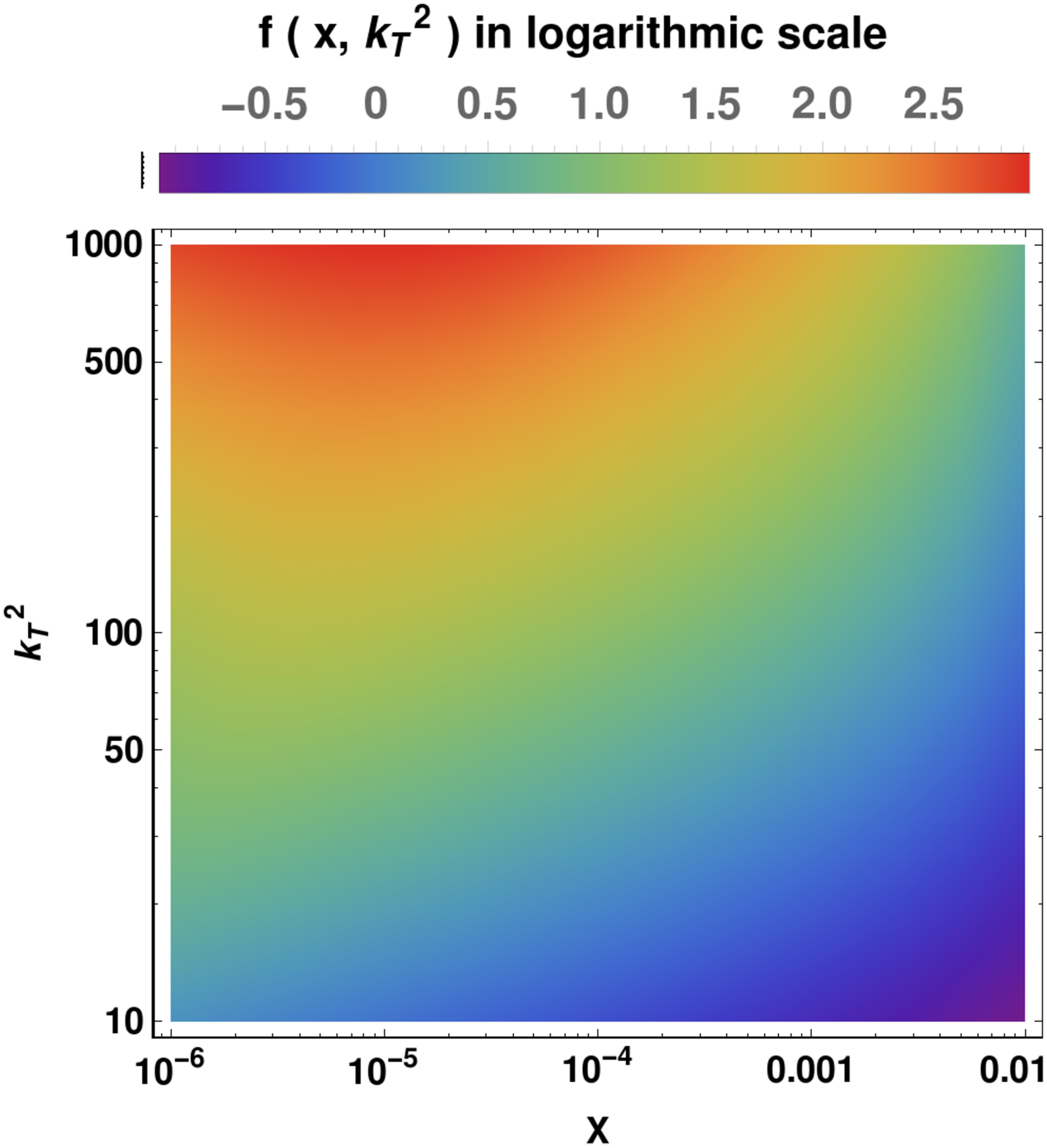}
	\caption{\label{6xx} Density plots showing $R$ sensitivity of unintegrated gluon distribution $f(x,k_T^2)$, sketched for two form of shadowing: conventional $R=5 \text{ GeV}^{-1}$ (left) and extreme $R=2 \text{ GeV}^{-1}$ (right).} 
\end{figure*}

Equation \eqref{41} serves as the parametric form of the solution for KC improved MD-BFKL equation. The input distribution is inclusive in the solution \eqref{41} itself, therefore, we do not have to depend on the experimental data fits for input distribution unlike we did in the previous section of $x$ and $k_T^2$ evolution. This parametric form of the solution actually helps us to explore the three-dimensional insight of the gluon evolution in $x$-$k_T^2$ phase space. However, using \eqref{41} one may also study $x$ and $k_T^2$ evolution separately by setting $x$ fixed for $k_T^2$ evolution or $k_T^2$ fixed for $x$ evolution.

\par In Fig.~\ref{5x} we have shown our solution of KC improved MD-BFKL equation \eqref{41} in three dimension. The kinematic region for our study is set to be $10^{-6}\leq x \leq10^{-2}$ and $1 \text{ GeV}^2\leq k_T^2 \leq10^3\text{ GeV}^2$. In the 3D surface, the suppression in the rise of the gluon distribution towards small $x$ due to shadowing correction is visible. However, a linear rise of the surface in the $k_T^2$ direction can be seen, attributed to the $1/k_T^2$ factor in the nonlinear term, which offset the effect of shadowing at large $x$.    

\par In Fig.~\ref{6y} we have shown density plots of $f(x,k_T^2)$ in $x\text{-}k_T^2$ domain to examine the sensitivity of $f(x,k_T^2)$ towards the parameter $\lambda$. Density plot allows us to visualize and distinguish the kinematic regions with high/low $f(x,k_T^2)$ in $x\text{-}k_T^2$ plane which is more informative then any ordinary 3D plot. In Fig.~\ref{6y} the plots are sketched for two canonical choices of $\lambda$ viz. $\lambda=0.4$ and $0.6$ corresponding to two $\alpha_s$ values 0.15 and 0.23.
Our solution seems to be very sensitive towards a small change in $\lambda$. An apparent 50\% change in $\lambda$ (0.4 to 0.6) suggests approximately around one order of magnitude rise in gluon distribution $f(x,k_T^2)$ for an approximate limit of $x$ and $k_T^2$: $10^{-6}\leq x\leq 10^{-5}$ and $50 \text{ GeV}^2 \leq k_T^2\leq 100 \text{ GeV}^2$. It is also observed that the range of high $k_T^2$ and very small-$x$ is the high gluon distribution $f(x,k_T^2)$ region where gluons are mostly populated. In Fig.~\ref{6xx} we have shown $R$ sensitivity of our solution for two choices of the shadowing parameter $R$ viz. $R= 5 \text{ GeV}^{-1}$ and $2 \text{ GeV}^{-1}$. A satisfactory shadowing effect is observed  from the comparison of the two plots. The extreme form of shadowing $(R=2 \text{ GeV}^{-1})$ is found to suppress atleast $10\text{-}20 \%$ magnitude of gluon density than the conventional shadowing $(R=5 \text{ GeV}^{-1})$ in the high $k_T^2$ and small-$x$ region.

\section{Equation of Critical line and prediction of saturation scale: a differential geometric approach}\label{critical}
\begin{figure*}[tbp]
	\centering 
	\includegraphics[width=.4\textwidth,clip]{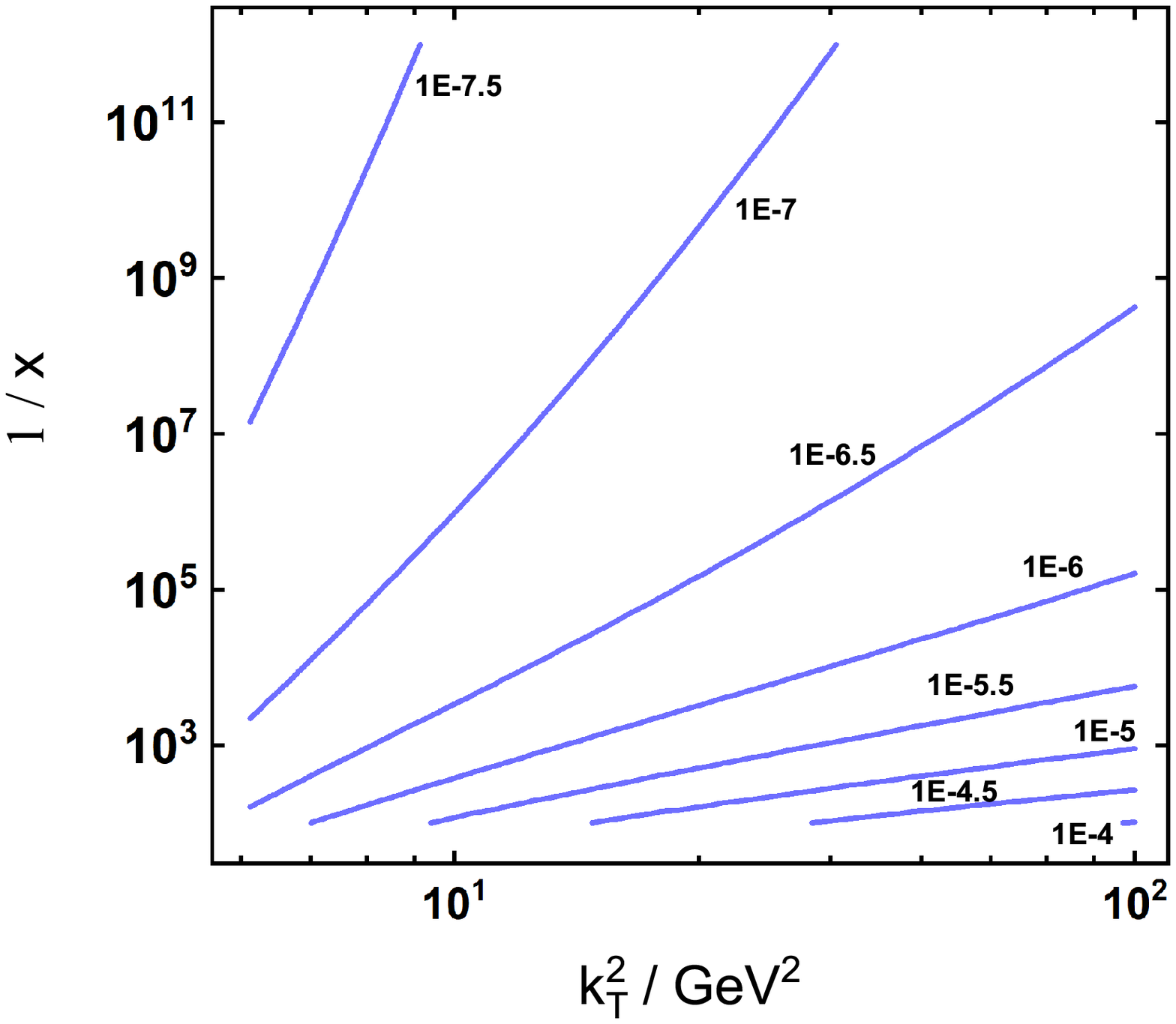}\hspace{1.5cm}
	\includegraphics[width=.4\textwidth,clip]{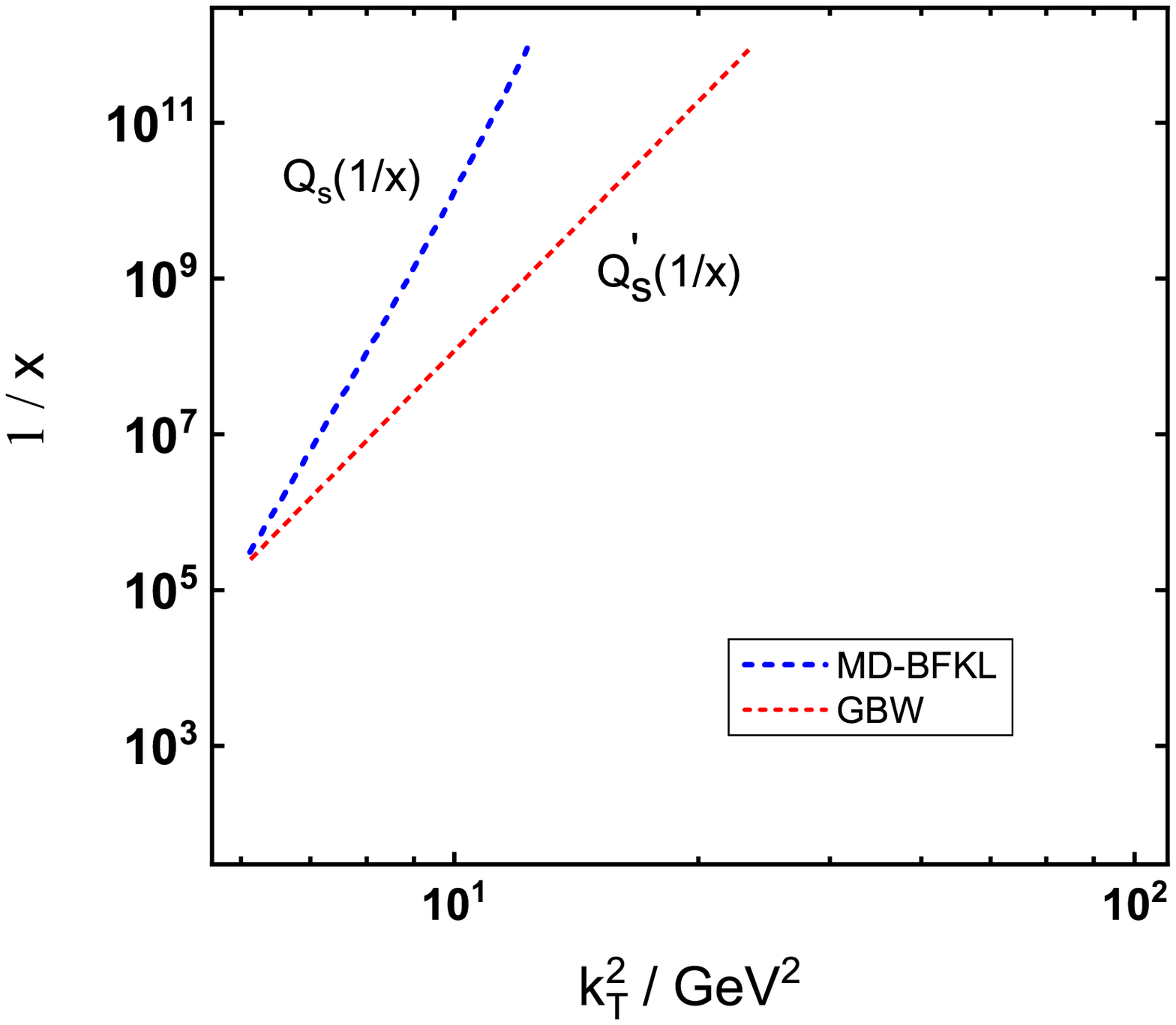}
	\caption{\label{6x} (a) Contour plot obtained by solving \eqref{35x} (left). Solid lines show contours of constant gradients along the curve. (b)Diagram showing critical boundary line separating high and low gluon density region (right). Blue dashed line is obtained from \eqref{38x} while red dashed line is corresponding to GBW model \cite{21}.  }
\end{figure*}
The so-called saturation physics allows one to study the high parton density region in the small coupling regime. The transition from the linear region to saturation region is characterized by the saturation scale. The saturation momentum scale $Q_s$ is the threshold transverse momentum for which non-linearity becomes visible. The boundary in $(x, Q^2)$ or $(x,k_T^2)$ plane along which saturation sets in is characterized by the critical line. An important feature of our analytical solution to the  KC improved MD-BFKL equation is the finding of the equation of the critical line which is supposed to mark the boundary between dilute and dense partonic system in $x\text{-}k_T^2$ phase space. 
\par Although the gluon saturation can be achieved only when $Q_s\sim Q$ (or $k_T^2$), the observables already become sensitive to $Q_s$ during the approach to saturation regime. This property is known as geometrical scaling in DIS inclusive event which means that instead of depending on $k_T^2$ and $x$ separately, the gluon distribution depends on a single dimensionless variable $\frac{k_T^2}{Q_s^2}$ i.e.
\begin{equation}
	\label{31s}
	\phi(x,k_T^2)\equiv\phi\left(\frac{k_T^2}{Q_s^2}\right).
\end{equation}
In recent years, this geometrical scaling property of DIS observables is studied very extensively for various frameworks \cite{42,56, 57, 58,84}. In \cite{56} an analysis of the saturation scale has been performed in the platform of resummed NLLx BFKL where the saturation scale was calculated via the relation
\begin{equation}
	\label{32s}
	-\frac{d\omega(\gamma_c)}{d\gamma_c}=\frac{\omega_s(\gamma_c)}{1-\gamma_c},
\end{equation}
which has been repeatedly derived in several literature \cite{59,60,61}. In \cite{42} the saturation scale $Q_s$ was obtained from the numerical solution of a nonlinear equation by finding the maximum of the momentum distribution of the gluon. Another approach for determination of $Q_s$ can be found in \cite{5} where a parameter $\beta$ is defined as the relative difference between the solutions to the linear and nonlinear equation,
\begin{equation}
	\label{33s}
	\beta=\frac{|f^\text{lin}(x,Q_s^2(x,\beta))-f^\text{lin}(x,Q_s^2(x,\beta))|}{f^\text{lin}(x,Q_s^2(x,\beta))}.
\end{equation} 
The crucial parameter $\beta$ actually depicts the percentage deviation that the non linear solution shows from the linear one and it lies in the order of $0.1 \text{-} 0.5$ (or $10\text{-}50 \%$ deviation).

\par Our approach towards studying geometrical scaling and critical line is primarily based on the basic understanding from differential geometry, in particular gradient of a function which is considered as the direction of steepest ascent of that function. Each component of the gradient gives us the rate of change of the function with respect to some standard basis i.e. it gives us an idea about how fast our function grows or decays or saturates with respect to the change of the variables. One important advantage of choosing gradient is that it is a two dimensional object since it does not possess any component along $f(x,k_T^2)$ axis in $\mathbb{R}^3$. This ensures that it does not have any direct dependence on the magnitude of gluon density $f(x,k_T^2)$, rather it depends on the rate at which $f(x,k_T^2)$ changes with respect to $x$ and $k_T^2$ change. This property of gradient actually helps us in distinguishing out the saturation region and linear region although the distribution function $f(x,k_T^2)$ has large variation in order of magnitude for different regions in $x\text{-}k_T^2$ phase space.
\par Recall that towards small-$x$, gluon evolution is suppressed due to shadowing effect as sketch in Fig.~\ref{5x}  which motivates us to evaluate the gradient of $f(x,k_T^2)$ particularly along $x$ basis. For simplicity we consider an unit vector $\vec{\nu}$ along $\tilde{\eta} \text{ }( = 1/x)$ basis, we can project the gradient $\nabla f(\tilde{\eta},k_T^2)$ along $\vec{\nu}$ via dot product $\nabla f(\tilde{\eta},k_T^2)$.$\vec{\nu}$. This scalar quantity can also be interpreted as the directional derivative along the direction $\vec{\nu}$,
\begin{equation}
	\label{34x}
	g\equiv \nabla_{\vec{\nu}} f(\tilde{\eta},k_T^2) = \nabla f(\tilde{\eta},k_T^2).\vec{\nu}.
\end{equation}
Taking the Euclidean norm yields
\begin{equation}
	\label{35x}
	g = \pm|\nabla_{\vec{\nu}} f(\tilde{\eta},k_T^2)|,
\end{equation}
where the negative (-) sign is for the descending function (or negative slope).
\par We obtained a family of contours (or level curves)  $\tilde{\eta}(k_T^2)$ in $\tilde{\eta}\text{-}k_T^2$ plane solving \eqref{35x} as shown in Fig.~\ref{6x}(a). Each contour depicts a constant gradient $g$ along the curves. The set of contours can also be identified as some set of possible saturation scales. As sketch in Fig.~\ref{6x}(a) the $\tilde{\eta}\text{-}k_T^2$ plane is divided into two regions: low gluon density region where the spacing between two consecutive contours is very small and high gluon density region where the spacing becomes very large compared to previous. This distinction in contour spacing in the two regions comes from the fact that the gradient changes very fast until the saturation is reached and then after reaching the saturation boundary gradient changes very slowly or almost freezes for further increase in $\tilde{\eta}(k_T^2)$ as can be seen in Fig.~\ref{6x}(a). In high gluon density region, the contour curves tend to $\tilde{\eta}(k_T^2)\rightarrow \infty $ towards high $k_T^2$. For low gluon density region, shadowing effects are negligible and the contours become almost parallel straight lines.

\par Let us try to find the equation of critical line which divides the two regions in $\tilde{\eta}\text{-}k_T^2$ space. The level curves of the function $g=\pm| \nabla_{\vec{\nu}} f(\tilde{\eta},k_T^2)| $ are two dimensional curves that can be obtained by setting $g=k$ where $k$ is a constant $(k\in \mathbb{R})$. Therefore, the equations of the level curves are given by 
\begin{equation}
	\label{36x}
	| \nabla_{\vec{\nu}} f(\tilde{\eta},k_T^2)|=\pm k.
\end{equation}
Now for a known initial saturation momentum scale $Q_{s0}(1/x_0)$ we can predict equation of  the critical line using  \eqref{35x} and \eqref{36x}. The equation of the critical line is the equation for the level curve of the function $g=\pm| \nabla_{\vec{\nu}} f(\tilde{\eta},k_T^2)| $ that passes through the point $(Q_{s0}(\tilde{\eta}_0),\tilde{\eta}_0)$. First we find the value of $k$ by plugging the point $(Q_{s0}(\tilde{\eta}_0),\tilde{\eta}_0)$ into \eqref{36x}
\begin{equation}
	\label{37x}
	k_{s0}=\pm| \nabla_{\vec{\nu}} f(Q_{s0}^2(\tilde{\eta}_0))|.
\end{equation}
Now the level curve passing through $(Q_{s0}(\tilde{\eta}_0),\tilde{\eta}_0)$ is obtained by setting
\begin{equation}
	\label{38x}
	| \nabla_{\vec{\nu}} f(\tilde{\eta},Q_s^2)|=| \nabla_{\vec{\nu}} f(Q_{s0}^2(\tilde{\eta}_0))|,
\end{equation}
which is the equation of the critical boundary. The knowledge of an appropriate initial saturation scale $Q_{s0}(1/x_0)$ allows one to separate out the linear and saturation region using \eqref{38x}. In Fig.~\ref{6x}(b) we have sketched a possible critical line obtained from \eqref{38x} for the choice of initial saturation scale $Q_{s0}^2(\eta_0)\backsimeq 2.8 \text{ GeV}^{2}$ at $\eta_0=10^6$ (or $x_0=10^{-6}$) which is taken from the calculation from the original saturation model by Golec-Biernat and Wusthoff \cite{21, 5}. A rough agreement between our prediction and that of GBW model is observed in Fig.~\ref{6x}(b). However, $Q_s$ given by \eqref{38x} is found to have weaker $x$ dependence than the one from GBW model $Q_s^{'2}$. The saturation scale has direct dependence on partons per unit transverse area. Smaller $x$ suggests larger parton density giving rise to a larger saturation momentum scale, $Q_s^2$. In other words the saturation scale $Q_s$ depends on $x$ in such a way that with decreasing $x$ one has to probe  smaller distances or higher $Q^2$ in order to resolve the dense partonic structure of the proton which is clear from our analysis.

\section{KC improved MD-BFKL prediction of HERA DIS data}\label{hera}
\subsection{DIS structure functions and reduced cross section}\label{hera1}
\begin{figure*}[tbp]
	\centering 
	\includegraphics[width=.23\textwidth,clip]{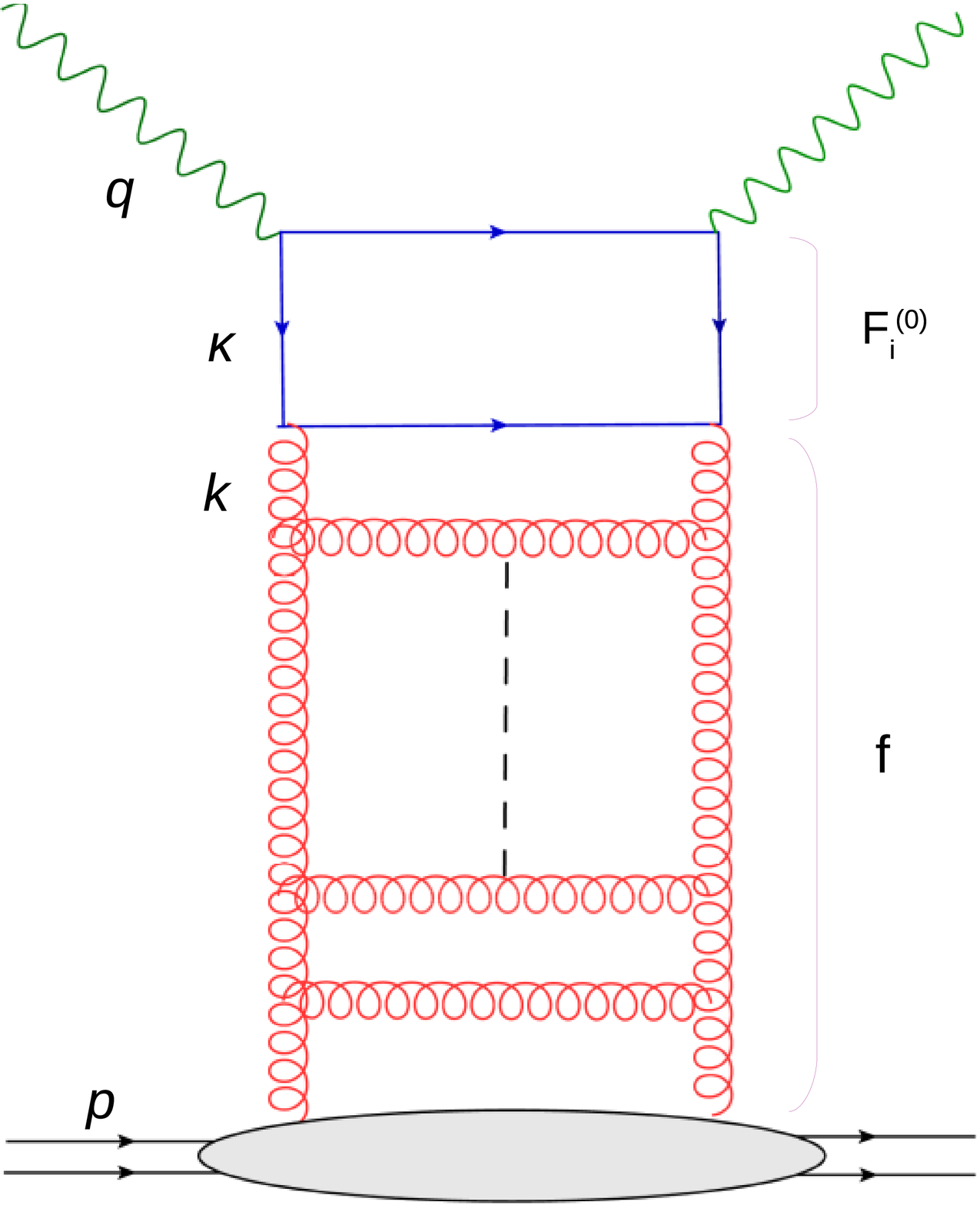}
	\hspace{2cm}
	\includegraphics[width=.23\textwidth,clip]{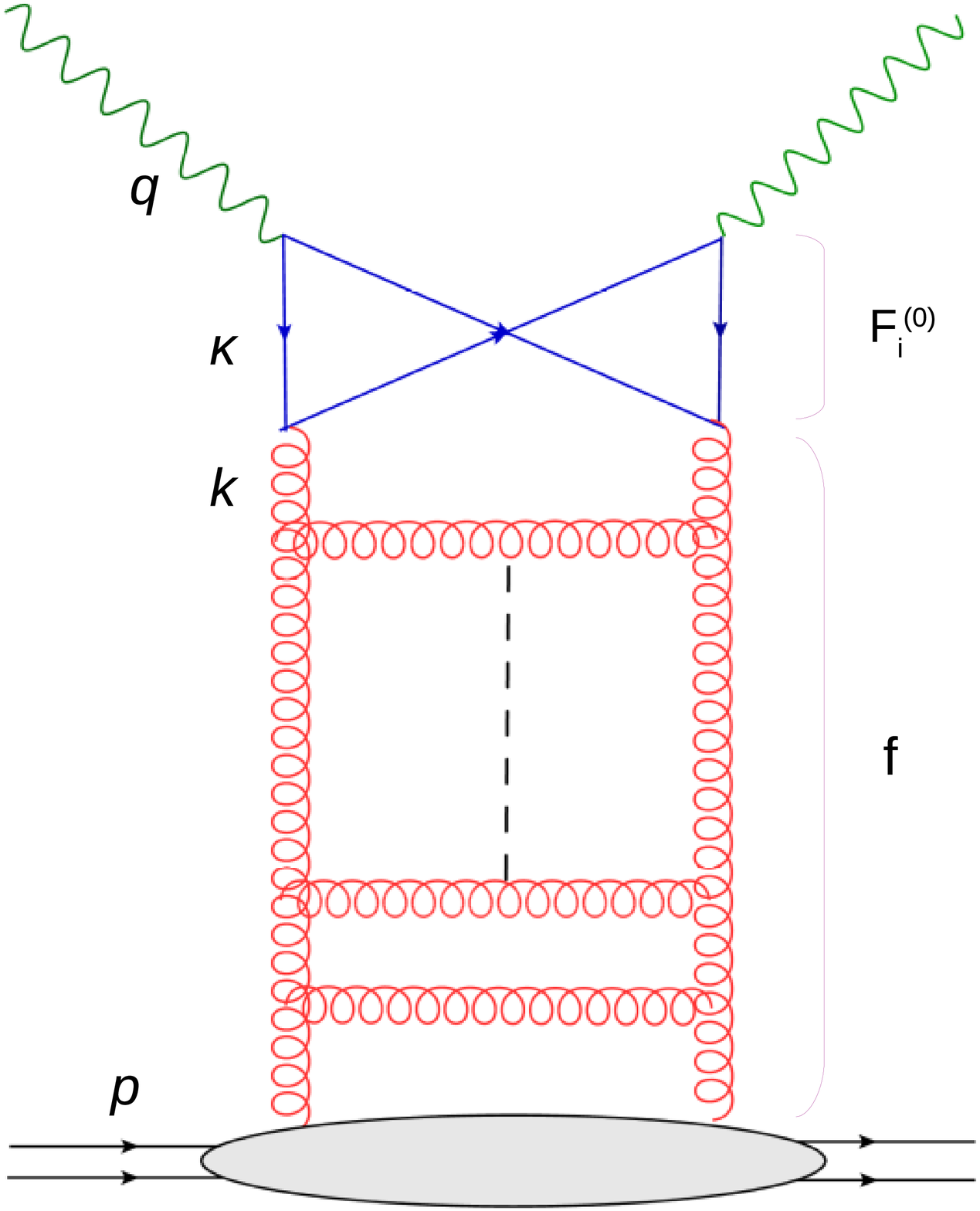}
	
	\caption{\label{d1} Diagrammatic representation of the factorization formula \eqref{42} where gluon couples to virtual photon through the (a) quark box (left) and (b) crossed box (right) diagrams. }
\end{figure*}

In this section we present a quantitative prediction of proton structure functions $F_2(x,Q^2)$ and longitudinal structure function $F_L(x,Q^2)$ as an outcome of our solution to the KC improved MD-BFKL equation. At small-$x$, the sea quark distribution is driven by gluons via. $g\rightarrow q\bar{q}$ process. This component from sea quark distribution can be calculated in perturbative QCD. The relevant diagram for this QCD process is shown in Fig.~\ref {d1} and the contribution to the transverse and longitudinal components of the structure functions can be written using the $k_T$-factorization theorem \cite{25,26}

\begin{equation}
	\label{42}
	F_\text{i}(x,Q^2)=\int_x^1\frac{dx^{'}}{x^{'}}\int\frac{dk_T^2}{k_T^4}f\left(\frac{x}{x^{'}},k_T^2\right)F_\text{i}^{(0)}(x^{'},k_T^2,Q^2),
\end{equation}
where i= T, L and $\frac{x}{x^{'}}$ is the fractional momentum carried by gluon which splits into $q\bar{q}$ pair. $F_\text{i}^{(0)}$ includes both the quark box and cross box contribution which comes from virtual gluon-virtual photon subprocess leading to $q\bar{q}$ production (Fig.~\ref{d1} ). The gluon distribution $f(\frac{x}{x^{'}},k_T^2)$ in \eqref{42} represents the sum of the gluon ladder diagrams in the lower part of the box as shown in Fig.~\ref {d1} is given by BFKL equation \cite{2}. Here we will study the effect of gluon shadowing using the  solution $f(x,k_T^2)$ of KC improved MD-BFKL in \eqref{42}.

\par The explicit expression for quark box contribution $F_\text{i}^{0}$ can be obtained by writing four momentum in terms of the convenient light like momenta $p$ and  $q^{'}\equiv q +xp$, where $x=\frac{Q^2}{2p.q}$ and $Q^2=-q^2$ like as usual (see Fig.~\ref{d1} ). Now we can decompose quark and gluon momentum in terms of sudakov variables
\begin{equation}
	\begin{split}
		&\kappa=\alpha p-\beta q^{'}+\kappa_T,\\
		&k= a p+bq^{'}+k_T.
	\end{split}
\end{equation}
The integration should be performed over the box diagram subject to quark mass-shell constraints \cite{8}

\begin{align}
	\begin{split}
		&(\alpha -x)2p.q(1-\beta)-\kappa_T^2=m_q^2,\\
		&(a-\alpha)2p.q\beta-(\kappa_T-k_T)^2=m_q^2,
	\end{split}
\end{align}
which leads to the box contribution \cite{29}
\begin{align}
	\label{43}
	\begin{split}
		&\tilde{F}_T^{(0)}(k_T^2,Q^2)=2 \sum_q e_q^2\frac{Q^2}{4\pi^2}\int_{0}^{1}d\beta\int d^2\kappa_T\alpha_s(\kappa_T)\\&\times\big\{[\beta^2+(1-\beta)^2]\bigg[ \frac{\kappa_T^2}{L_1^2}-\frac{\kappa_T.(\kappa_T-k_T)}{L_1L_2}\bigg]+\frac{m_q}{L_1^2}-\frac{m_q^2}{L_1L_2} \big\},
		\end{split}
	\end{align}
	\begin{align}
	\label{44}
	\begin{split}
		\tilde{F}_L^{(0)}(k_T^2,Q^2)=&2 \sum_q e_q^2\frac{Q^4}{\pi^2}\int_{0}^{1}d\beta\int d^2\kappa_T\alpha_s(\kappa_T)\\&\times\beta^2(1-\beta)^2\bigg( \frac{1}{L_1^2}-\frac{1}{L_1L_2}\bigg),
		\end{split}
\end{align}
where the denominators $L_\text{i}$ are
\begin{equation*}
	\begin{split}
		&L_1=\kappa_T^2+\beta(1-\beta)Q^2+m_q^2,\\
		&L_2=(\kappa_T-k_T)^2+\beta(1-\beta)Q^2+m_q^2.
	\end{split}
\end{equation*}

Note that $\tilde{F}_\text{i}^{(0)}(k_T^2,Q^2)\equiv \int_x^1\frac{dx^{'}}{x^{'}}F_\text{i}^{(0)}(x^{'},k_T^2,Q^2)$ i.e. the $x^{'}$ integration of \eqref{42} is implicit in the $d^2k_T^{'}$ and $d\beta$ integration in \eqref{43} and \eqref{44}. Now plugging \eqref{43} and \eqref{44} in the $k_T$-factorization formula \eqref{42} we get
\begin{align}
	\label{45}
	\begin{split}
		F_T(x,Q^2)=&2 \sum_q e_q^2\frac{Q^2}{4\pi^2}\int_{k_0^2}^\infty\frac{dk_T^2}{k_T^4} \int_{0}^{1}d\beta\int d^2\kappa_T\alpha_s(\kappa_T)\\&\times\big\{[\beta^2+(1-\beta)^2]\bigg[ \frac{\kappa_T^2}{L_1^2}-\frac{\kappa_T.(\kappa_T-k_T)}{L_1L_2}\bigg]\\
		&+\frac{m_q}{L_1^2}-\frac{m_q^2}{L_1L_2} \big\}f(\frac{x}{x^{'}},k_T^2),
		\end{split}
	\end{align}
\begin{align}
\begin{split}	
	\label{46}
		F_L(k_T^2,Q^2)=&2 \sum_q e_q^2\frac{Q^4}{\pi^2}\int_{k_0^2}^\infty\frac{dk_T^2}{k_T^4}\int_{0}^{1}d\beta\int d^2\kappa_T\alpha_s(\kappa_T)\\&\times\beta^2(1-\beta)^2\bigg( \frac{1}{L_1^2}-\frac{1}{L_1L_2}\bigg)f(\frac{x}{x^{'}},k_T^2).
	\end{split}
\end{align}
\begin{figure*}[tbp]
	\label{}
	\centering 
	\includegraphics[trim=21 0 21.5 0,width=.45\textwidth,clip]{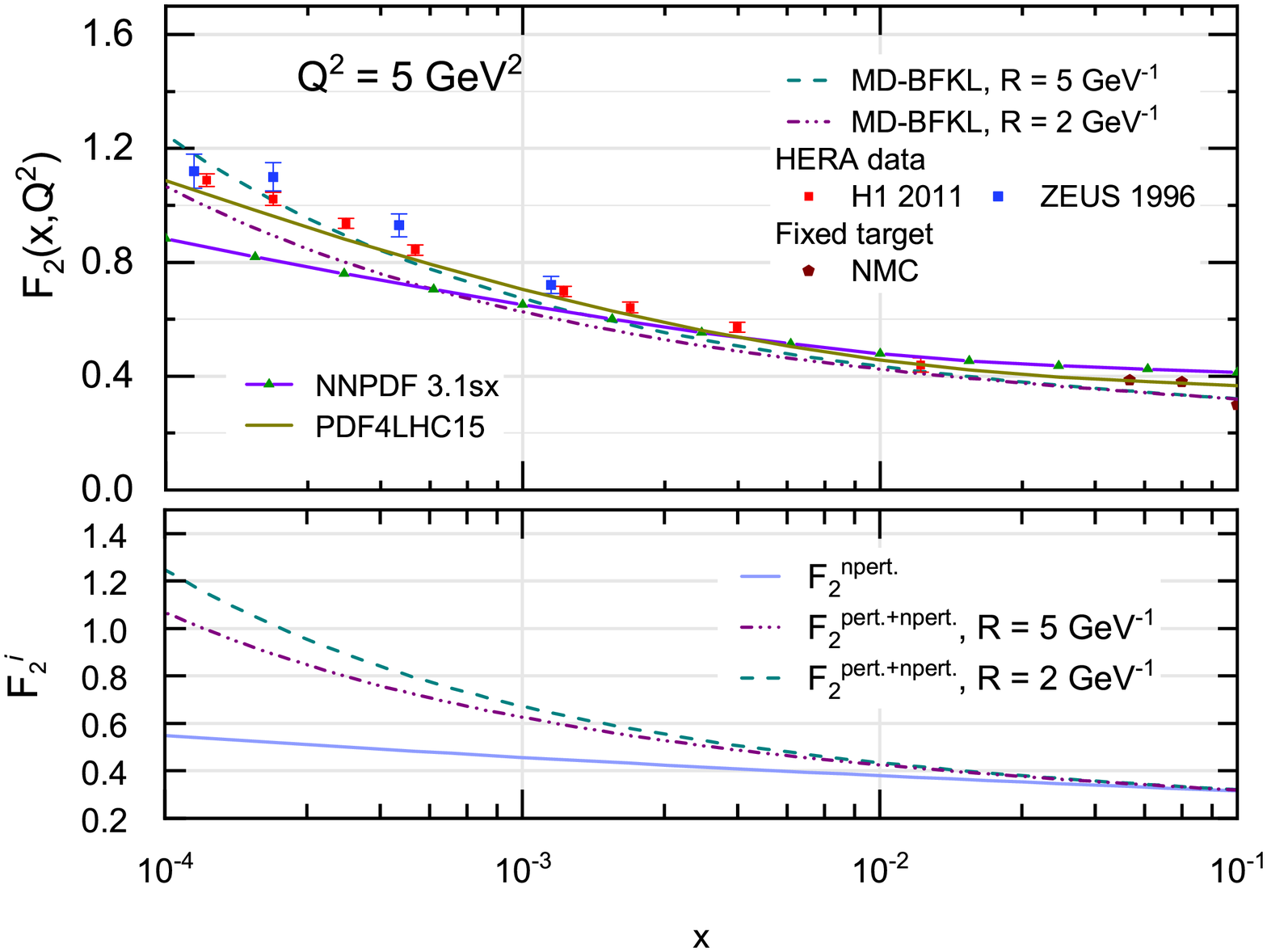}\hspace{5mm}
	\includegraphics[trim=21 0 21.5 0,width=.45\textwidth,clip]{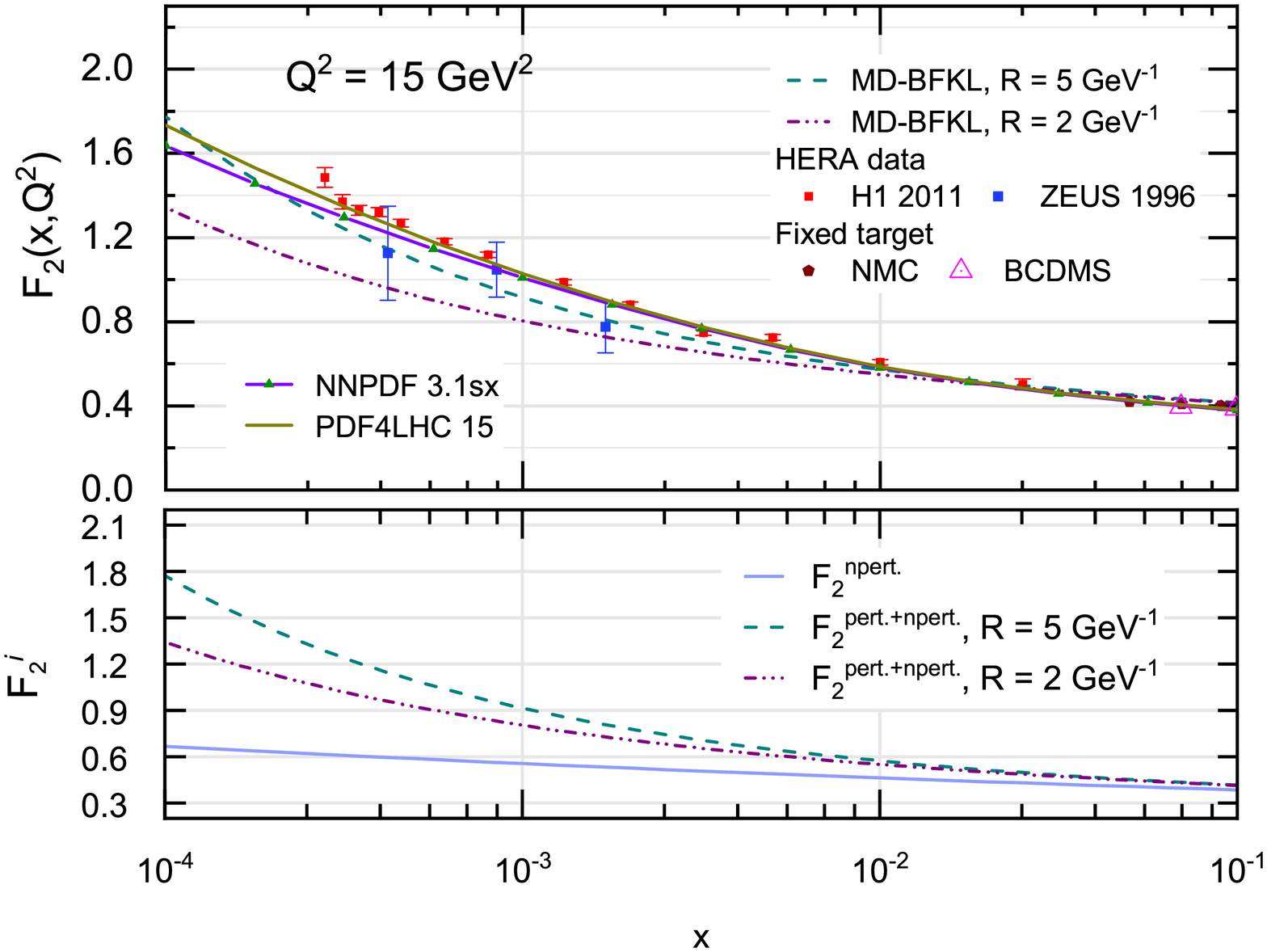}\\
	\includegraphics[trim=21 0 21.5 0,width=.45\textwidth,clip]{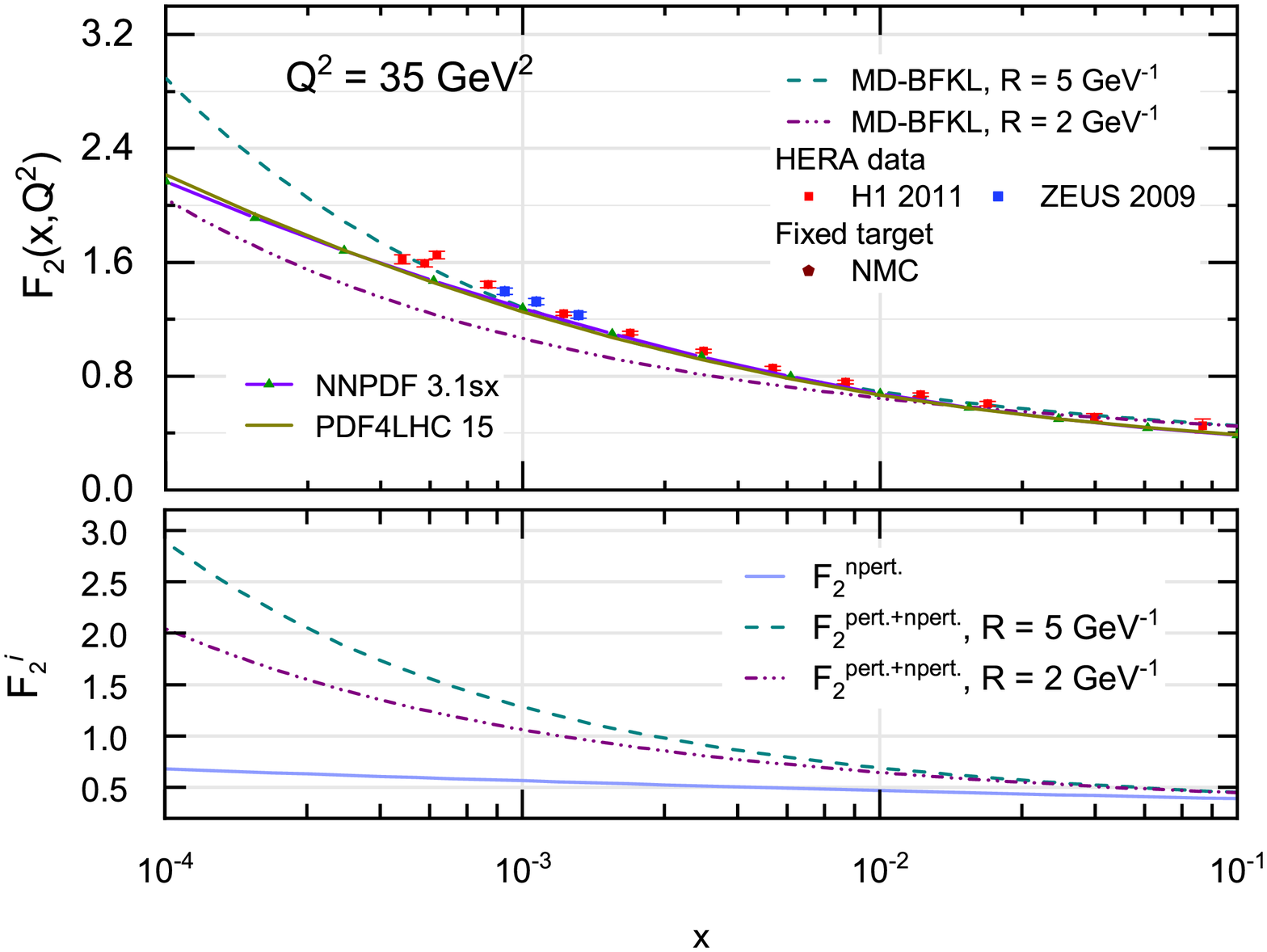}\hspace{5mm}
	\includegraphics[trim=21 0 21.5 0,width=.45\textwidth,clip]{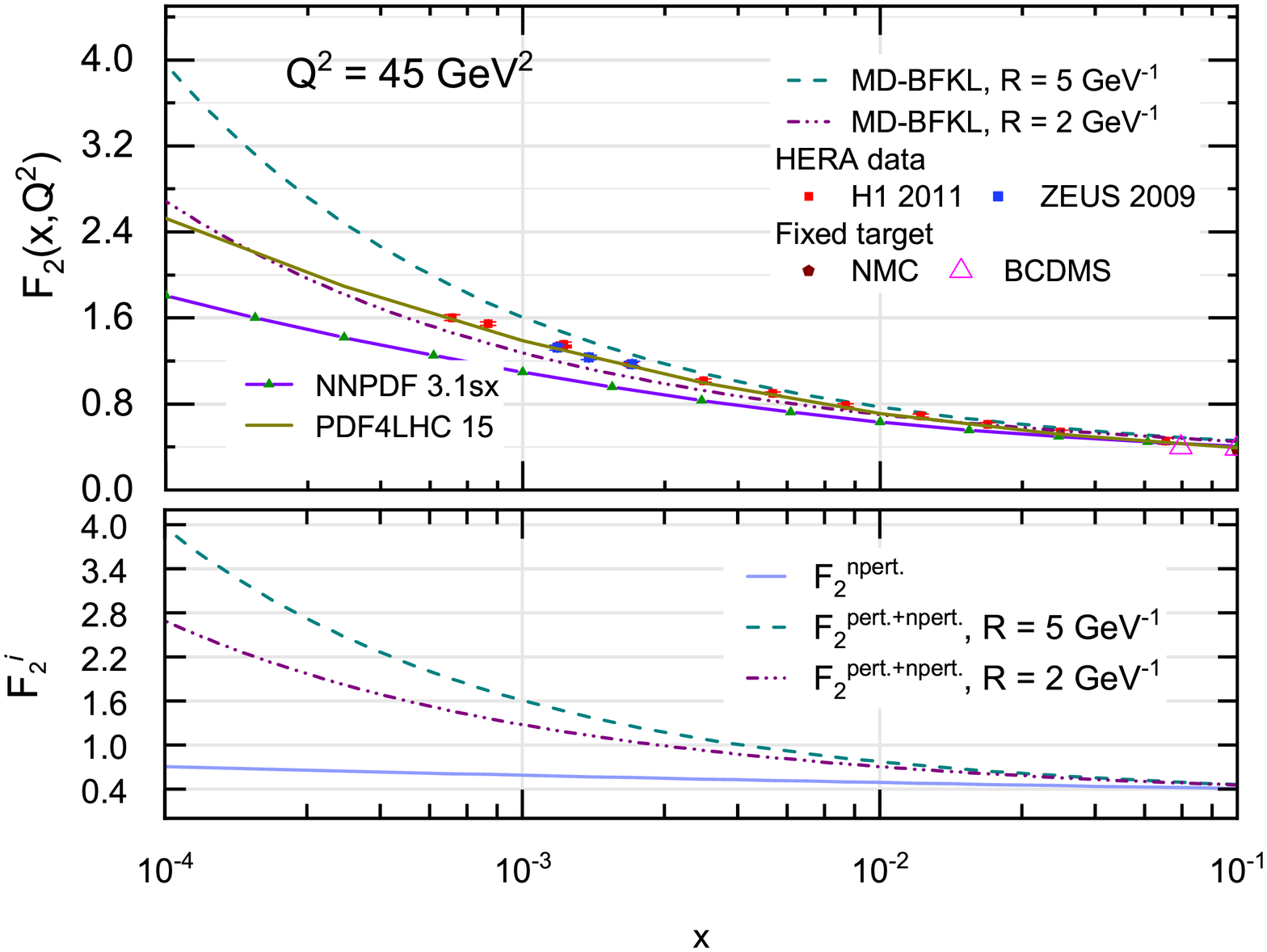}

	\caption{\label{7x} Prediction for proton structure function F$_2$ obtained for two choices of shadowing: conventional ($R= 5 \text{ GeV}^{-1}$) and "hotspot" ($R= 2 \text{ GeV}^{-1}$). Data are taken from HERA (H1 \cite{32} and ZEUS \cite{33}) as well as fixed target experiment (NMC \cite{31} and BCDMS \cite{34}). Data-sets from global parametrization groups NNPDF3.1sx \cite{36} and PDF4LHC 15 \cite{35} is also included. The background contribution is given by \eqref{49} with $F_2^{BG}(x_0=0.1)\approx$ 0.316, 0.384, 0.391 and 0.406 corresponding to four $Q^2$ values viz. $5 \text{ GeV}^2, 15 \text{ GeV}^2, 35 \text{ GeV}^2$ and $45 \text{ GeV}^2$. Separate plots of $F_2^{\text{npert.}}$ vs. $F_2^{\text{npert.+pert.}}$ is also sketched.
	}
\end{figure*}
Equations \eqref{45} and \eqref{46} serve as the basic tool for calculating structure functions at small-$x$ provided that the gluon distribution $f(x,k_T^2)$ is known. An analytical approach towards the calculation of $F_2$ and $F_L$ can be found in literature \cite{27,28} for fixed coupling case using linear BFKL equation. In the literature \cite{9}, structure functions are calculated taking gluon distribution $f(x,k_T^2)$ from the numerical solution of unitarized BFKL equation for running coupling consideration. It has been seen that the analytical approach \cite{28} considerably overestimates the actual (numerical) solution \cite{9}. This is because of the fact that analytical approach neglects terms down only by powers of $\ln (1/z)^{-1}$ as well as it does not accommodate running of strong coupling. However, our approach is a semi-analytical one, in the sense that we take gluon distribution from our analytical solution while further integrations are performed numerically. This indeed allows us to check the feasibility of our analytical solution in describing DIS data. 
\par In \eqref{42}  $m_q$ denotes the quark mass and it is taken to be $m_q=1.28 \text{ GeV}$ for charm quark while  massless ($m_q=0$) for light quarks (u, d and s). Our phenomenology is limited for light quark therefore putting $m=0$ in \eqref{43} and replacing the quark transverse momentum by $\kappa_T=\kappa_T^{'}+(1-\lambda)k_T$ we obtain \cite{28}
\begin{equation}
	\label{47}
	\begin{split}
		&\tilde{F}_T^{(0)}(k_T^2,Q^2)=\\& \sum_q e_q^2\frac{\alpha_s}{\pi}Q^2k_T^2\int_{0}^{1}d\beta\int_{0}^{1}d\lambda\int_0^{\infty} d\kappa_T^{'2}[\beta^2+(1-\beta)^2]\\&\times\lambda\frac{(2\lambda-1)\kappa_T^{'2}+(1-\lambda)[\lambda(1-\lambda)k_T^2+\beta(1-\beta)Q^2]}{[\kappa_T^{'2}+\lambda(1-\lambda)k_T^2+\beta(1-\beta)Q^2]^3}.
	\end{split}
\end{equation}
After integrating over $\kappa_T^{'2}$ one can arrive at
\begin{equation}
	\label{48}
	\begin{split}
		\tilde{F}_T^{(0)}(k_T^2,Q^2)=& \sum_q e_q^2\frac{\alpha_s}{\pi}Q^2k_T^2\int_{0}^{1}d\beta\int_{0}^{1}d\lambda\\&\times\frac{[\lambda^2+(1-\lambda)^2][\beta^2+(1-\beta)^2]}{\lambda(1-\lambda)k_T^2+\beta(1-\beta)Q^2}.
	\end{split}
\end{equation}
Equation \eqref{47} and \eqref{48}   are written in terms of Feynman integral which actually eliminates the azimuthal dependence and reduces the two fold integral $d^2\kappa_T$ of \eqref{43} to single integral $\pi d\kappa_T^{'2}$ . From \eqref{48} it is clear that $\tilde{F}_T^{(0)}(k_T^2,Q^2)$ or $F(x^{'},k_T^2,Q^2)$ possess the dimension of $k_T^2$. Therefore, $F(x^{'},$$k_T^2,$ $Q^2)$ or more conveniently $F(x^{'},k_T^2,Q^2)/k_T^2$ may be considered as the structure function of an off mass shell gluon of approximate virtuality $k_T^2$. In \cite{27} differential structure functions have been studied for fixed coupling and it is found that the ratio between longitudinal $F_L$ and transverse structure function $F_T$ is 2:9 for fixed coupling approximation. We have considered this ratio directly in our calculation of longitudinal structure function. Finally, we have taken an assumption $f(x/x^{'}.k_T^2)\rightarrow f(x,k_T^2)$ i.e. we ignore the $x^{'}$ dependence of $f(x/x^{'}.k_T^2)$ which is reasonable in LLx accuracy since 
\begin{equation*}
	(\ln\frac{x}{x^{'}})^n=(\ln x)^n[1+O(1/\ln x)].
\end{equation*}
The advantage of taking this assumption is that we do not have to impose the possible constraint \cite{8} coming from $x/x^{'}<1$ on the region of integration. In principle, the factorization formula \eqref{42} require to be run down to $k_T^2=0$. The integral itself is infrared finite as both the functions $f(\frac{x}{x^{'}},k_T^2)$ and $F_i^{(0)}(x^{'},k_T^2,Q^2)$ vanish at $k_T^2=0$. However, BFKL dynamics is based on perturbative QCD which is not expected to hold the nonperturbative small $k_T^2$ physics. On the other hand, for small $k_T^2$ the gluon distribution vanishes linearly with the decrease in $k_T^2$ on account of gauge invariance \cite{9} making the contribution small. Therefore, we have neglected this small contribution from small $k_T^2$ region in our calculations of unintegrated gluon distribution $f(x,k_T^2)$.

\begin{figure*}[tbp]
	\label{}
	\centering 
	\includegraphics[trim=21 0 21.5 0,width=.45\textwidth,clip]{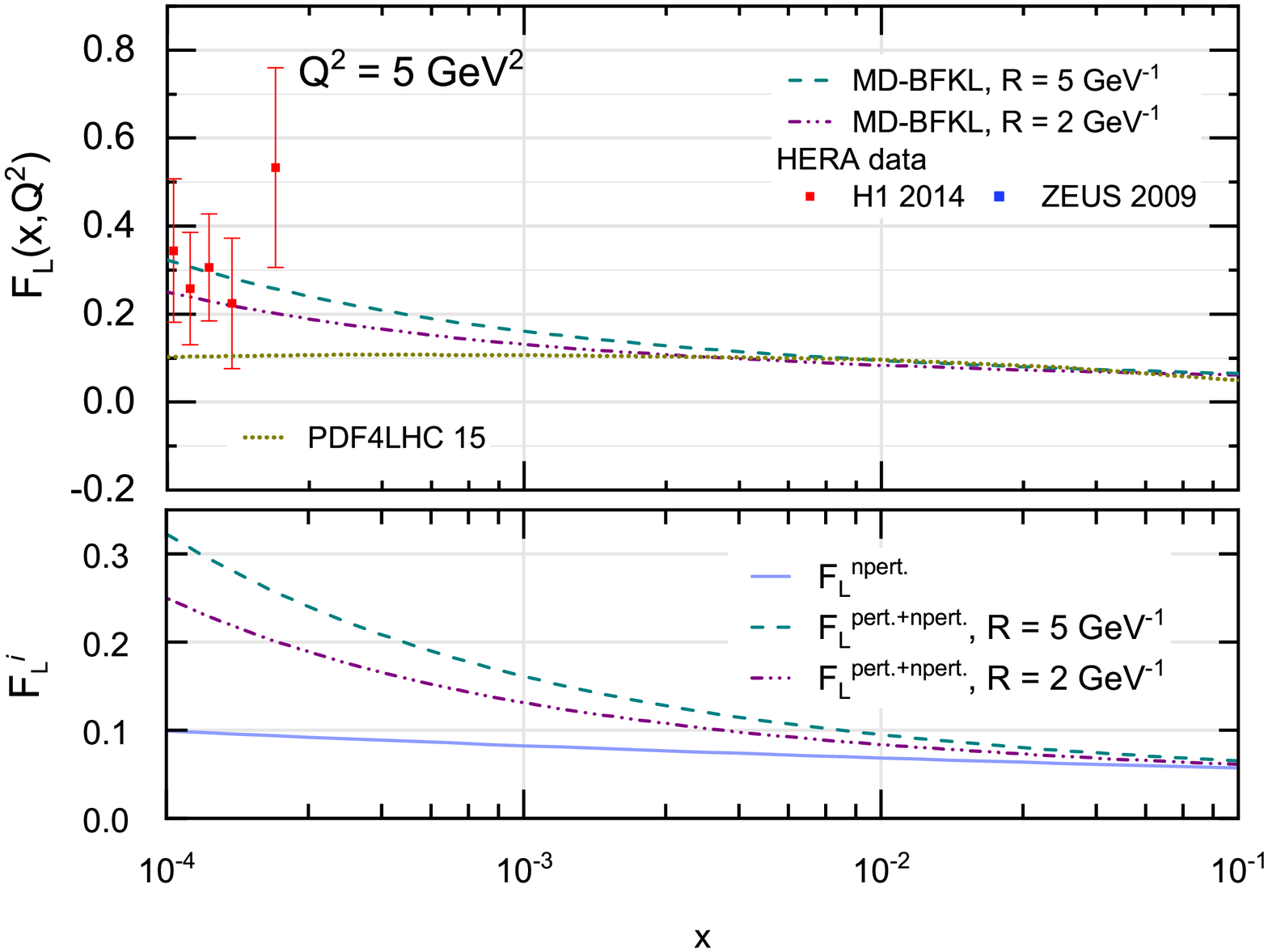}\hspace{5mm}
	\includegraphics[trim=21 0 21.5 0,width=.45\textwidth,clip]{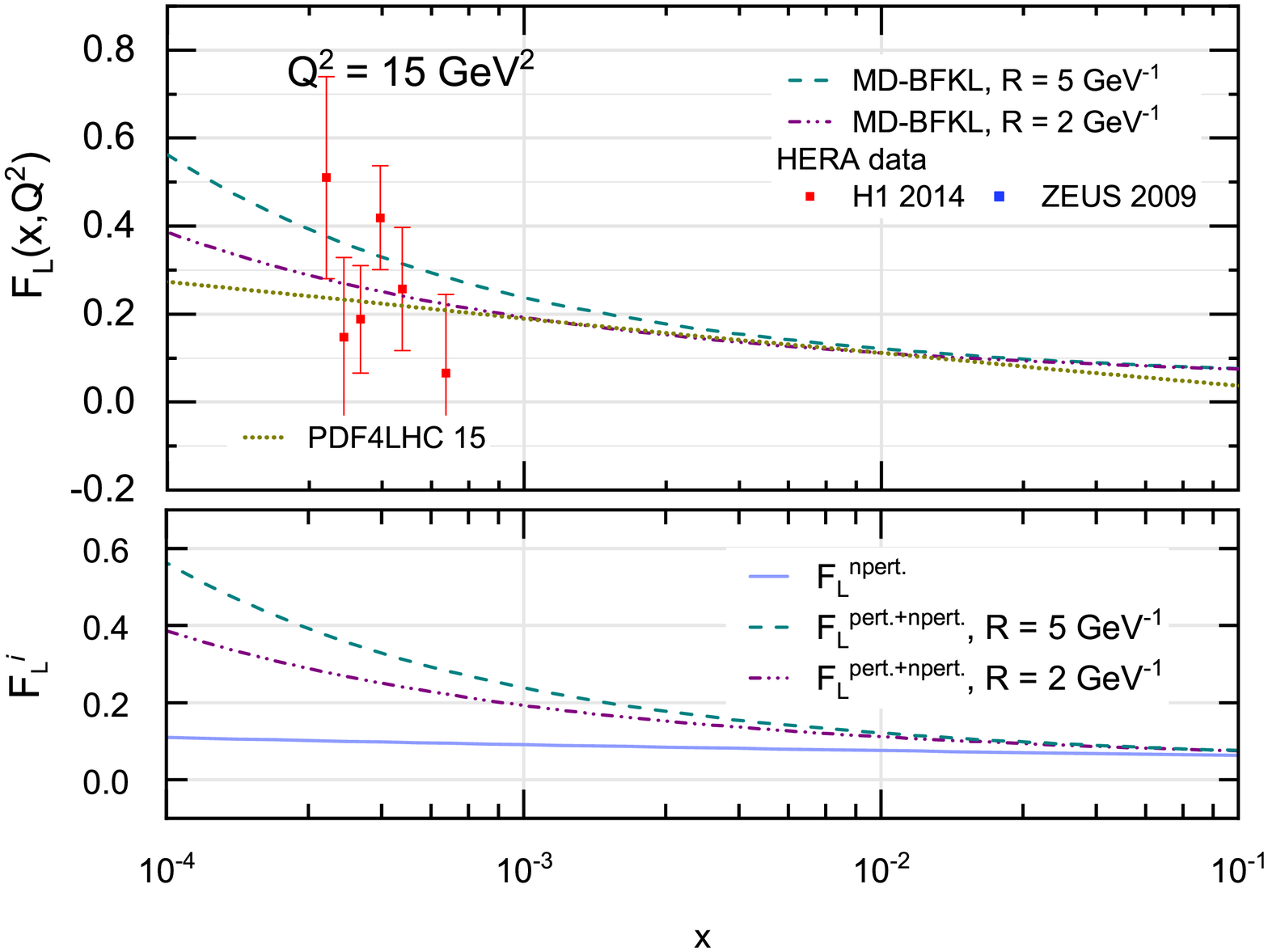}\\
	\includegraphics[trim=21 0 21.5 0,width=.45\textwidth,clip]{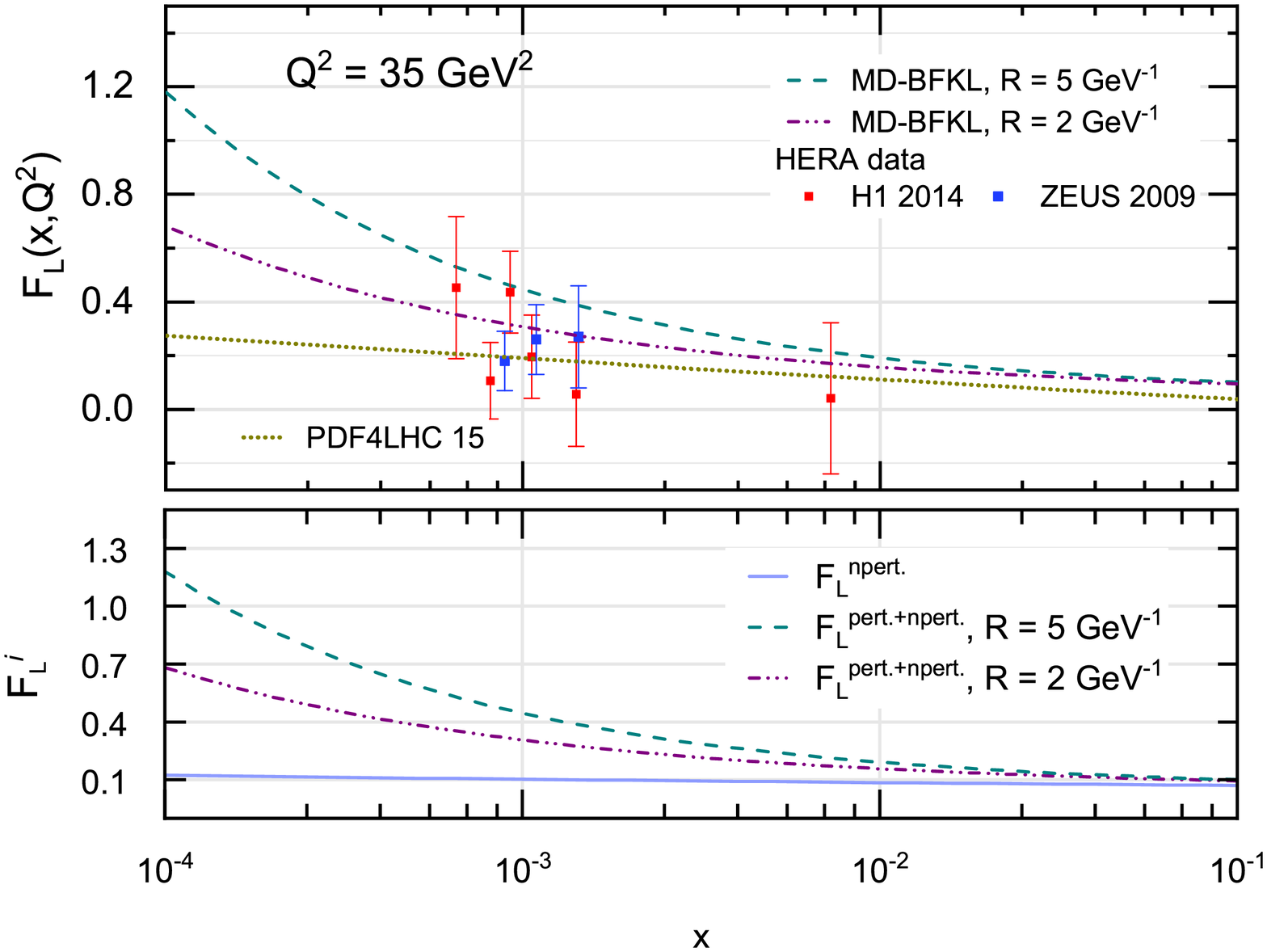}\hspace{5mm}
	\includegraphics[trim=21 0 21.5 0,width=.45\textwidth,clip]{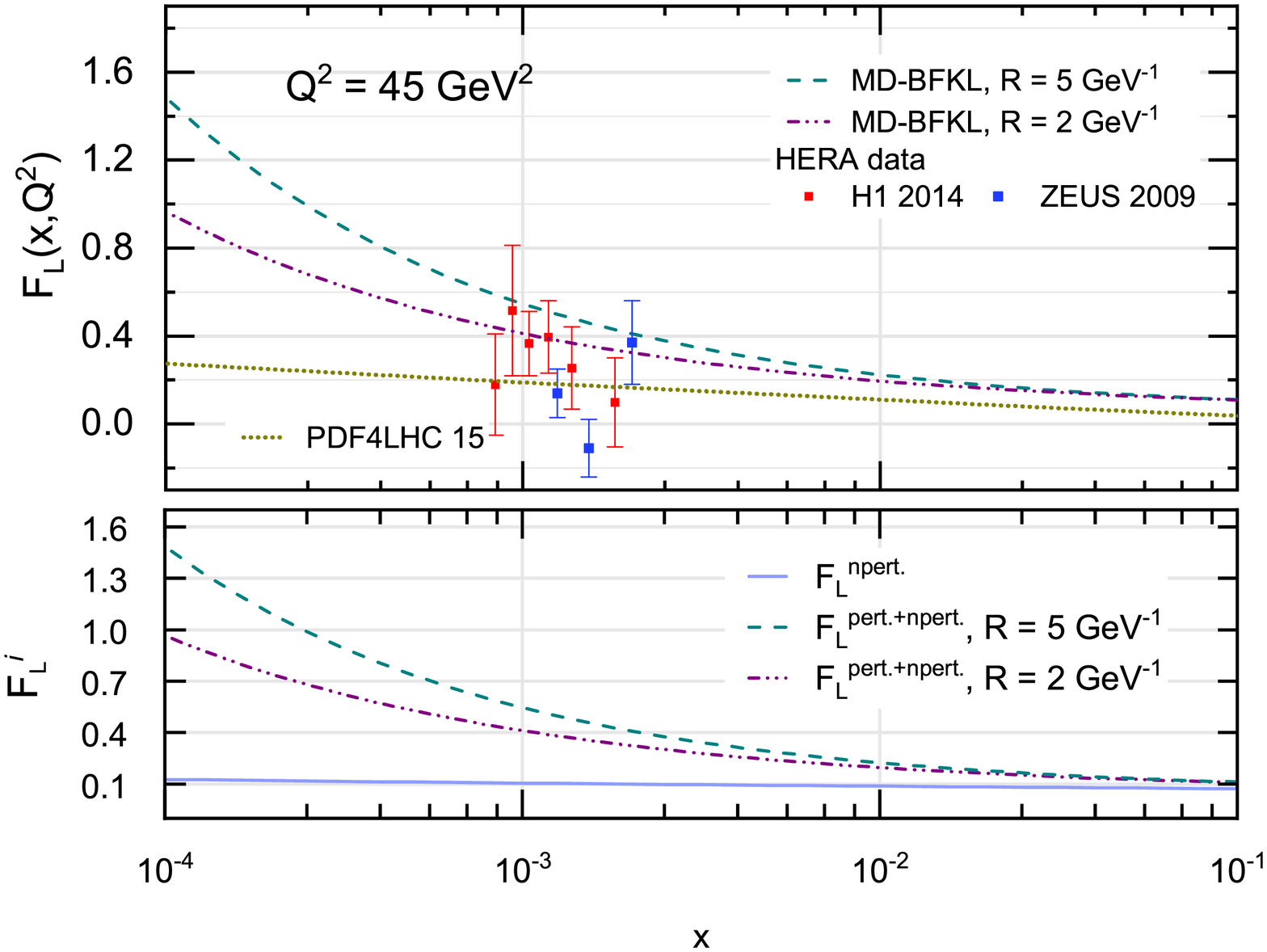}

	\caption{\label{77x} Prediction for proton structure function $F_L$ obtained for two choices of shadowing: conventional ($R= 5 \text{ GeV}^{-1}$) and "hotspot" ($R= 2 \text{ GeV}^{-1}$). Data are taken from HERA H1 \cite{49} and ZEUS \cite{33}. Data-sets from global parametrization group PDF4LHC 15 \cite{35} is also included. The background contribution is given by \eqref{49} with $F_L^{BG}(x_0=0.1)\approx$ 0.057, 0.063, 0.071 and 0.073 corresponding to four $Q^2$ values viz. 5 $\text{ GeV}^2$, 15 $\text{ GeV}^2$, 35 $\text{ GeV}^2$ and 45 $\text{ GeV}^2$. Separate plots of $F_2^{\text{npert.}}$ vs. $F_2^{\text{npert.+pert.}}$ is also sketched. }
\end{figure*}
\begin{figure}[tbp]
	\centering 
	\includegraphics[width=.48\textwidth,clip]{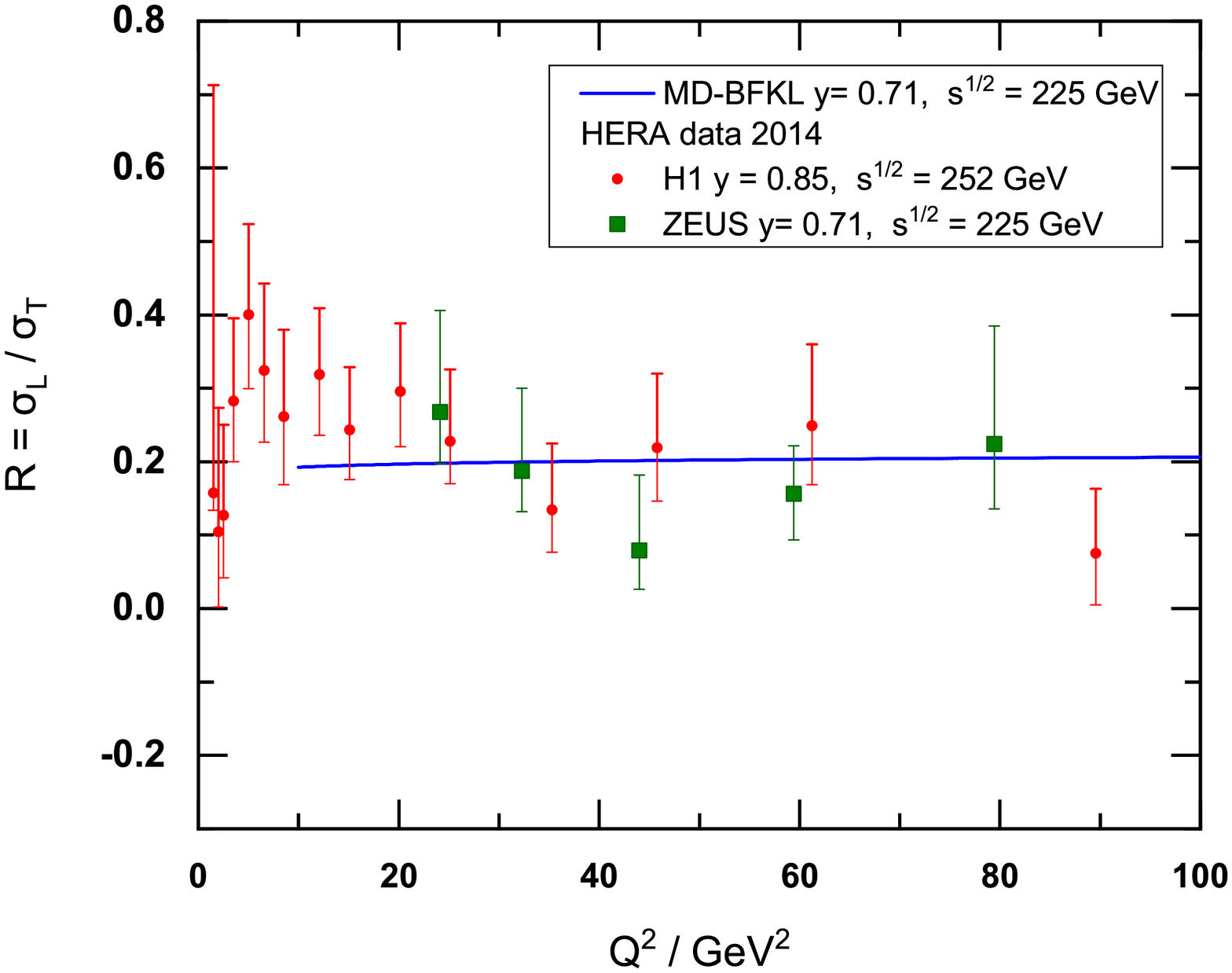}
	\caption{\label{Rx} pQCD prediction from KC improved MD-BFKL for the ratio $R(Q^2)$ in the kinematic range $2 \text{ GeV}^2\leq Q^2\leq 100 \text{ GeV}^2$. The data \cite{49} are from H1  ($\sqrt{s}= 252\text{ GeV}$) and ZEUS ($\sqrt{s}= 225\text{ GeV}$) experiment. The theoretical calculation are for $\sqrt{s}= 225\text{ GeV}$ analogous to c.m.s. of ZEUS .
	}	
\end{figure}

\par Before proceeding to the realistic estimation of the structure functions, we add on the "background" some non-BFKL contribution, $F_\text{i}^{\text{BG}}$ to $F_\text{i}$  since the above prediction of structure function is not enough for describing DIS data \cite{9}. This is because \eqref{45} and \eqref{46} represent only the LLx gluon contribution and they are not the only contributions to the DIS structure functions. Although gluonic contribution is dominant at small-$x$, towards higher $x$ their effect becomes weak and we cannot neglect other non-BFKL contribution. For instance, we assume that $F_\text{i}^{\text{BG}}$ evolves like $x^{-0.08}$ motivated from soft pomeron intercept $\alpha_{P}(0)=1.08$ \cite{30}. To be precise we use 
\begin{equation}
	\label{49}
	F_\text{i}^{\text{BG}}(x,Q^2)=F_\text{i}(x_0,Q^2)\left(\frac{x}{x_0}\right)^{-0.08}.
\end{equation}
An intelligent way of calculating this non-BFKL contribution is to choose $x_0$ at some high $x$ and then take $F_\text{i}^{\text{BG}}(x_0,Q^2)$ from data \cite{31} which is also listed in the figure caption. The recent HERA DIS data taken for comparison with our result can be found in \cite{32,49} from H1 collaboration and in \cite{33} from ZEUS collaboration. In the text \cite{32} from H1 collaboration inclusive neutral current $e^{\pm}p$ scattering cross section data collected during the years (2003-2007) is presented. The beam energies $E_p$ of corresponding H1 experiment run are $920$, $575$ and $460 \text{ GeV}^2$. Corresponding $F_L$ data of H1 is taken from \cite{49} where measurement are performed at c.m.s. energies $\sqrt{s}=$ 225 and 252 $\text{GeV}$. On the other hand, in \cite{33} from ZEUS collaboration reduced cross sections for $ep$ scattering for different c.m.s. energies viz. $318$, $251$ and $225\text{ GeV}$ is presented. The fixed target data from NMC \cite{31} and BCDMS \cite{34} collaboration which exist for $x>10^{-2}$ is also shown in the Figure \ref{7x}. Finally, we have included data from global parameterization groups viz. NNPDF3.1sx \cite{36} and PDF4LHC15 \cite{35} for comparison.

\begin{figure*}[tbp]
	\centering 
	\includegraphics[width=.32\textwidth,clip]{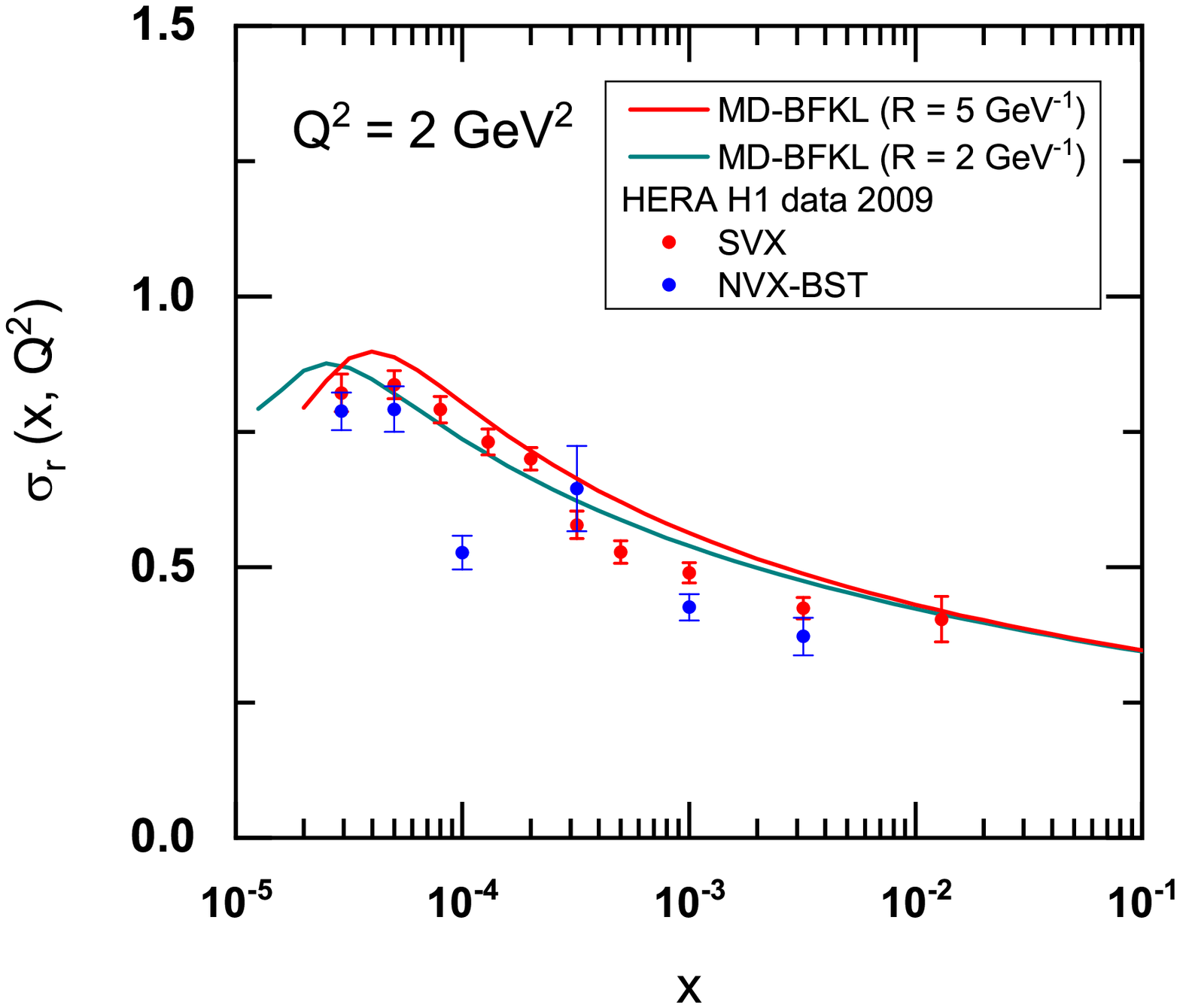}
	\includegraphics[width=.32\textwidth,clip]{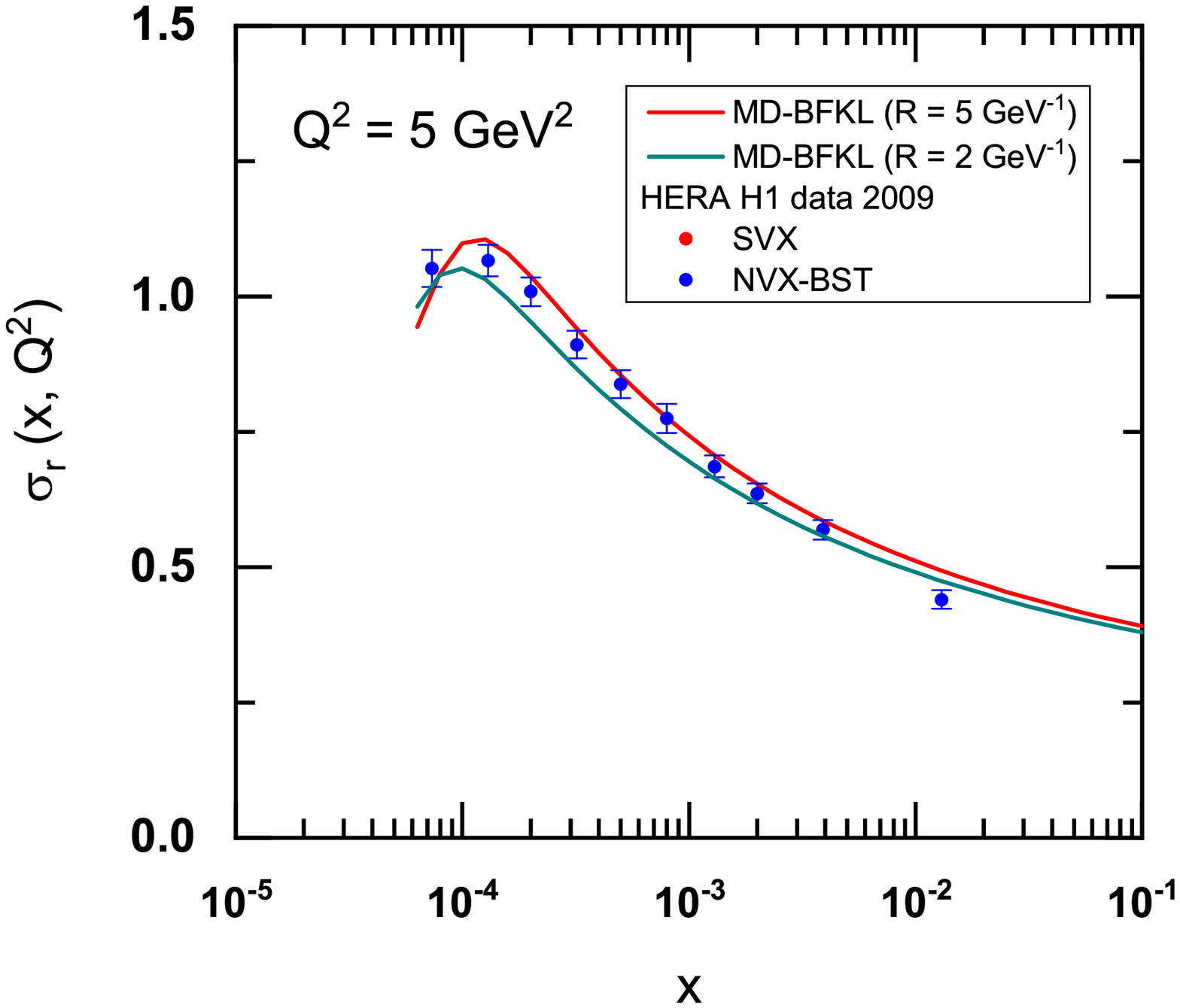}
	\includegraphics[width=.32\textwidth,clip]{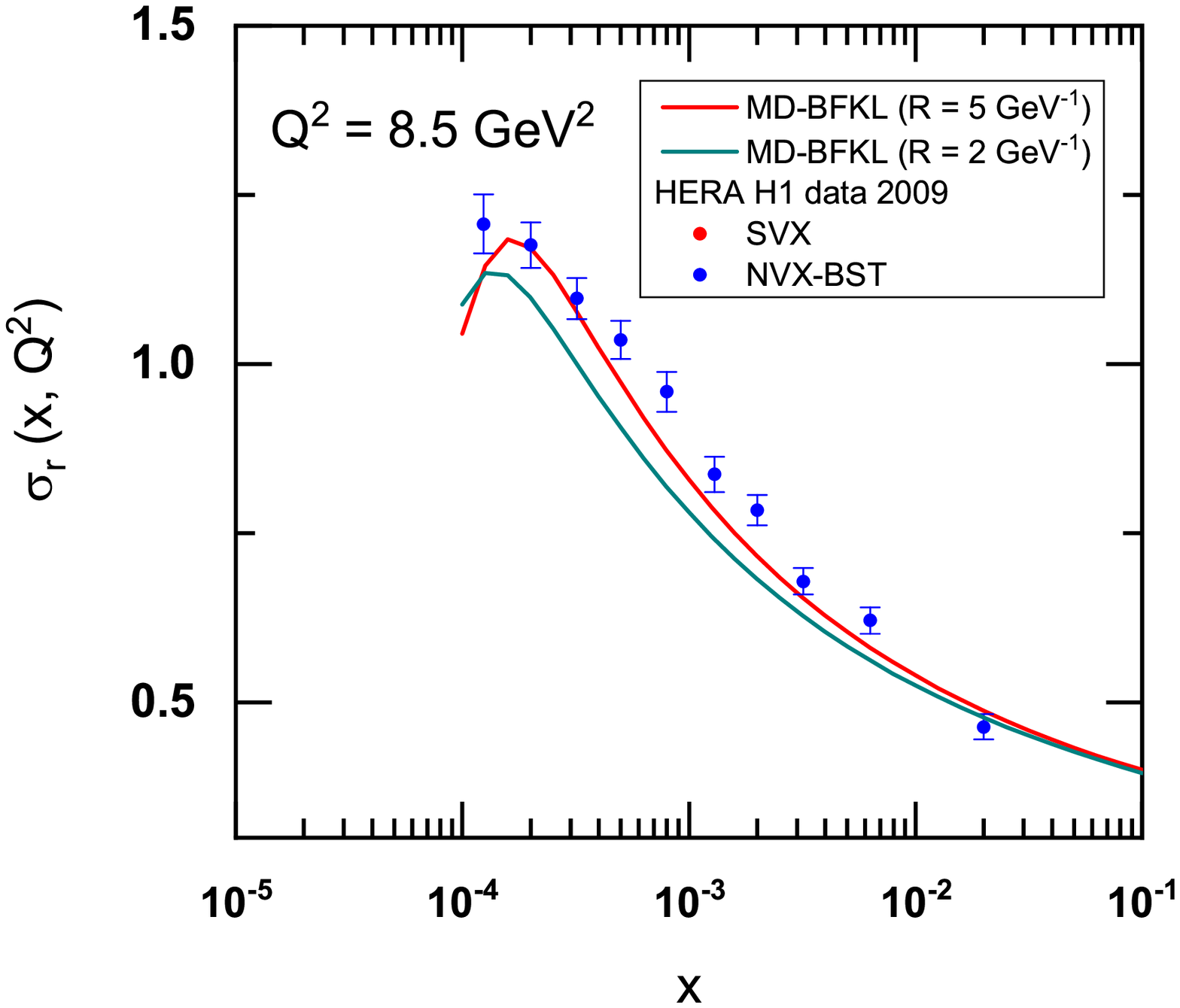}\\
	\includegraphics[width=.32\textwidth,clip]{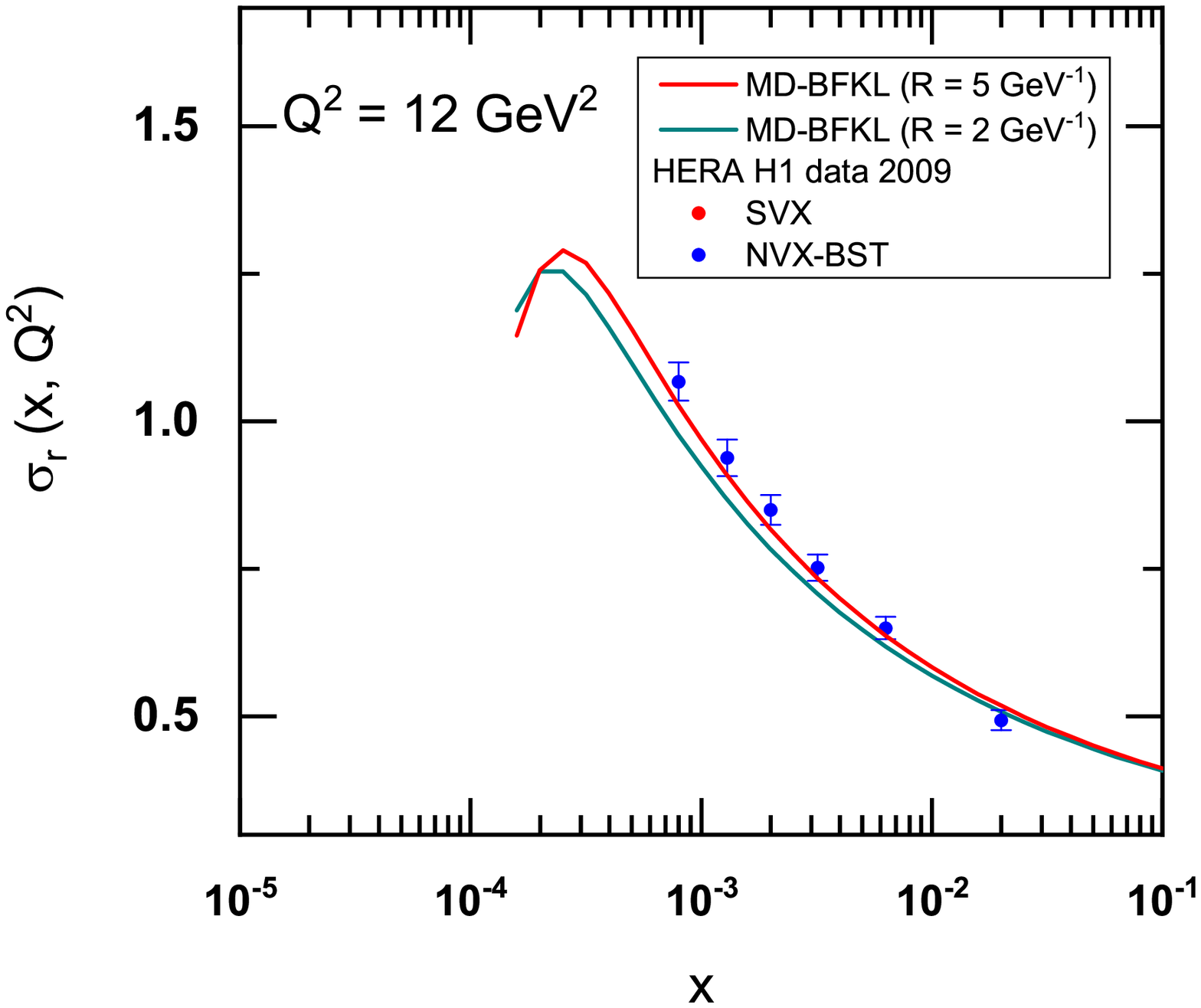}	
	\includegraphics[width=.32\textwidth,clip]{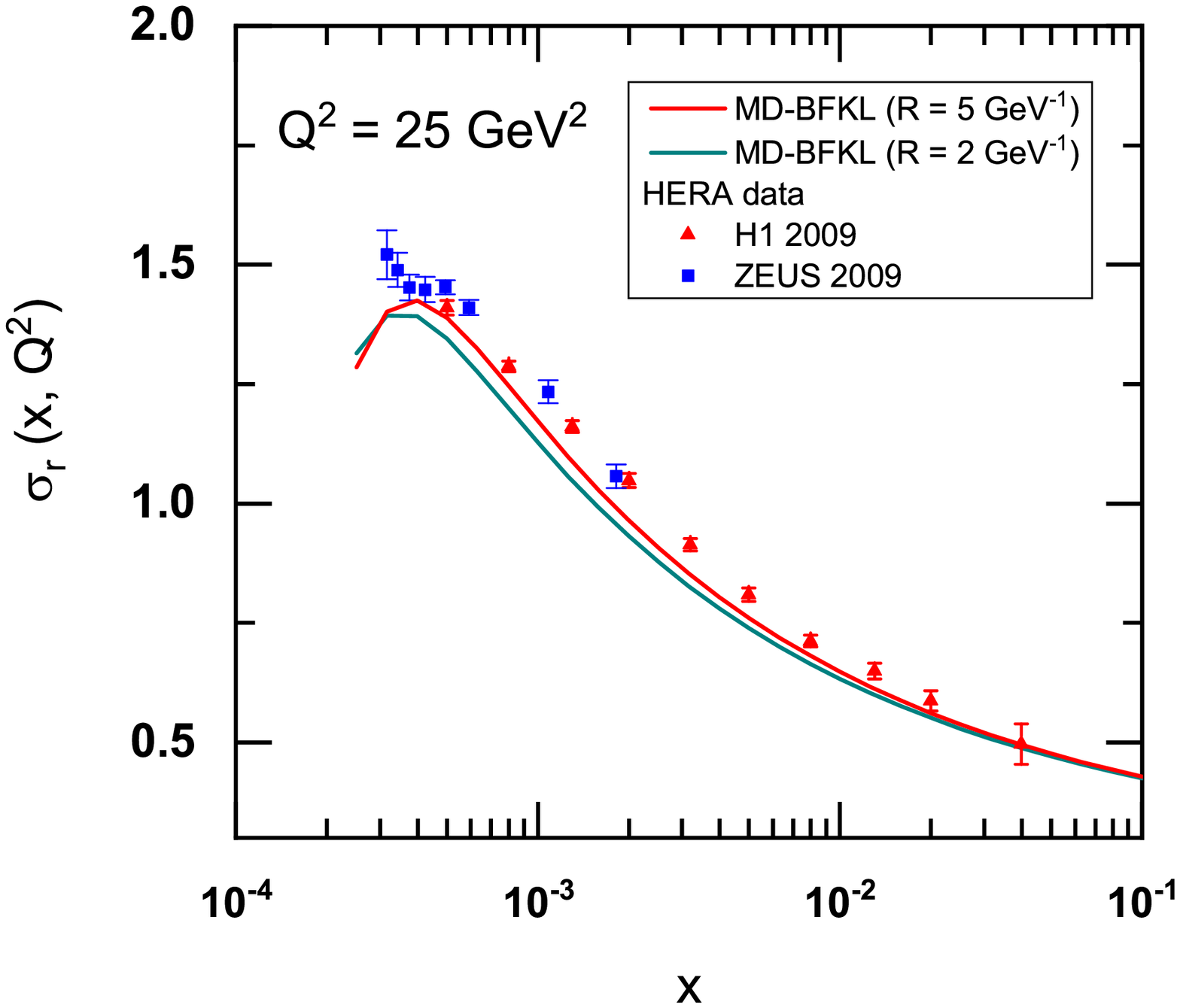}
	\includegraphics[width=.32\textwidth,clip]{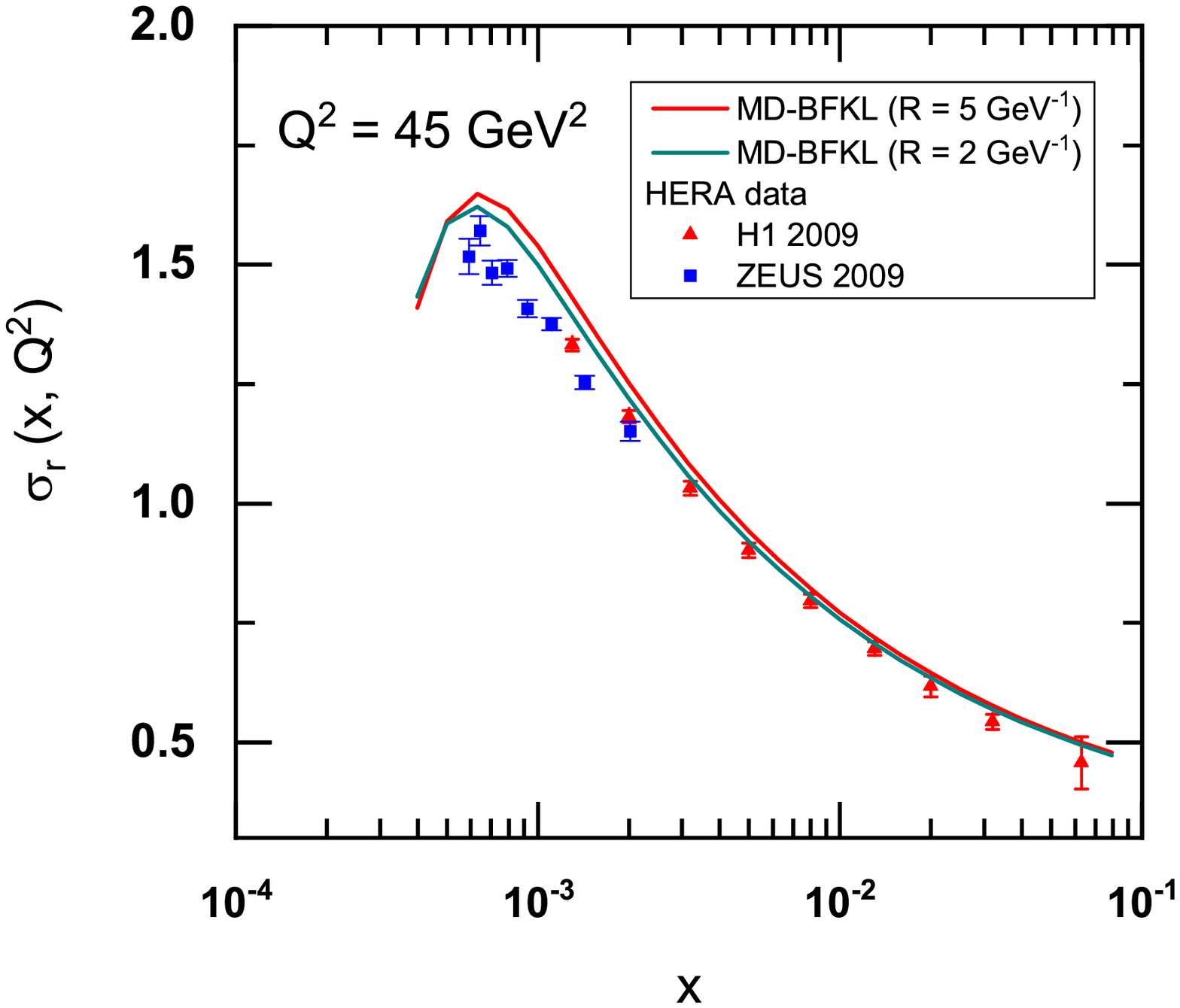}
	\caption{\label{8x} Theoretical prediction from KC improved MD-BFKL for reduced cross section $\sigma_r (x,Q^2)$. The data are from H1 \cite{40,83} ($\sqrt{s}\approx$ 318, 300 and 252 $\text{GeV}^2$) and ZEUS \cite{33} ($\sqrt{s}\approx$ 318 $\text{GeV}^2$). Theoretical calculations are for $\sqrt{s}= 318 \text{ GeV}^2$.}
\end{figure*}
\par The $x$ dependence of the structure functions $F_L$ and $F_2$ is shown in Fig.~\ref{7x} and Fig.~\ref{77x} for the four $Q^2$ values $5 \text{ GeV}^2$, $15 \text{ GeV}^2$, $35 \text{ GeV}^2$ and $45 \text{ GeV}^2$. The nonperturbative contribution, $F_{2/L}^{\text{npert.}}$ (or $F_{2/L}^{\text{BG}}$) given by \eqref{49} contrasted with total (nonpert. +pert.) contribution,  $F_{2/L}^{\text{npert.+pert.}}$. QCD prediction shows a satisfactory agreement with DIS data for both structure functions $F_2$ and $F_L$. It is clear that perturbative contribution is insignificant towards large $x$  ($\geq 10^{-2}$), while for small $x$ around $x\leq 10^{-3}$ this contribution becomes visibly important. Interestingly both data and theory for $F_2$ structure function seem to preserve the $x^{-\lambda}$ singular behavior rather than showing taming due to net shadowing effect in the asymptotic limit $x\rightarrow0$. It is difficult to observe the existence of any sizable shadowing effect for $x>10^{-3}$. However, for $x<10^{-3}$, recalling that the shadowing term is proportional to $1/R^2$ it is seen that for $R=2\text{ GeV}^{-1}$ (gluons in hotspots) the rise of structure function becomes slower than that of $R=5\text{ GeV}^{-1}$ ($R=R_H$, hadron radius) as expected. But even the extreme form of the gluon shadowing ($R=2\text{ GeV}^{-1}$) suppresses $F_2$ only by about $10\%$ or less at $10^{-3}$. It just depicts that shadowing has the negligible impact on structure functions even towards smaller $x$. This is because of the fact that in this low $x$ regime gluons are expected to drive the sea quark distribution via $g\rightarrow q\bar{q}$. Therefore, a similar sea quark contribution to the structure function $F_2$ in addition to the gluon contribution can be expected in low $x$ regime. On the other hand from Fig.~\ref{77x} it is clear that for $x>10^{-3}$, the size of the longitudinal structure function is negligibly small while for very small-$x$ regime ($x<10^{-3}$) , $F_L$  grows eventually. This is in accord with our expectation since measurement of $F_L$ directly probes the gluonic content of the proton which is dominant in small-$x$ regime.

\par The proton structure functions $F_2$ and $F_L$ are of complementary in nature. These are related to the $\gamma^{*}p$ cross sections of longitudinally and transversely polarized photons $\sigma_L$ and $\sigma_T$ as $F_L\propto\sigma_L$ and $F_2\propto (\sigma_L+\sigma_T)$.  Since $\sigma_L$ and $\sigma_T$ are positive, that imposes a restriction on $F_L$ and $F_2$ i.e. $0\leq F_L\leq F_2$. To circumvent the need of a proper relationship between the structure functions and  polarized cross sections to study $\gamma^*p$ cross section one can define a ratio between the structure functions or the equivalent cross section ratio
\begin{equation}
	\label{R}
	R=\frac{\sigma_L}{\sigma_T}=\frac{F_L}{F_2-F_L}.
\end{equation}
The advantage behind this formulation is that this is independent of any normalization factors.

\par In Fig \ref{Rx} a phenomenological comparison between data and theory is shown to illustrate $Q^2$ dependence of the ratio $R(x,Q^2)$. The HERA data taken for comparison can be found in \cite{49} where $ep$ cross section data measured for two center of mass energies $\sqrt{s}=225$ and $252\text{ GeV}$ is enlisted. From Fig.~\ref{Rx} it  is clear that the HERA data is in general agreement with the QCD expectation. Available H1 data for very low $Q^2$ $(Q^2\leq 5\text{ GeV})$ is also sketched in Fig.~\ref{Rx}. We have excluded the very low $Q^2$ region in our calculation of $R$ since we are bounded to stick in the perturbative region. However, a more realistic study in this transition region is performed in Sect.~\ref{hera2}. 

\par Now we try to predict reduced cross section $\sigma_r$ for $ep$ scattering process from the knowledge of structure functions that we have discussed above. The inclusive deep-inelastic differential cross section for $ep$ scattering can be represented in terms of three structure functions $F_2$, $F_L$ and $xF_3$. The structure functions have direct dependence with DIS kinematic variables, $x$ and $Q^2$ only, while the cross section has additional dependence with inelasticity $y=Q^2/s x$. The inclusive cross section for neutral current $ep$ scattering is given by
\begin{equation}
	\label{50}
	\begin{split}
		\frac{d^2 \sigma^{e^{\pm}p}}{dx dQ^2}=\frac{2\pi\alpha^2}{x Q^4}Y_{+}\left(F_2-\frac{y^2}{Y_+}F_L\mp\frac{Y_{-}}{Y_{+}}xF_3\right),
	\end{split}
\end{equation}
where $Y_{\pm}=1\pm(1-y)^2$. The cross section for exchanged virtual boson $(\text{Z} \text{ or } \gamma^{*})$ is related to $F_2$ and $F_3$ in which contribution from both longitudinal and transverse boson polarization state exists. On the other hand, only the longitudinal polarized virtual boson exchange processes contribute to $F_L$ which has a significant impact on higher order QCD though it vanishes at lowest order. At small momentum transfer $Q^2$ $(\text{i.e. } Q^2\ll M_Z^2\approx800\text{ GeV}^2)$, interaction of massless photon is dominant over the exchange of heavy Z boson. Thus, the contribution from Z boson exchange and the influence of the term $xF_3$ is negligible at low and moderate $Q^2$. Therefore, in this range of $Q^2$, in one photon exchange approximation, the differential cross section formula \eqref{50} can be written as 

\begin{equation}
	\label{51}
	\sigma_r\equiv\frac{d^2 \sigma^{e^{\pm}p}}{dx dQ^2}\frac{x Q^4}{2\pi\alpha^2}\frac{1}{Y_{+}}=\left(F_2-\frac{y^2}{Y_+}F_L\right),
\end{equation}
where $\sigma_r$ is the reduced cross section. Note that \eqref{51} is symmetric under charge exchange i.e. identical for both $e^{+}p$ and $e^{-}p$ processes unlike \eqref{50}. Additionally, it is independent  of incoming electron  helicity state. Thus, at low and moderate $Q^2$ $(\leq m_Z^2)$ which is our region of study, the knowledge of $F_2$ and $F_L$ is enough to predict the reduced cross section. We can also express reduced cross section in terms of the ratio $R(x,Q^2)$ defined in \eqref{R} replacing $F_L$ by $F_L=\frac{R}{1+R}F_2$ which yields
\begin{equation}
	\label{51x}
	\sigma_r=F_2(x,Q^2)\left[1-\frac{y^2}{Y_+}\frac{R}{1+R}\right].
\end{equation}

\par The $x$ dependence of $e^{\pm}p$ reduced cross section $\sigma_r$ calculated for center of mass energy $\sqrt{s}=318\text{ GeV}$ is shown in Fig.~\ref{8x}. Our theoretical expectation is compared with HERA H1 measurement \cite{40,83} and ZEUS  ($\sqrt{s}=318\text{ GeV}$) \cite{33}. The available H1 data for low $Q^2$ ($\leq12\text{ GeV}^2$) measured at $\sqrt{s}=$ 318 $\text{GeV}$ (SVX) and 300 $\text{GeV}$ (NVX-BST) is taken from \cite{40}, while the same for high $Q^2$ ($>12\text{ GeV}^2$) is taken from \cite{83}.
The cross section measurement of SVX is found to be slightly higher than that of NVX-BST as expected because of the increase in center of mass energy. Both theory and data agree well for our phenomenology range $Q^2\leq 100\text{ GeV}^2$. The distinction in $\sigma_r$ due to the two forms of shadowing $R=5\text{ GeV}^{-1}$ and $R=2\text{ GeV}^{-1}$ is more prominent for smaller $Q^2$ values. For each $Q^2$, starting at some high $x$ the reduced cross section first increases as $x\rightarrow0$ and then an abrupt fall in cross section can be observed in both theory and data at very small-$x$ region ($x<10^{-4}$) . For all $Q^2$, this region of $x$ corresponds to the highest inelasticity $y\approx0.65$ ($y=Q^2/sx$) and thus characteristic turn over of cross section at $y\approx0.65$ can be attributed to the influence of $F_L$.  In simple words, towards very small-$x$ (or high $y$) the monotonic rise of $F_2$ is suppressed by the contribution from longitudinal structure function $F_L$ thereby causing an overall fall in the cross section. For low inelasticity $y<0.65$, the contribution from the longitudinal structure function is small on the other hand, structure function $F_2$ exhibits a steady increase as $x\rightarrow 0$. Therefore, in the region where $x$ is not so small, the growth of the cross section is found to be power like as expected.

\subsection{Virtual photon-proton cross section prediction in transition region}\label{hera2}
\par Traditionally the photon-proton interaction is classified into two separate processes depending upon the photon virtuality $Q^2$ viz. photoproduction (at low $Q^2$) and deep inelastic scattering (at high $Q^2$). DIS is considered  as a basic tool for exploring pQCD where the point like virtual photon directly probes the partonic contents of the proton. On the other hand, photoproduction is completely nonperturbative phenomena defined in the limit of vanishing $Q^2$ where real (or quasi-real) photons interact with the proton more likely a hadron-hadron collision. The experiment at HERA collider provides a unique opportunity to study both the processes photoproduction and DIS on their respective kinematic domains. The $Q^2$ dependence of the proton structure functions are well described by pQCD over a wide range of $x$ and $Q^2$ \cite{37,38} in accordance with HERA data. However, for $Q^2\lesssim 2\text{ GeV}^2$ (photo production region) the pQCD breakdowns since the higher order contributions to the perturbative expansion becomes very large. In this region, data can be only described by non-perturbative phenomenological models \cite{39}. Our present study is especially focused on the transition region $(2\text{ GeV}^2\leq Q^2\leq10\text{ GeV}^2)$ from photoproduction to deep inelastic scattering. To measure the photon-proton cross section in the transition region two dedicated runs were performed in the years 1999 (Nominal vertex "NVX'99") and 2000 (Shifted Vertex "SVX'00") by H1 experiment at HERA. The published data can be found in \cite{40}.

\par Recall the neutral current $ep$ double differential cross section formula \eqref{51} defined in the region $Q^2\ll M_Z^2$. In this region massive boson $(M_Z)$ exchange is neglected and only one photon exchange is considered, thereby the role of incoming electron reduces to be a source of virtual photon interacting proton. Thus, we can recast the formula \eqref{51} in terms of photon-proton reaction. In fact the structure function $F_2$ and $F_L$ in \eqref{51} related to the longitudinally and transversely polarized photon-proton scattering cross sections $\sigma_L$ and $\sigma_T$ by the relations
\begin{align}
	\label{52}
	&F_L = \frac{Q^2}{4\pi^2\alpha}(1-x)\sigma_L,\\
	\label{53}
	&F_2 = \frac{Q^2}{4\pi\alpha}(1-x)(\sigma_L+\sigma_T),
\end{align}
which hold good at low $x$. Considering \eqref{52} and \eqref{53} the reduce cross section in \eqref{51} can be written as
\begin{equation}
	\label{54}
	\sigma_r = \frac{Q^2(1-x)}{4\pi^2\alpha}\sigma_{\gamma^*p}^{\text{eff}},
\end{equation}
where \begin{align}
	\label{55}
	\sigma_{\gamma^*p}^{\text{eff}}=\sigma_T+(1-\frac{y^2}{Y_+})\sigma_L
\end{align}
is the effective virtual photon-proton cross section. Note that the expression for  $\sigma_{\gamma^*p}^{\text{eff}}$ is similar to the total cross section $\sigma_{\gamma^*p}^{\text{tot}}$ which is linear combination of $\sigma_L$ and $\sigma_T$ i.e.  
$\sigma_{\gamma^*p}^{\text{tot}}=\sigma_L+\sigma_T$. In fact the total cross section $\sigma_{\gamma^{*}p}^{\text{tot}}$ and $\sigma_{\gamma^*p}^{\text{eff}}$ can be regarded as the same quantity at low inelasticity $y$ i.e. $\sigma_{\gamma^*p}^{\text{eff}} \xrightarrow{y\rightarrow0}\sigma_{\gamma^*p}^{\text{tot}}$ which differ only in the region of high y.

\begin{figure}[tbp]
	\centering
	\includegraphics[width=.4\textwidth,clip]{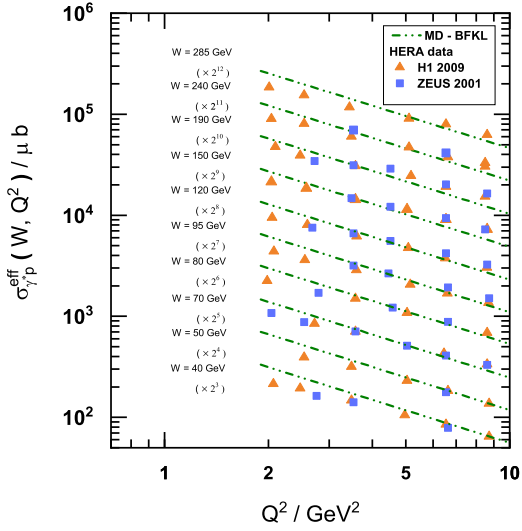}
	\caption{\label{7xx} QCD prediction from KC improved MD-BFKL for virtual photon-proton cross section $\sigma_{\gamma^{*}p}^\text{eff}$ as a function of $Q^2$ at different values of $W$. The cross section for different $W$ values are multiplied by factor multiple of 2 indicated in the figure. The included data are from  H1 \cite{40} ($\sqrt{s}= 318\text{ GeV}$) and ZEUS  \cite{81} ($\sqrt{s}= 300\text{ GeV}$). }	
\end{figure}
\par Figure \ref{7xx} shows the measurement of the virtual photon-proton cross section $\sigma_{\gamma^{*}p}^\text{eff}$ as a function of $Q^2$ corresponding to different values of $W$.  The total cross section $\sigma_{\gamma^*p}^{\text{tot}}$ is often expressed as a function of $Q^2$ and invariant mass $W$. The standard relation between $W$, $x$ and $Q^2$ is $W=\sqrt{Q^2(1-x)/x}$. Since $Q^2\approx syx$, for small-$x$ we can have an approximate relationship between $W$ and $y$ i.e. $W^2\simeq sy$ which we have used in our calculations. HERA measurement for $\sigma^{\text{eff}}_{\gamma^{*}p}$ from H1 \cite{40} ($\sqrt{s}= 318\text{ GeV}$) and ZEUS  \cite{81} ($\sqrt{s}= 300\text{ GeV}$) are included for comparison with our theoretical prediction. We have chosen  $\sqrt{s}= 318\text{ GeV}$ for our calculation analogous to H1 measurement. The precision of the data is such that their errors are hardly visible.  Both the HERA data and theoretically measured cross sections for different values of $W$ are multiplied by the factors multiple of 2 as indicated in the figure. For $Q^2\gtrsim3 \text{ GeV}^2$, an excellent agreement between the theory and HERA data can be observed for the wide range of $W$, while for $Q^2<3 \text{ GeV}^2$, the slop of the QCD prediction seems unsatisfactory. This indicates the inadequacy of perturbative QCD at very low $Q^2$ ($<3 \text{ GeV}^2$) and roughly provides a lower bound of $Q^2$ to our theory.

\section{Conclusion}\label{con}

\par In conclusion, we have presented a phenomenological study on the behavior of unintegrated gluon distribution at small-$x$ and moderate $k_T^2$ region particularly $10^{-6}\leq x\leq 10^{-2}$ and $2 \text{ GeV}^2\leq k_T^2\leq 1000 \text{ GeV}^2 $ which is also the accessible kinematic range to experiments performed at HERA $ep$ collider. In the beginning, we have improved the MD-BFKL equation supplementing so-called kinematic constraint on it. Then we have solved this unitarized BFKL equation analytically in order to study $x$ and $k_T^2$ dependence of unintegrated gluon distribution function. Our prediction of gluon distribution is contrasted with that of modified BK equation as well as global datasets NNPDF 3.1sx and CT 14. The $x$ evolution of gluon distribution shows the singular $x^{-\lambda}$ type behavior of gluon evolution tamed by shadowing correction. We found that for intense shadowing $(R= 2 \text{ GeV}^{-1})$, towards smaller x, certainly from $x=10^{-4}$, the gluon distribution emerges from rapid BFKL growth which indicates the dominance of gluon shadowing. While in case of conventional shadowing $(R= 5 \text{ GeV}^{-1}) $ an appreciable modification of BFKL behavior can only be seen from $x= 10^{-5}$. The $k_T^2$ dependence of unintegrated gluon distribution is also studied. Although obvious suppression due to net shadowing correction is seen in our studies, however no significant saturation phenomenon is observed. The reason is the nonlinear contribution term is suppressed by the factor $1/k_T^2$ at large values of $k_T^2$.

\par We have obtained a more general solution of KC improved MD-BFKL equation implementing a pre-defined input distribution on it. This has allowed us to visualize the gluon evolution in three dimensions $\mathbb{R}^3$ into $x\text{-}k_T^2$ phase space. We have also shown the sensitiveness of unintegrated gluon distribution towards $R$ and $\lambda$ using density plots in $x\text{-}k_T^2$ plane.

\par An important achievement of obtaining an analytical solution in this work is its implication on qualitative studies on geometrical scaling which is presented in Sect. \ref{critical}. Starting from a basic concept of differential geometry and knowledge of level curves we have managed to obtain an equation of the critical boundary which is supposed to separate low and high gluon density regions.
\par In Sect. \ref{hera} we have studied the small-$x$ dependence of the structure function $F_2$ and $F_L$ obtained via $k_T$-factorization formula $F_i=f\otimes F_i^{(0)}$. Here the unintegrated gluon distribution $f(x,k_T^2)$ is taken from our analytical solution of KC improved MD-BFKL equation setting boundary condition at $x_0=0.01$. Surprisingly the quantitative size of the shadowing correction to $F_2$ and $F_L$ is found to be very small and the structure functions seem to hold the singular $x^{-\lambda}$ behavior. Even at intense shadowing condition $(R= 2 \text{ GeV}^{-1})$ the $F_2$ structure function is found to be suppressed only by $10\%$. In addition we have measured $e^{\pm}p$ reduced cross section as well as equivalent $\gamma^*p$ longitudinal to transverse cross section ratio $R(x,Q^2)$ from the knowledge of $F_2$ and $F_L$. Our results are compared with recent high precision HERA measurements. The comparison reveals a good agreement between our theory and DIS data.
\par In Sect. \ref{hera2} we have examined the virtual photon-proton effective cross section, particularly in the transition region from photoproduction to deep inelastic scattering. The quantity $\sigma_{\gamma^{*}p}^\text{eff}$ serves the role of the total cross section if small inelasticity $y$ is concerned. We have compared our predicted $\sigma_{\gamma^{*}p}^\text{eff}$ with the HERA data of two dedicated runs "NVX'99" and "SVX'00" by H1 as well as ZEUS for the transition region. Our theoretical prediction shows well-consistency with HERA data particularly upto $Q^2\sim3 \text{ GeV}^2$ in the transition region.

\par In summary, there are several attractive features of our present study. First, we have able to predict a wide range of physical quantities, starting right from our solution for unintegrated gluon density. Secondly, all analysis is performed in terms of relatively small numbers of parameters. Moreover, the idea developed in this work for studying geometrical scaling can be implemented in any other framework. Finally, a very significant feature of this analysis is that we have considered two extreme possibilities of shadowing which can be distinguished by experimental data. In the end, we conclude that, as per feasibility towards HERA DIS data is concerned, the KC improved MD-BFKL equation could be a reliable framework for exploring high energy physics over a wide range of $x$ and $k_T^2$ which is also relevant for LHC probe and future collider phenomenology.

\section*{Acknowledgments}
We thank Prof.~Johannes Bluemlein, DESY (Hamburg, Germany) and Prof.~Dieter Schildknecht, Bielefeld university (Germany)  for their valuable comments. Two of us ( P.P. and M.L.) acknowledge Department of Science and Technology (DST), India (grant DST/INSPIRE Fellowship /2017 /IF160770) and Council of Scientific and Industrial Research (CSIR), New Delhi (grant 09/796(0064)2016-EMR-I) respectively for the financial assistantship.

\bibliographystyle{spphys}
\bibliography{111.bib}

\end{document}